\pdfoutput=1
\documentclass{aa}
\usepackage{graphicx}
\usepackage{natbib}
\usepackage{bm}
\usepackage{color}
\graphicspath{{./fig/}{./png/}}
\newcommand{\EQ}{\begin{equation}}
\newcommand{\EE}{\end{equation}}
\newcommand{\EQA}{\begin{eqnarray}}
\newcommand{\EEA}{\end{eqnarray}}
\newcommand{\brac}[1]{\langle #1 \rangle}
\newcommand{\pd}{\partial}

\newcommand{\ve}[1]{\boldsymbol{#1}}
\newcommand{\mean}[1]{\overline{#1}}
\newcommand{\meanv}[1]{\overline{\bm #1}}

\newcommand{\nut}{\nu_{\rm t}}

\newcommand{\urms}{u_{\rm rms}}

\newcommand{\Rs}{R_{\odot}}

\newcommand{\kef}{k_{\rm f}}

\newcommand{\chiSGS}{\chi_{\rm SGS}}
\newcommand{\chiSGSm}{\mean\chi_{\rm SGS}}

\newcommand{\Ta}{{\rm Ta}}

\newcommand{\Pm}{{\rm Pm}}
\newcommand{\Rm}{{\rm Rm}}
\newcommand{\Pra}{{\rm Pr}}

\newcommand{\Ra}{{\rm Ra}}

\newcommand{\Roc}{{\rm Ro}_{\rm c}}
\newcommand{\Rey}{{\rm Re}}
\newcommand{\Co}{{\rm Co}}

\newcommand{\qij}{Q_{ij}}

\newcommand{\qrp}{Q_{r\phi}}
\newcommand{\qtp}{Q_{\theta\phi}}
\newcommand{\mrp}{M_{r\phi}}
\newcommand{\mtp}{M_{\theta\phi}}

\newcommand{\LamV}{\Lambda_{\rm V}}
\newcommand{\LamH}{\Lambda_{\rm H}}
\newcommand{\Mij}{M_{ij}}

{}
\def\onethird{{\textstyle{1\over3}}}
\def\onehalf{{\textstyle{1\over2}}}

\def\blue{\textcolor{black}}

\newcommand{\Fig}[1]{Fig.~\ref{#1}}
\newcommand{\Eq}[1]{Eq.~(\ref{#1})}
\newcommand{\Eqs}[2]{Eqs.~(\ref{#1}) and~(\ref{#2})}

\newcommand{\s}{\,{\rm s}}
\newcommand{\m}{\,{\rm m}}
\begin{document}
\authorrunning{Karak et al.}
\titlerunning{Magnetically controlled stellar differential rotation}
   \title{Magnetically controlled stellar differential rotation near the transition from solar to anti-solar profiles}

   \author{B. B. Karak
          \inst{1}
          \and
          P. J. K\"apyl\"a
          \inst{2,3}
           \and
          M. J. K\"apyl\"a
          \inst{3}
           \and
          A. Brandenburg
          \inst{1,4}
           \and
          N. Olspert
          \inst{3}
           \and
          J. Pelt
          \inst{5}
          }

   \offprints{\email{bbkarak@nordita.org}
              }

   \institute{NORDITA, KTH Royal Institute of Technology and Stockholm University, Roslagstullsbacken 23, SE-10691 Stockholm, Sweden
         \and Department of Physics, Gustaf H\"allstr\"omin katu 2a (PO Box 64), FI-00014 University of Helsinki, Finland
         \and ReSoLVE Centre of Excellence, Department of Information and Computer Science, Aalto University, PO Box 15400, FI-00076 Aalto, Finland
         \and Department of Astronomy, Stockholm University, SE-10691
              Stockholm, Sweden
         \and Tartu Observatory, 61602 T\~{o}ravere, Estonia}

   \date{Received ? / Accepted ?, $ $Revision: 1.361 $ $}

   \abstract{
Late-type stars rotate differentially owing to anisotropic turbulence
in their outer convection zones.
The rotation is called solar-like (SL) when the equator rotates fastest and
anti-solar (AS) otherwise.
Hydrodynamic simulations show a transition from SL to AS rotation
as the influence of rotation on convection is reduced, but the opposite
transition occurs at a different point in the parameter space.
The system is bistable, i.e., SL and AS rotation profiles can both be stable.
   }{
We study the effect of a dynamo-generated magnetic field
on the large-scale flows, particularly on the possibility of bistable 
behavior of differential rotation.
   }{
We solve the hydromagnetic equations numerically in a rotating spherical
shell that typically covers $\pm75^\circ$ latitude (wedge geometry)
for a set of different
radiative conductivities controlling the relative importance of convection.
We analyze the resulting differential rotation, meridional circulation,
and magnetic field and compare the corresponding modifications of the
Reynolds and Maxwell stresses.
   }{
In agreement with earlier findings, our models display
SL rotation profiles when the rotational influence on convection
is strong and a transition to AS when the rotational influence decreases.
We find that dynamo-generated magnetic fields help to produce
SL differential rotation compared to the hydrodynamic simulations. 
We do not observe any bistable states of differential rotation.
In the AS cases we get coherent single-cell meridional circulation, 
whereas in SL cases we get multi-cellular patterns.
In both cases, we obtain poleward circulation near the surface
with a magnitude close to that observed in the Sun.
In the slowly rotating cases we find \blue{activity} cycles, but
no clear polarity reversals, whereas in the more rapidly rotating
cases irregular \blue{variations} are obtained.
Moreover, both differential rotation and meridional circulation have
significant \blue{temporal} variations that are similar
in strength to those of the Sun.
   }{
Purely hydrodynamic simulations of differential rotation and meridional
circulation are shown to be
of limited relevance as magnetic fields, self-consistently generated
by dynamo action, significantly affect the flows.
     }

   \keywords{   convection --
                turbulence --
                Sun: magnetic fields
                Sun: rotation --
                stars: rotation
               }

   \maketitle

\section{Introduction}

Differential rotation is an important ingredient for the generation of
stellar magnetic fields.
The internal rotation rate of the Sun has been mapped by
helioseismology, revealing that
the angular velocity within the convection zone mildly increases (decreases)
as a function of radius at low (high) latitudes and that the radial
shear is concentrated in shallow layers at the base of the convection
zone and near the surface \cite[e.g.,][]{Brownea89,Schouea98,TCDMT03}.
Thus the solar equator rotates faster than its poles.
This kind of rotation profile is called {\it solar-like} (SL)
differential rotation.
The opposite case where the equator rotates slower than the poles,
is referred to as {\it anti-solar} (AS) differential rotation.
Due to the difficulties in observing
slowly rotating stars that might possess AS differential rotation, it
is not clear how common it is in main-sequence stars. However, it has
been observed in some K giants \cite[e.g.,][]{SKW03,WSW05,KKKea14}.

Historically, the differential rotation and magnetic fields of the Sun and other stars
have been modeled by two approaches -- mean-field models and global convection simulations.
In the mean-field approach, small-scale turbulence is
parameterized by expressing {\it the Reynolds stress} in the momentum equation in terms of the
mean velocity, {\it the turbulent electromotive force} in the
induction equation in terms of the mean magnetic field, and
{\it the turbulent heat flux} in the entropy equation in terms of the mean
entropy.
These parameterizations involve turbulent transport coefficients that
need to be calculated for highly turbulent flows of stellar
interiors. Analytical approaches, such as first-order smoothing,
involve approximations that are ill-suited for stellar conditions and
may yield inaccurate results.
A numerical method for determining the turbulent transport coefficients relevant
for the electromotive force is the test-field method \citep{SRSRC05,SRSRC07}, but for
angular momentum or heat transport no similar methods have been developed yet.
This means that the turbulent
transport coefficients used in mean-field models are often based on
educated guesses or they are even used as free parameters.
Despite these shortcomings, hydrodynamical mean-field models are
capable of producing SL differential rotation
\citep{BMT92,KR95,Re05,KO11} as well as basic properties of the rotation in
some other stars \citep{KR11,KO11,HY11}. However, obtaining AS differential
rotation is less straightforward for mean-field models
\citep[e.g.,][]{KR04}.
At the same time, mean-field dynamo models also reproduce some features of solar and 
stellar magnetic cycles either by including turbulent inductive
effects \citep[e.g.,][]{KKT06,PK11}, or by applying the Babcock-Leighton
process in the so-called flux transport dynamo models
\citep[e.g.,][]{CSD95,DC99,Ka10,KKC14,MD14}.

On the other hand, there have been some successes in modelling the differential rotation and magnetic fields 
using global convection simulations, mainly in recent years
\citep{MBT06,GCS10,RCGBS11,KMB12,KMCWB13,ABMT13,WKMB13}.
However, due to the extreme parameter regimes of the Sun,
realistic simulations are not possible at present.
Nevertheless, the simulations are able to reproduce solar values of the
Coriolis number $\Co$, which measures the relative importance of rotation
and turbulent convection.
For certain values of $\Co$, but with different values for other
parameters such as the fluid and magnetic Reynolds and Prandtl
numbers, simulations
occasionally produce AS differential rotation
\citep[e.g.,][]{MCBB11,KMB14}, poleward migration of the large-scale
magnetic fields \citep{Gi83,KKBMT10,NBBMT13}, no clear magnetic cycles
\citep{BBBMT10}, or sometimes even no appreciable large-scale
contribution to the magnetic field \citep{BMT04}.

According to mean-field hydrodynamics, differential rotation is generated
from the anisotropy of the Reynolds stress which is parameterized in terms
of the so-called $\Lambda$-effect \citep{R80,R89}.
A radially increasing (SL) angular velocity results if horizontal turbulent
velocities dominate over vertical ones, while AS rotation follows if radial
motions (even laminar ones) are dominant.
The importance of the $\Lambda$-effect depends on the rotational
influence on the turbulence, i.e., the value of $\Co$,
which is the ratio of the convective turnover time to the rotation period. 
At large $\Co$, the SL rotation is more favorable and 
the transition from SL to AS rotation depends on the Coriolis number 
\citep{BP09,Chan10,KMB11,KMGBC11,GSKM13,GWA13,GYMRW14}.
Using Boussinesq convection, \cite{GYMRW14} discovered that near the transition from AS to SL rotation,
both states are possible, depending on the initial conditions of the simulations.
This has been independently verified by \cite{KMB14} in fully compressible convection simulations. 
If this discovery were to apply to the Sun, this might
have important consequences, because young rapidly rotating stars,
which preferably possess SL rotation, 
slowly spin down due to loss of the angular momentum
and can persist in the SL rotation state
even when their rotation is slow.
Some doubts have already been expressed by \cite{FF14}, who found that the
bistability disappears when magnetic fields are present.

In the magnetohydrodynamic case the situation is more complicated than
in the case of pure hydrodynamics.
A dynamically significant magnetic field, which possibly varies
cyclically, introduces extra time dependent effects into the system,
capable of influencing the fluid flow both through large-scale effect
\cite[Malkus-Proctor effect; see][]{MP75,BMT92} and small-scale effects
\cite[Reynolds and Maxwell stresses and therefore the $\Lambda$-effect;
see][]{KRK94}.
Therefore, the magnetic field tries to destabilize the equilibrium
states of the rotation.
To explore to what extent the presence of a dynamically significant
large-scale magnetic field affects the bistable nature of the differential
rotation we perform several simulations with the same setup as in
\cite{KMB14}, but including magnetic fields.  Similar to their work,
we perform two types of simulations.  In one of them we run the
simulations from scratch, i.e., with an initially rigid rotation
profile.  Then we take either a SL state or an AS state and vary the
rotational influence by varying the radiative heat conductivity to
identify the transition.
We analyse the \blue{activity cycles} using diagnostic tools of stellar
activity in Sect.~\ref{sec:Diagnostic}. 
Next we measure the temporal variations of the Lorentz forces 
(both from large-scale and small-scale contributions) 
to understand the temporal variations of the large-scale
flows observed in the simulations (Sect.~\ref{sec:modulation}).
Finally, we compute the contributions of Reynolds stress, 
Maxwell stress, and the stresses from the azimuthally averaged
mean flow and mean magnetic field to the angular
momentum balance (Sect.~\ref{sec:stress}).
Then we study the influence of the magnetic field on the angular
momentum transport by comparing the results with the hydrodynamic simulations.

\section{The Model} \label{sect:model}

\subsection{Basic equations}
\label{sec:basic}

Our model is similar to many earlier studies \citep{KMB12,KMCWB13,CKMB14}.
The hydrodynamic part of this model has been used in \cite{KMB14}.
We model a spherical wedge with 
radial, latitudinal, and longitudinal extents $r_0
\leq r \leq r_1$, $\theta_0 \leq \theta \leq \pi-\theta_0$, and $0
\leq \phi \leq \phi_0$, respectively. Here, $r_0=0.72\,R_\odot$ and
$r_1=0.97\,R_\odot$ are the positions of the bottom and top of the
computational domain, $R_\odot$ is the radius of the Sun,
$\theta_0=\pi/12$ is the colatitude of the polar cap,
and $\phi_0=\pi/2$ is the longitudinal extent.
The following hydromagnetic equations are solved.
\begin{equation}
\frac{\pd \bm A}{\pd t} = {\bm u}\times{\bm B} - \mu_0\eta {\bm J},
\end{equation}
\begin{equation}
\frac{D \ln \rho}{Dt} = -\bm\nabla\cdot\bm{u},
\end{equation}
\begin{equation}
\frac{D\bm{u}}{Dt} = \bm{g} -2\bm\Omega_0\times\bm{u}+\frac{1}{\rho}
\left(\bm{J}\times\bm{B}+\bm\nabla \cdot 2\nu\rho\bm{\mathsf{S}}-\bm\nabla p\right),
\end{equation}
\begin{equation}
T\frac{D s}{Dt} = -\frac{1}{\rho}\bm\nabla \cdot
\left({\bm F^{\rm rad}}+ {\bm F^{\rm SGS}}\right) +2\nu \bm{\mathsf{S}}^2 + \frac{\eta\mu_0}{\rho}\bm{J}^2,
\label{equ:ss}
\end{equation}
where ${\bm A}$ is the magnetic vector potential,
${\bm B} =\bm\nabla\times{\bm A}$ is the magnetic field,
${\bm J} =\bm\nabla\times{\bm B}/\mu_0$ is the current density
with $\mu_0$ being the vacuum permeability,
$\bm{u}$ is the velocity,
$D/Dt = \pd/\pd t + \bm{u} \cdot \bm\nabla$ is the advective derivative,
$\rho$ is the density,
$s$ is the specific entropy,
$T$ is the temperature, $p$ is the pressure,
$\nu$ is the constant kinematic viscosity,
$\bm{g}=-GM_\odot\bm{r}/r^3$ is the gravitational acceleration
with $M_\odot$ being the mass of the Sun,
$\bm\Omega_0=(\cos\theta,-\sin\theta,0)\Omega_0$ is the angular velocity vector,
$\mathsf{S}_{ij}=\onehalf(u_{i;j}+u_{j;i})
-\onethird \delta_{ij}\bm\nabla\cdot\bm{u}$
is the rate of strain tensor,
where the semicolons denote covariant differentiation.
The radiative and subgrid scale (SGS) heat fluxes are given by
\begin{equation}
{\bm F^{\rm rad}}=-K\ve{\nabla} T\quad\mbox{and}\quad
{\bm F^{\rm SGS}} =-\chiSGS \rho  T\ve{\nabla} s,
\end{equation}
respectively. 
Here $K$ is the radiative heat conductivity and $\chiSGS$ is the turbulent heat
diffusivity.
The latter represents the unresolved
convective transport of heat \citep{KMCWB13}.
The fluid obeys the ideal gas law $p=(\gamma-1)\rho e$, where
$\gamma=c_{\rm P}/c_{\rm V} =5/3$ is the ratio of specific heats at
constant pressure and volume, and $e = c_{\rm V} T$
is the specific internal energy.

\subsection{Initial and boundary conditions}
\label{sec:initcond}

The initial hydrostatic state is isentropic, so the temperature is given by
\begin{equation}
\frac{\pd T}{\pd r}=-\frac{GM_\odot/r^2}{c_{\rm V}(\gamma-1)(n_{\rm ad}+1)},
\end{equation}
where $n_{\rm ad}=1.5$ is the polytropic index
and the value of $\pd T/\pd r$ at $r=r_0$ is fixed. The
density stratification follows from hydrostatic equilibrium.
The initial state chosen is not in thermodynamic equilibrium but closer to the
final convecting state to reduce the needed computational time
to reach a thermally relaxed state.
The heat conductivity profile is chosen such that
radiative diffusion is responsible for supplying the energy flux into
the system. Radiative diffusion becomes progressively less efficient
toward the surface \citep{KMB11}.
As in \cite{KMCWB13,KMB14},
this is achieved by taking a depth-dependent polytropic index
$n(r)=\delta n (r/r_0)^{-15}+n_{\rm ad}- \delta n$
for the radiative conductivity $K(r)=K_0[n(r)+1]$, where the
reference conductivity is $K_0=(\mathcal{L}/4\pi)c_{\rm
  V}(\gamma-1)(n_{\rm ad}+1)\rho_0\sqrt{GM_\odot R_\odot}$, with
$\mathcal{L}$ being the non-dimensional luminosity.
Note that $n=n_{\rm ad}$ at the bottom and $n \to$ $n_{\rm ad} -\delta n$ towards the surface.
Hence, $K$ decreases toward the surface like
$r^{-15}$ such that the value of $\delta n$ regulates the flux that is
carried by convection \citep{BCNS05,KMB14}.

Along with the imposed energy flux at the bottom boundary 
$F_{\rm b}=-(K \pd T/\pd r)_{r=r_0}$, the values of $\Omega_0$, $\nu$, $\eta$, and
$\chiSGSm=\chiSGS$ at the middle of the convection zone $r=r_{\rm m}=0.845\, \Rs$ are defined.
The turbulent heat conductivity $\chiSGS$ is piecewise constant above
$r>0.75\Rs$ with $\chiSGS=\chiSGSm$ in $0.75\Rs < r <0.95\Rs$, and
$\chiSGS=1.35\chiSGSm$ at $r \ge 0.95\Rs$. At $r < 0.75\Rs$, $\chiSGS$ tends
smoothly to zero \cite[see Fig.~1 of][]{KMB11}.
We fix $\chiSGS$ in such a way that at $r=r_1$ it corresponds to
$5\times10^8\m^2\s^{-1}$ in physical units. We also assume the density
and temperature at $r=r_0$ to have solar values, $\rho_0=200$ kg~m$^{-3}$ 
and $T_0=2.23\times10^6$~K.

Radial and latitudinal boundaries are assumed to be impenetrable and
stress free for the flow, whereas for the magnetic field we assume
a radial field condition on the outer radial boundary and perfect conductor
conditions on the lower radial and
latitudinal boundaries; see \cite{KMCWB13} for details.
Density and entropy have vanishing first derivatives
on the latitudinal boundaries.
A black body condition with
$\sigma T^4  = -K\nabla_r T - \chiSGS \rho T \nabla_r s,$
where $\sigma$ is related to the Stefan--Boltzmann constant, is
applied on the upper radial boundary.
However the value of $\sigma$ is modified to attain the desired values
of surface temperature and energy flux.
Moreover, we choose $\sigma$ in such a way
that in the initial non-convecting
state the flux at the surface carries the total luminosity through the boundary.
We use small-scale low amplitude Gaussian noise as initial condition
for the velocity and magnetic fields.

As discussed by \cite{KMCWB13,KMB14}, in
our fully compressible simulation the time step is severely limited if we
were to use the solar luminosity. This would imply both huge Rayleigh and
small Mach numbers.
This problem is avoided by taking an about $10^6$ times higher
luminosity in the simulation than in the Sun. However, the convective
velocity $u$ becomes then in our simulation 100 times larger than in
the Sun because the convective energy flux $F_{\rm conv}$ scales as $\rho u^3$.
Therefore to achieve the same rotational influence on the flow as in the Sun,
we need to increase $\Omega$ by the same factor. Consequently, we have the 
relations:
$\Omega_{\rm sim}=(L_0/L_\odot)^{1/3} \Omega_\odot \quad {\rm and}  \quad {\bm u}_{\rm sim}=(L_0/L_\odot)^{1/3}{\bm u}_\odot,$
where $L_0$ is the luminosity in simulation, $L_\odot\approx3.84\times10^{26}$~W 
is the solar luminosity, and
$\Omega_\odot$ is the average solar rotation rate $\approx2.7\times10^{-6}$~s$^{-1}$. 
This allows us to quote values for angular velocity, meridional
circulation, and magnetic field in physical units that can be
compared with solar values and with results of other groups.
However, we often quote ratios between different quantities that are
obviously non-dimensional and therefore not affected.
All computations are performed with the
{\sc Pencil Code.}\footnote{http://pencil-code.google.com}

\subsection{Dimensionless parameters and diagnostics}
\label{sec:para}

First, we define the non-dimensional input parameters.
The luminosity parameter $\mathcal{L}$ and the normalized pressure scale
height at the surface $\xi$ are given by
\begin{equation}
\mathcal{L} = \frac{L_0}{\rho_0 (GM_\odot)^{3/2} R_\odot^{1/2}}, \quad \xi = \frac{(\gamma-1) c_{\rm V}T_1}{GM_\odot/R_\odot},
\label{eq:lumino}
\end{equation}
where $T_1$ is the temperature at the surface.
The influence of rotation is measured by the Taylor number,
\begin{equation}
{\Ta=(2\Omega_0 \Delta r^2/\nu)^2,}
\end{equation}
where $\Delta r=r_1-r_0$ is the thickness of the convecting shell.
The fluid, magnetic, and SGS Prandtl numbers are defined as
\begin{equation}
\Pra=\frac{\nu}{\chi_{\rm m}},\quad \Pm=\frac{\nu}{\eta},\quad \Pra_{\rm SGS}=\frac{\nu}{\chiSGSm},
\end{equation}
respectively, where $\chi_{\rm m}=K(r_{\rm m})/c_{\rm P} \rho_{\rm m}$
is the thermal diffusivity and $\rho_{\rm m}$ is the density,
both evaluated $r=r_{\rm m}$.
Furthermore, we define the non-dimensional viscosity,
\begin{equation}
\tilde{\nu}=\frac{\nu}{\sqrt{GM_\odot R_\odot}},
\end{equation}
and the Rayleigh and convective Rossby numbers \citep{Gi77},
\begin{eqnarray}
\Ra\!=\!\frac{GM_\odot(\Delta r)^4}{\nu \chiSGSm R_\odot^2} \bigg(\!-\frac{1}{c_{\rm P}}\frac{{\rm d} s}{{\rm d}r}\! \bigg)_{r_{\rm m}}\!\!, \ \ \Roc = \left( \frac{\Ra}{\Pra_{\rm SGS} \Ta}  \right)^{1/2}\!\!,\;\;\;
\end{eqnarray}
where the entropy gradient of the non-convecting hydrostatic solution
is evaluated in the middle of the convection zone, $r=r_{\rm m}$.
We also quote the initial density contrast
$\Gamma_\rho^{(0)}\equiv\rho(r_0)/\rho(r_1)$.

As diagnostic quantities we define the fluid and magnetic Reynolds
numbers, and the Coriolis number as
\begin{equation}
\Rey=\frac{\urms}{\nu \kef},\quad \Rm=\frac{\urms}{\eta \kef}=\Pm \Rey,\quad \Co=\frac{2\Omega_0}{\urms \kef},
\label{eq:Coriolis}
\end{equation}
where $\urms=\sqrt{(3/2)\brac{u_r^2+u_\theta^2}_{r\theta\phi t}}$ is
the volume and time averaged rms velocity during the time
when the simulation is thermally relaxed.
We exclude $u_\phi$ from $\urms$ as it is dominated by
the differential rotation, and use $\kef=2\pi/\Delta r$ as an estimate of the
wavenumber of the largest eddies.
The Taylor number can also be written as $\Ta=\Co^2\Rey^2(\kef R_\odot)^4$.

We define mean values as averages over longitude and time and
denote these by an overbar. Sometimes we also perform additional
averaging over latitude and/or radius which we always mention
explicitly.

\begin{table*}[t!]
\centering
\caption[]{Summary of the runs.}
      \label{tab:runs}
      \vspace{-0.5cm}
     $$
         \begin{array}{p{0.05\linewidth}ccccccrrccccccl}
           \hline
           \noalign{\smallskip}
Run & \Ra  & \Pra & \delta n  & \Roc & \Rey & \Co & \Delta_\Omega^{(\theta)} & \Delta_\Omega^{(r)} & \tilde{E}_{\rm kin}[10^{-7}] &
E_{\rm mer}/E_{\rm kin} & E_{\rm rot}/E_{\rm kin} & \tilde{L}_{\rm rad} & \tilde{L}_{\rm conv} & {\rm DR} & {\rm \blue{ activity~cycle}}\\ \hline
A  & 3.93\cdot10^5 & 39.9 & 2.5 &0.73& 33 & 1.34 & -0.184& -0.185 & 0.25 &3.77\times10^{-3} & 0.173 & 0.09 & 0.95 & {\rm AS} & {\rm \blue{fairly}~regular}\\ 

B  & 3.54\cdot10^5 & 20.3 & 2.25 &0.69& 32 & 1.35 & -0.158&-0.143 & 0.24 &2.87\times10^{-3} & 0.134 & 0.19 & 0.84 & {\rm AS}& {\rm \blue{fairly}~regular} \\ 

BC  & 3.54\cdot10^5 & 15.6 & 2.1 &0.67& 32 & 1.38 & -0.151&-0.134 & 0.23 &2.69\times10^{-3} & 0.155 & 0.24 & 0.77 & {\rm AS}& {\rm \blue{fairly}~regular} \\ 

C  & 3.16\cdot10^5 & 13.6 & 2.0  &0.65& 30 & 1.44 &  0.033& 0.014 & 0.27 &1.05\times10^{-3} & 0.341 & 0.28 & 0.70 & {\rm SL}& {\rm intermittent} \\ 

D  & 2.92\cdot10^5 & 11.3 & 1.85 &0.63& 26 & 1.67 &  0.118&  0.059& 0.26 &0.67\times10^{-3}& 0.487 & 0.33 & 0.57 & {\rm SL} & {\rm irregular}\\ 

E  & 2.77\cdot10^5 & 10.2 & 1.75 &0.61& 25 & 1.75 &  0.111 & 0.057 & 0.23 &0.81\times10^{-3}&0.468 & 0.37 & 0.52 & {\rm SL}& {\rm irregular} \\ 
                                                                             
\hline   

D0 & 2.92\cdot10^5 & 11.3 & 1.85 &0.63& 26 & 1.67 &  0.118&  0.059& 0.26 &0.67\times10^{-3}& 0.487 & 0.33 & 0.57 & {\rm SL} & {\rm irregular}\\ 

D1 & 3.16\cdot10^5 & 13.6 & 2.00 &0.65& 30 & 1.46 &  0.013&  0.008& 0.29 &1.53\times10^{-3}& 0.333 & 0.28 & 0.71 & {\rm SL} & {\rm irregular} \\ 

D2 & 3.31\cdot10^5 & 15.7 & 2.10 &0.67& 31 & 1.40 & -0.069& -0.074& 0.21 &1.70\times10^{-3}& 0.121 & 0.24 & 0.76 & {\rm AS} & {\rm irregular} \\ 

D3 & 3.47\cdot10^5 & 18.5 & 2.20 &0.68& 32 & 1.38 & -0.153& -0.140& 0.23 &2.69\times10^{-3}& 0.143 & 0.20 & 0.81 & {\rm AS} & {\rm \blue{fairly}~regular} \\ 

D4 & 3.62\cdot10^5 & 22.5 & 2.30 &0.70& 32 & 1.36 & -0.167& -0.158& 0.24 &3.17\times10^{-3}& 0.149 & 0.17 & 0.86 & {\rm AS} & {\rm \blue{fairly}~regular} \\ 

\hline   

A0 & 3.93\cdot10^5 & 39.9 & 2.5 &0.73& 33 & 1.34 & -0.184& -0.185 & 0.25 & 3.77\times10^{-3} & 0.173 & 0.09 & 0.95 & {\rm AS} & {\rm \blue{fairly}~regular}\\ 

A1 & 3.85\cdot10^5 & 33.3 & 2.45&0.72& 32 & 1.36 & -0.172& -0.184 & 0.25 & 3.57\times10^{-3} & 0.163 & 0.11 & 0.92 & {\rm AS} & {\rm \blue{fairly}~regular} \\ 

A2 & 3.78\cdot10^5 & 28.5 & 2.4 &0.71& 32 & 1.36 & -0.135& -0.144 & 0.23 & 2.93\times10^{-3} & 0.122 & 0.13 & 0.89 & {\rm AS} & {\rm irregular} \\ 

A3 & 3.62\cdot10^5 & 22.3 & 2.3 &0.70& 32 & 1.38 & -0.154& -0.150 & 0.23 & 2.95\times10^{-3} & 0.126 & 0.17 & 0.85 & {\rm AS} & {\rm irregular} \\ 

A4 & 3.47\cdot10^5 & 18.3 & 2.2 &0.68& 32 & 1.38 & -0.141& -0.133 & 0.23 & 2.79\times10^{-3} & 0.124 & 0.20 & 0.81 & {\rm AS} & {\rm irregular} \\ 

A5 & 3.31\cdot10^5 & 15.5 & 2.1 &0.67& 31 & 1.41 & -0.075& -0.073 & 0.23 & 2.04\times10^{-3} & 0.142 & 0.24 & 0.76 & {\rm AS} & {\rm irregular} \\ 

A6 & 3.16\cdot10^5 & 13.4 & 2.0 &0.65& 30 & 1.44 &  0.066&  0.017 & 0.24 & 1.40\times10^{-3} & 0.212 & 0.28 & 0.71 & {\rm SL} & {\rm irregular} \\ 

A7 & 3.00\cdot10^5 & 11.9 & 1.9 &0.63& 29 & 1.53 &  0.091&  0.055 & 0.31 & 0.58\times10^{-3}& 0.499 & 0.31 & 0.62 & {\rm SL} & {\rm irregular} \\ 

A8 & 2.85\cdot10^5 & 10.7 & 1.8 &0.62& 27 & 1.61 &  0.121&  0.073 & 0.29 & 0.61\times10^{-3}& 0.553 & 0.35 & 0.54 & {\rm SL} & {\rm irregular} \\ 

\hline   
B$^\prime$ & 3.54\cdot10^5 & 20.3 & 2.25 &0.69& 32 & 1.34 & -0.154&-0.149 &0.21 &3.66\times10^{-3} & 0.117 & 0.19 & 0.82 & {\rm AS}& {\rm fairly~regular} \\ 
B$^{\prime\prime}$  & 3.54\cdot10^5 & 20.3& 2.25 &0.69& 32& 1.36 &-0.132&-0.140&0.25 &2.66\times10^{-3} & 0.128 & 0.19 & 0.85 & {\rm AS}& {\rm fairly~regular} \\ 
\hline
         \end{array}
     $$
\tablefoot{
In all runs, $\Pm=1$, $\chiSGSm=3.7\cdot10^8\m^2\s^{-1}$ by taking $\chiSGS(r_1)=5\cdot10^8\m^2\s^{-1}$, 
$\nu=9.3\cdot10^7\m^2\s^{-1}$, $\Pra_{\rm SGS}=0.25$, $\mathcal{L}=3.85\cdot10^{-5}$,
$\Ta=2.98\cdot 10^{6}$, $\xi=0.0325$ which gives $\Gamma_\rho^{(0)}\approx 12$,
$\Omega_0/\Omega_\odot=1$,
and the grid resolution is $128\times256\times128$. The volume and
time averaged total kinetic energy in units of
$GM_\odot\rho_0/R_\odot$ is $\tilde{E}_{\rm kin}=\brac{\onehalf \rho
  \bm{u}^2}$ and the kinetic energies of the meridional circulation
and the differential rotation are $E_{\rm
  mer}=\onehalf\brac{\rho(\mean{u}_r^2+\mean{u}_\theta^2)}$ and
$E_{\rm rot}=\onehalf\brac{\rho\mean{u}_\phi^2}$, respectively.
$\tilde{L}_{\rm rad}$ and $\tilde{L}_{\rm conv}$ are the fractions of total
flux transported by radiative conduction and resolved convection at the middle of the convection zone.
Here, `DR' stands for differential rotation.
Runs~B$^\prime$ and B$^{\prime\prime}$ are same as Run~B, but with larger 
($\pm84^\circ$) and smaller ($\pm66^\circ$) latitudinal extents, respectively.
}\end{table*}

\section{Results} \label{sect:res}

First, we perform a set of simulations for
different values of the radiative conductivity starting from the initial
condition described in Sect.~\ref{sec:initcond}.
These are referred to as Runs~A--E in Table~\ref{tab:runs}.
Except for the allowance of a magnetic field, all other input parameters
of our Runs~A--E are identical to those of \cite{KMB14}.
However, we have an additional Run~BC in this set, whose radiative
conductivity lies between those of Runs~B and C.
It turns out that Runs~A--BC produce AS differential rotation, while
Runs~C--E produce SL differential rotation. 
Next, we perform two further sets of runs where we use Runs~A and D
with AS and SL differential rotation, respectively, as progenitors to
study the possibility of bistability of the rotation profile.

\subsection{Energy fluxes in our dynamo runs}

The radiative conductivity in our model is controlled by the parameter
$\delta n$, regulating the fractional flux that convection has to
transport. Increasing $\delta n$ reduces the radiative flux and
increases the convective flux and thus $\urms$. Having $\Omega_0$
fixed, changing $\delta n$ affects the rotational influence on the
convection via the convective velocities \citep[for
further details see][]{KMB14}.
Hence, different values of $\delta n$ in Runs A--E imply different
values of $\Co$.
Thus, Run~E is more rotationally dominated than Run~A.
For Run~A with $\delta n = 2.5$, the convective flux dominates
over the other fluxes and the radiative
flux transports a very small fraction of the luminosity. 
For the definitions of the fluxes we refer to Eqs.~(26)--(31) of
\cite{KMCWB13}.

\begin{figure}[t]
\centering
\includegraphics[width=0.475\textwidth]{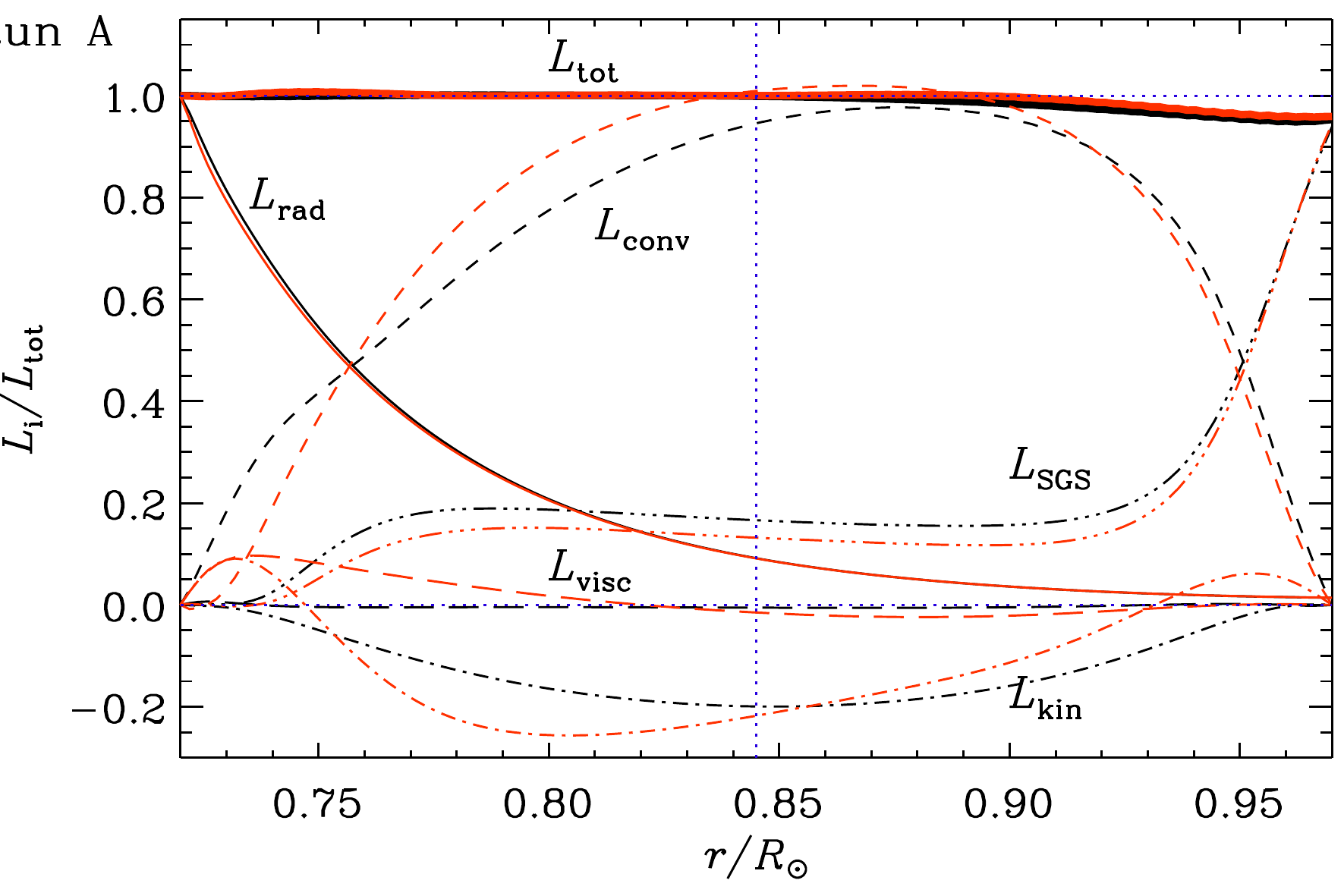}
\includegraphics[width=0.475\textwidth]{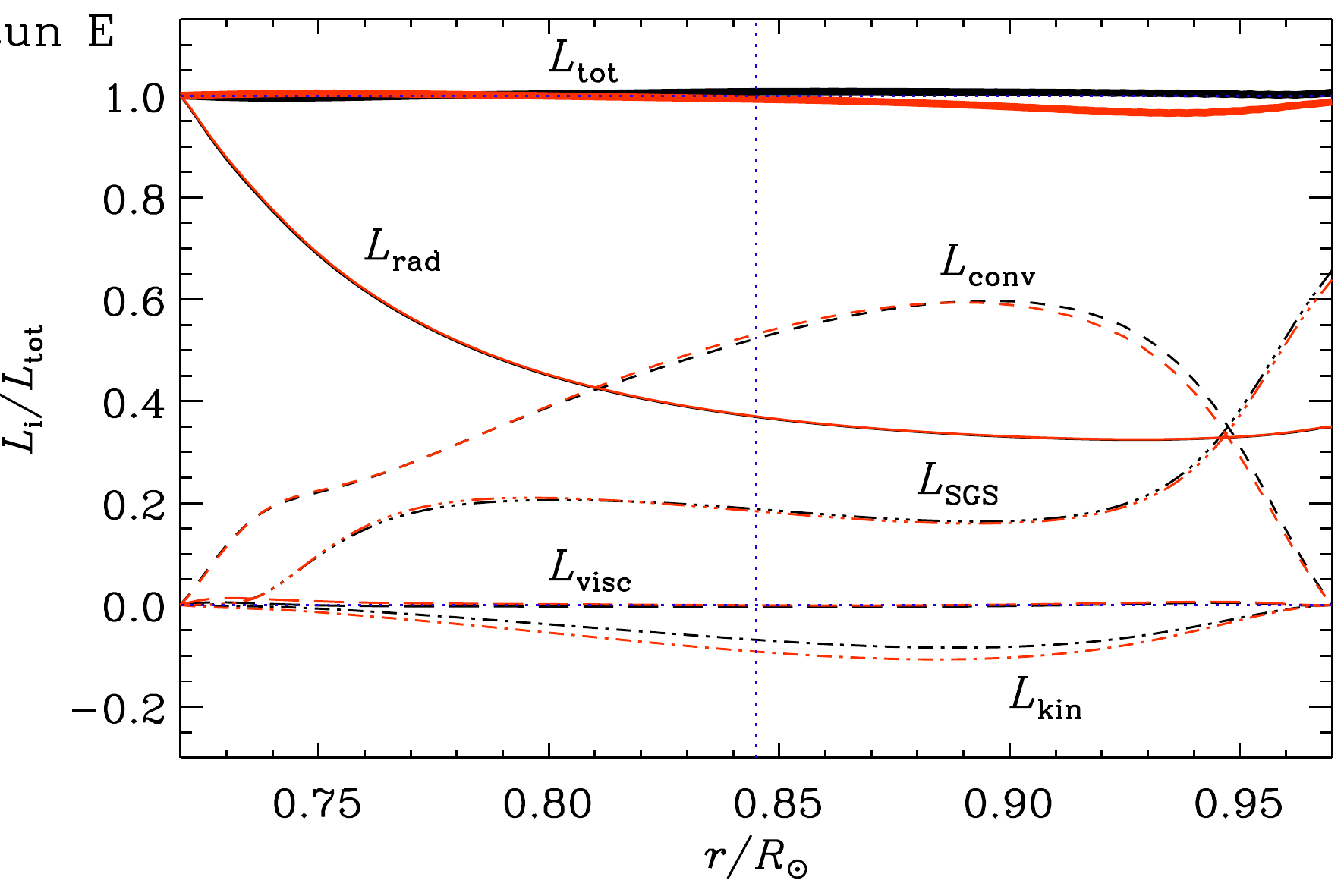}
\caption{The contributions of different energy fluxes of Runs A (top) 
and E (bottom). The black (red) lines correspond to
magnetohydrodynamic (hydrodynamic) case. 
Thin solid: radiative, dashed: convective, long dashed: viscous, 
dash-dotted: kinetic energy, dash-triple-dotted: SGS,
and thick solid: total. Dotted horizontal lines indicate the zero and unity values.
The blue vertical dotted line indicates the position of the middle of the 
convection zone $r=r_{\rm m}$.
}\label{fig:flux}
\end{figure}

In the statistically stationary state, the total luminosity
$L_{\rm tot}(r)=4\pi r^2 {\cal F}_{\rm tot}(r)$ is constant, where
${\cal F}_{\rm tot}$ is the time averaged total energy flux.
In \Fig{fig:flux} we show the radial dependence of the contributions
from radiation ($L_{\rm rad}$) and convection ($L_{\rm conv}$),
as well as kinetic ($L_{\rm kin}$), viscous ($L_{\rm visc}$),
and subgrid scale ($L_{\rm SGS}$) energy fluxes
in the convection zone for Runs~A and E.
For comparison we also show the fluxes from
the corresponding hydrodynamic simulations of \cite{KMB14} with red lines.
We see that for Run~A, the convective, kinetic, and
SGS energy fluxes have decreased in the lower part of the convection zone in
the magnetic case. In Run~E there is very little change in the SGS
flux and only a small reduction of the convective and kinetic energy
fluxes is visible.
In Table~\ref{tab:runs} we show the fractions of the radiative and the
convective fluxes at the middle of the convection zone for all the runs.
In Runs~A, B, and BC, the convective flux is above $75\%$
and the radiative flux is less than $25\%$.
By contrast, Runs~C, D, and E have a convective flux of less than $70\%$
and the radiative flux is larger than $25\%$. 

\begin{figure}[t]
\centering
\includegraphics[width=0.47\textwidth]{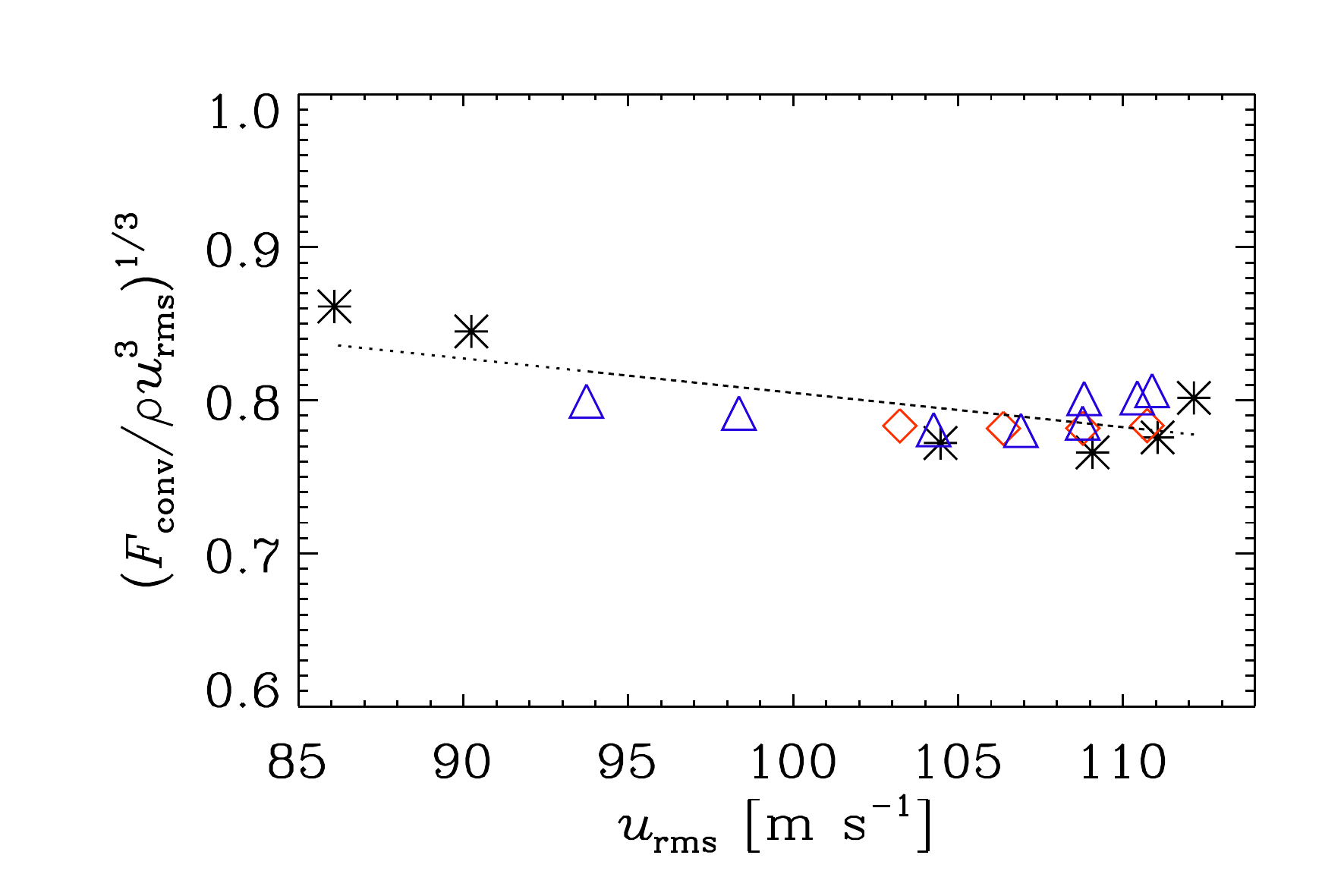}
\caption{Variation of ${(F_{\rm conv}/\rho \urms^3)}^{1/3}$ averaged over the whole convection zone as a function of $\urms$ from different runs.
Black asterisks: Runs~A--E, red diamonds: Runs~E1--E4,
and blue triangles: Runs~A1--A8. 
The dotted line shows the linear
dependence between $\urms$ and ${(F_{\rm conv}/\rho \urms^3)}^{1/3}$.
}\label{fig:urmsFconv}
\end{figure}

We find that the rms-velocity is \blue{compatible with} a one-third power
proportionality to the convective energy flux -- at least in the narrow
range of parameters studied here; see Fig.~\ref{fig:urmsFconv}.
This is in agreement with the scaling used in connection with the
artificially high luminosities used in our simulations \citep[see
also][]{BCNS05}.

\subsection{Differential rotation}\label{sec:dr}

\begin{figure*}[t]
\centering
\includegraphics[width=0.16\textwidth]{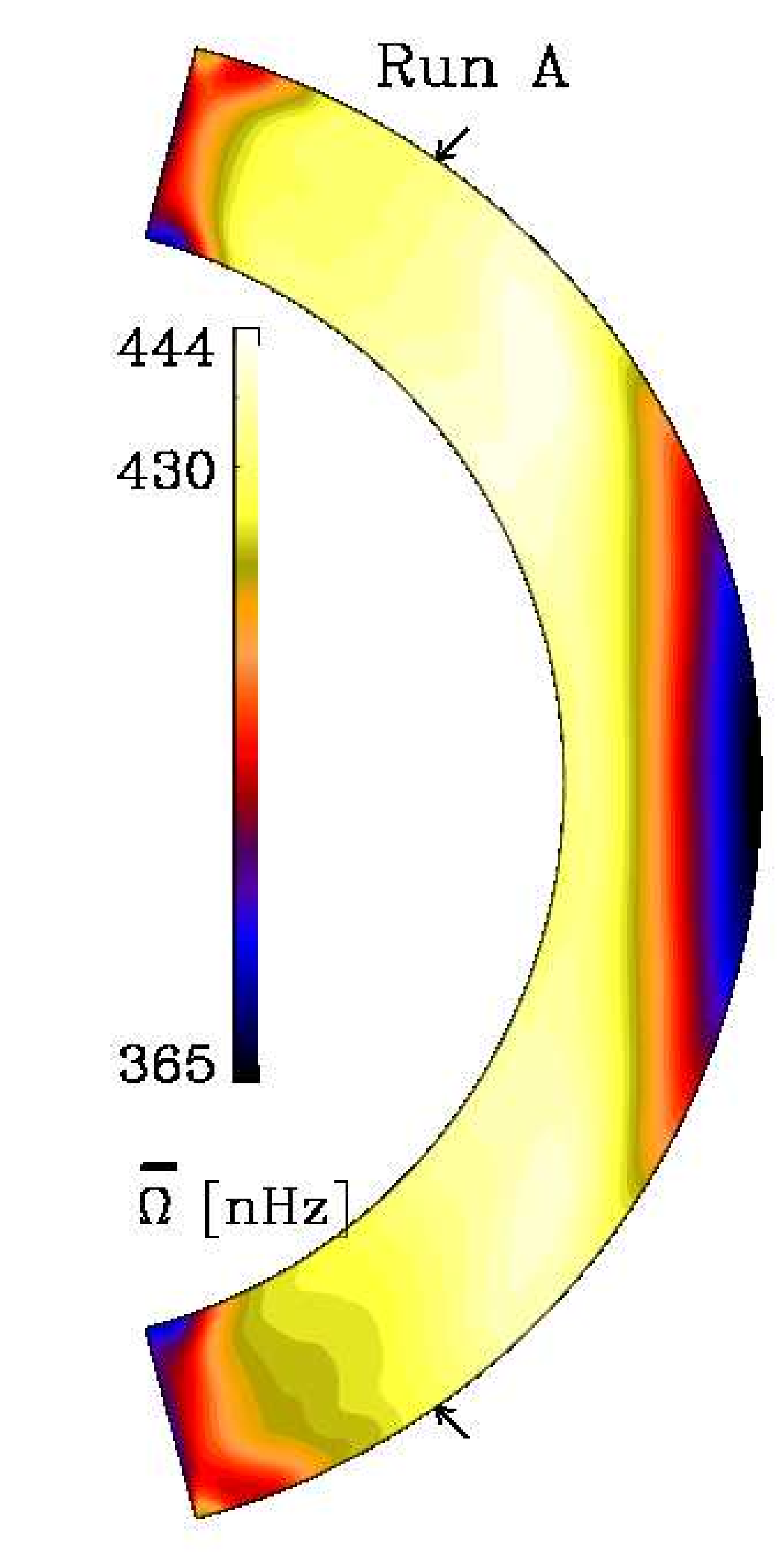}
\includegraphics[width=0.16\textwidth]{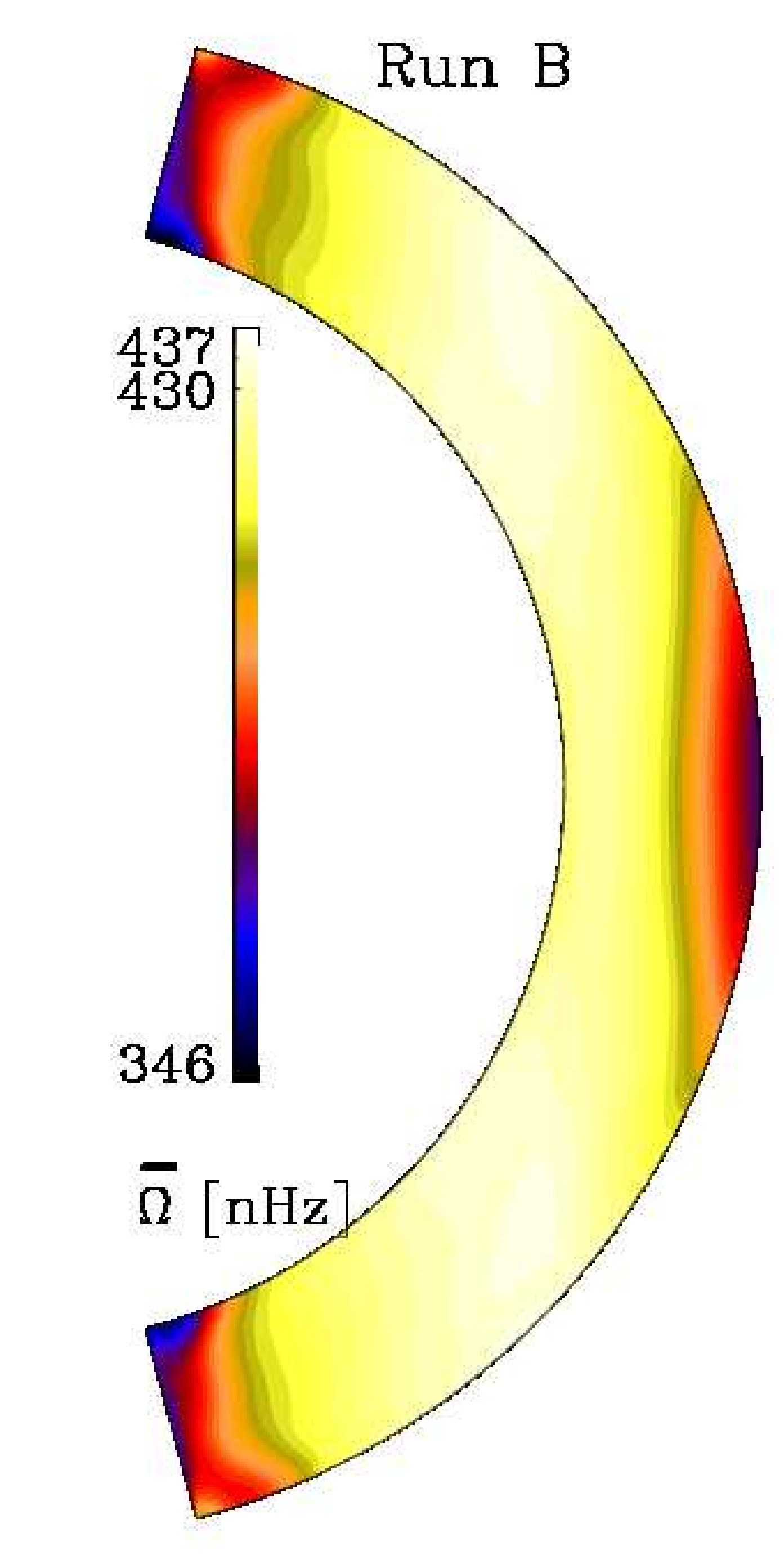}
\includegraphics[width=0.16\textwidth]{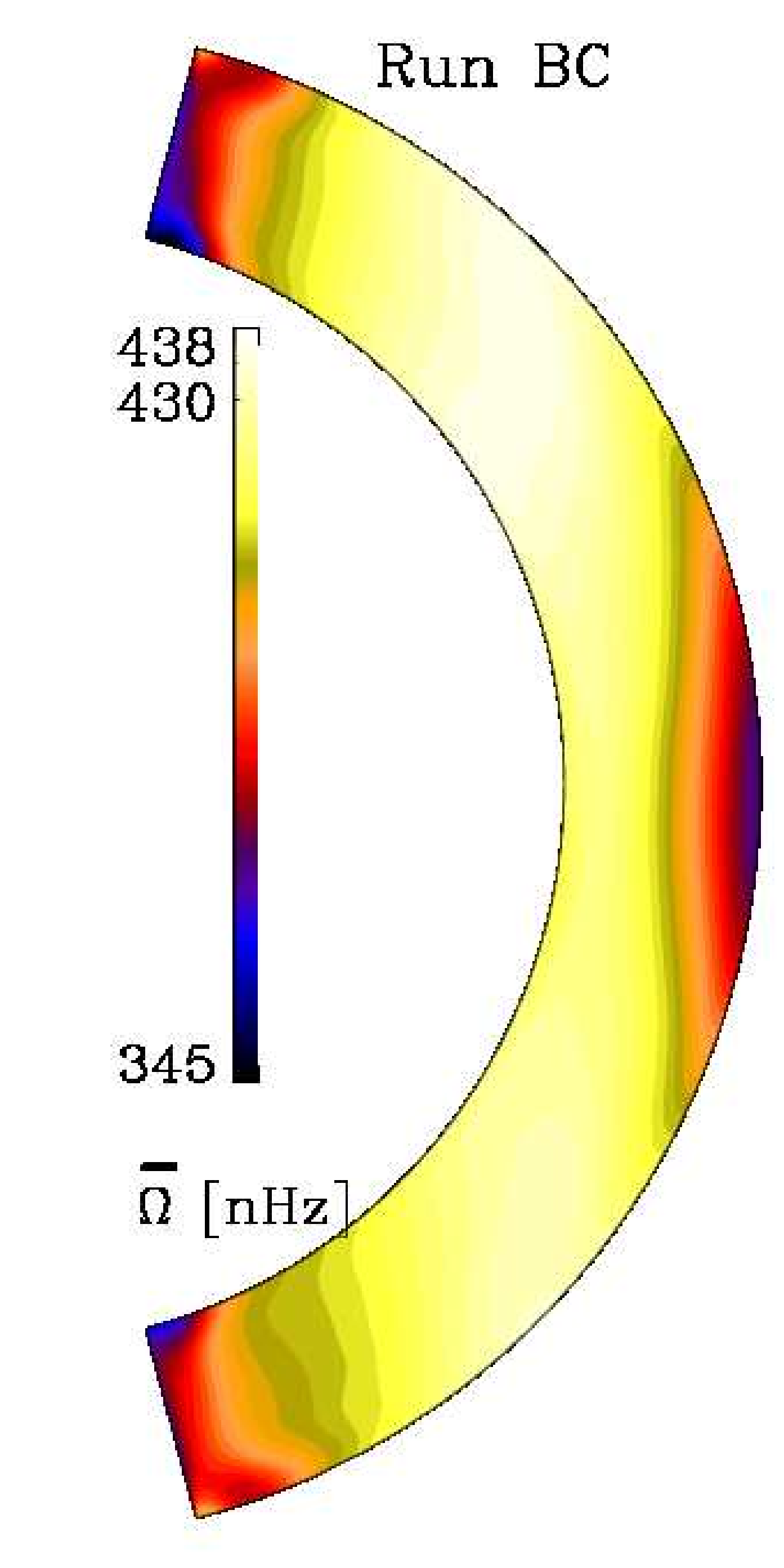}
\includegraphics[width=0.16\textwidth]{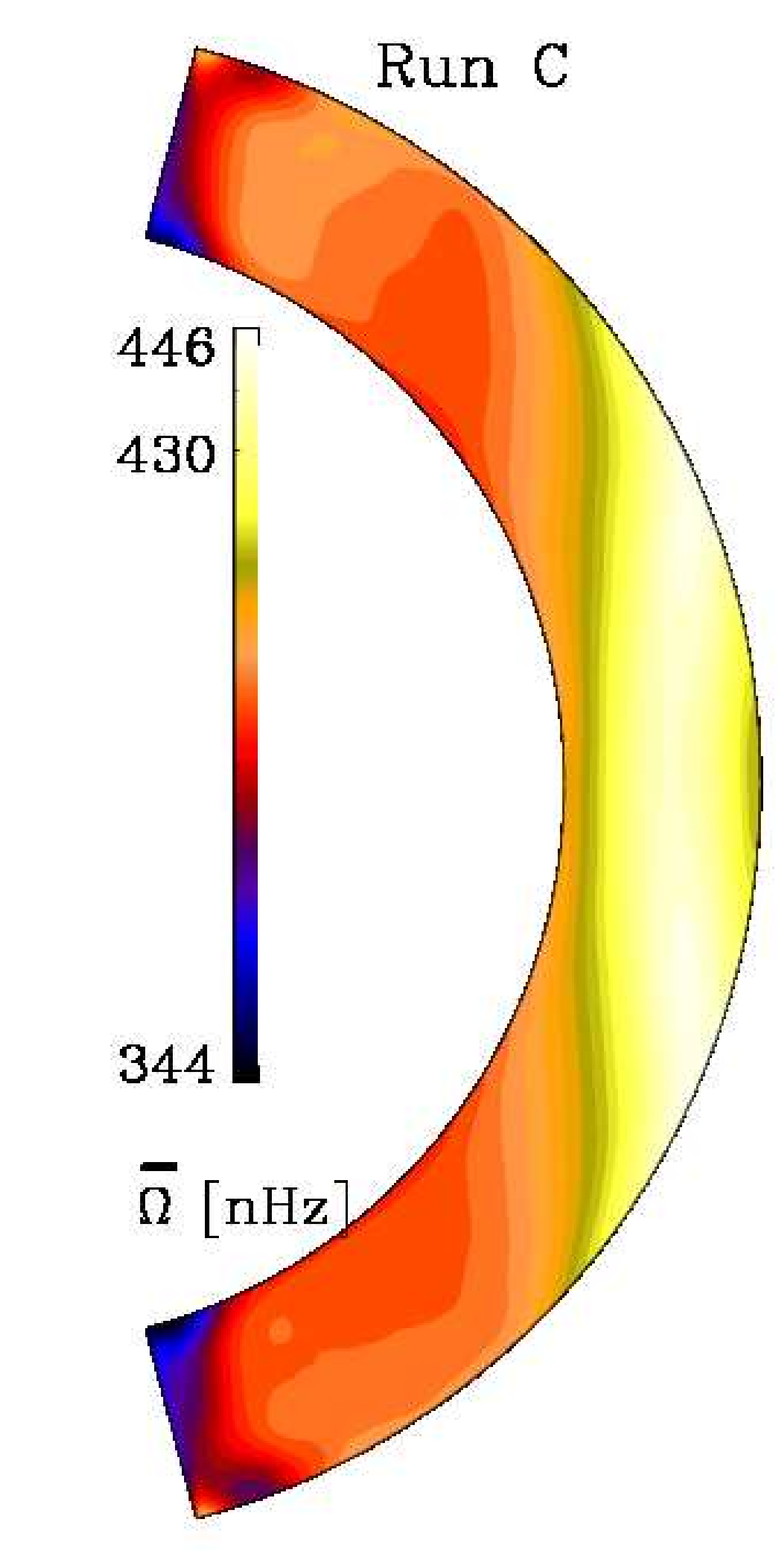}
\includegraphics[width=0.16\textwidth]{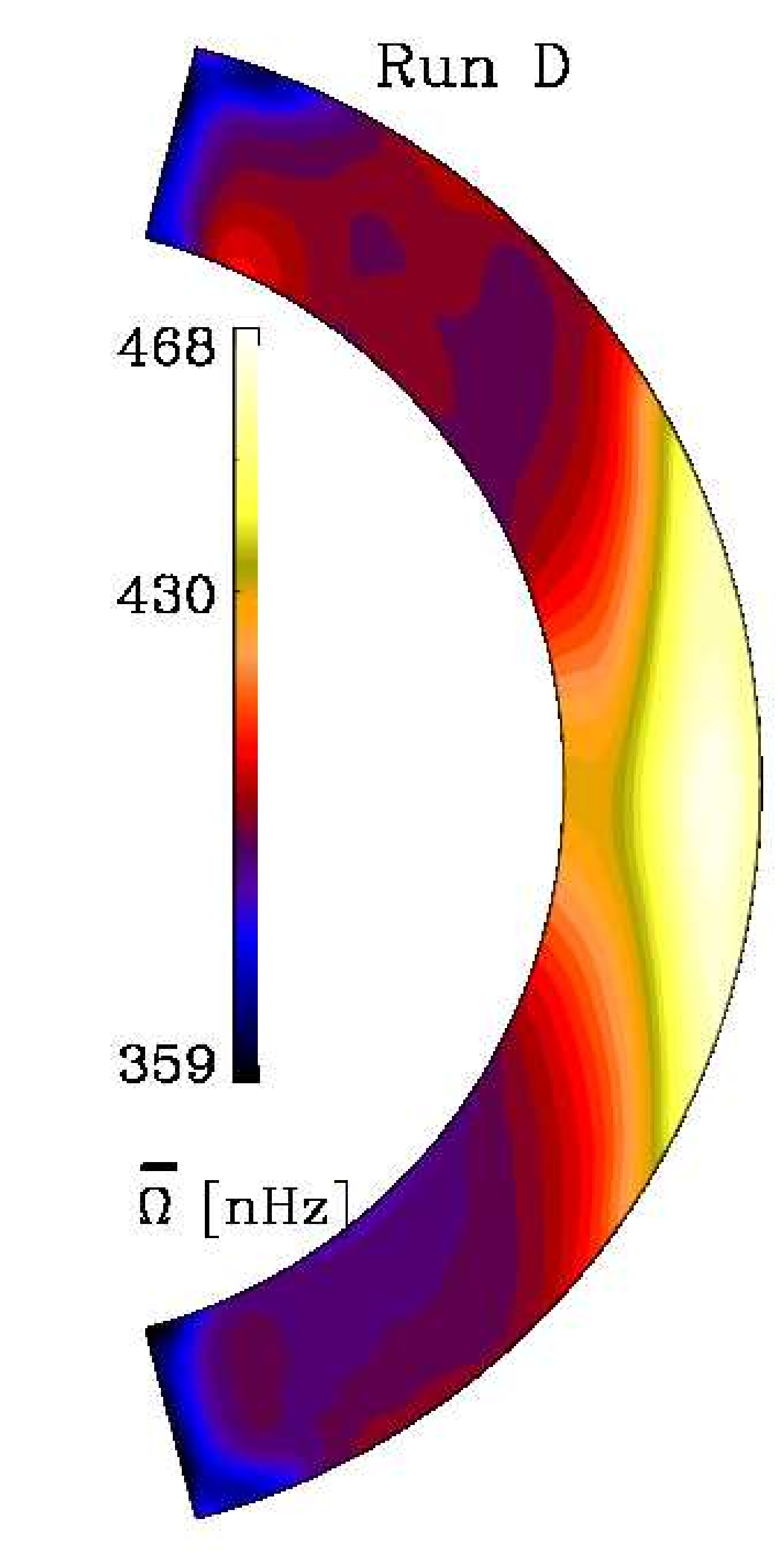}
\includegraphics[width=0.16\textwidth]{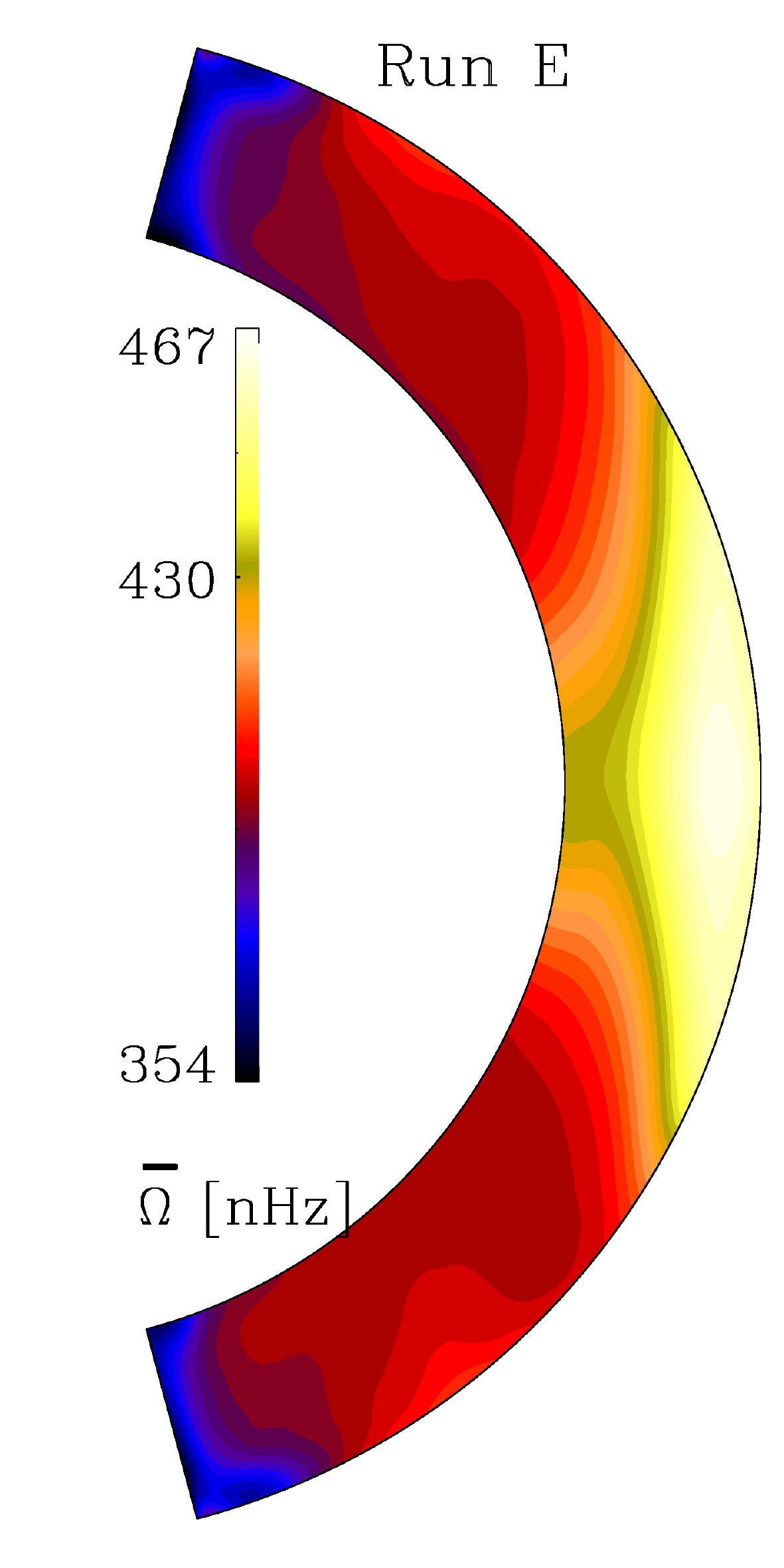}
\caption{The distribution of angular velocity (in nHz) $\mean\Omega$ in the meridional
plane from Runs~A--E.
$\mean\Omega$ is computed from $\Omega$ first by the longitudinal average
and then the time average over the last few \blue{cycles}.
The arrows in the leftmost panel show the co-latitudes at which the
latitudinal differential rotation is computed in \Eq{equ:pDRt}.
}\label{fig:om}
\end{figure*}

\begin{figure}[t]
\centering
\includegraphics[width=0.16\textwidth]{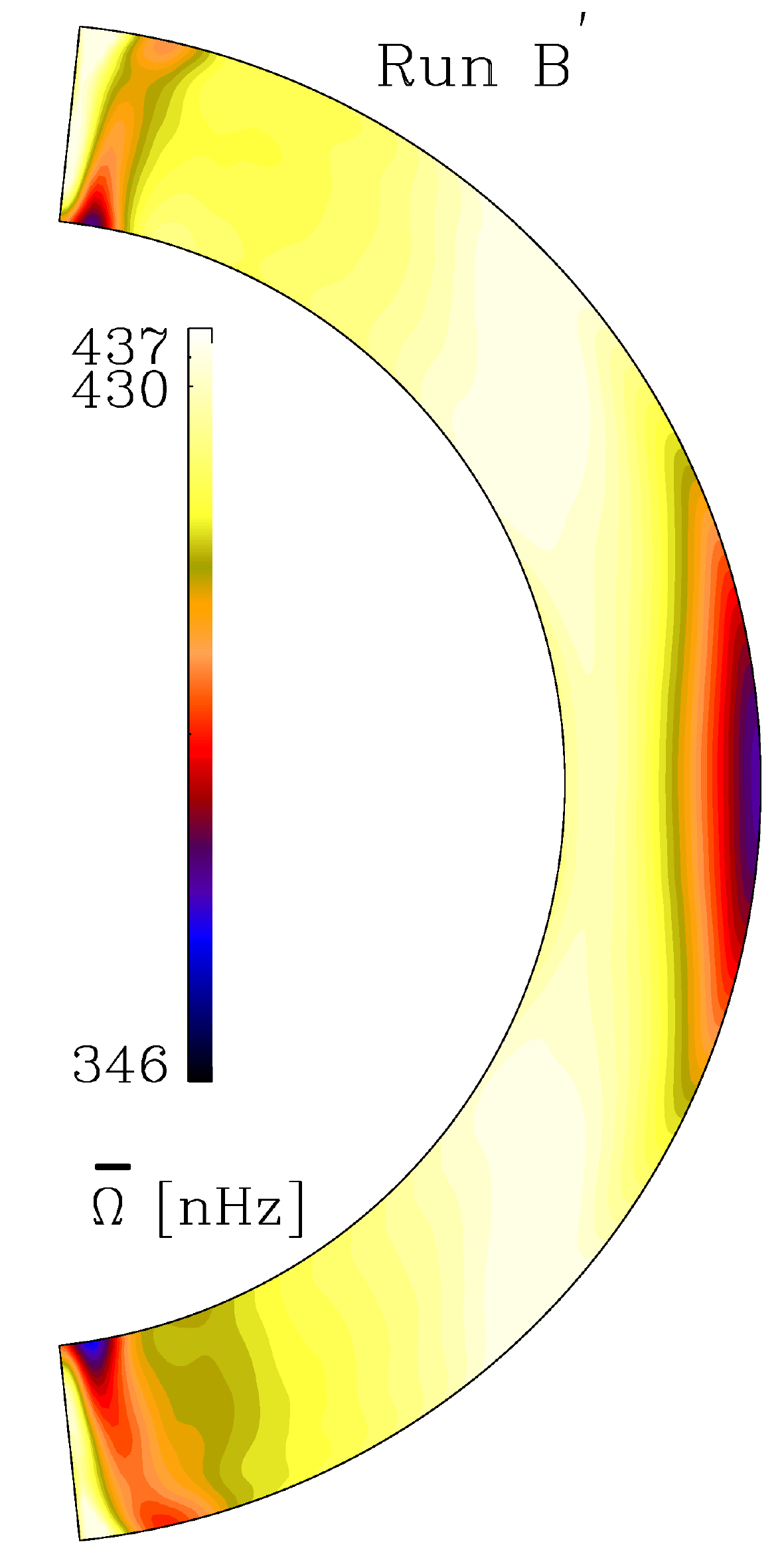}\includegraphics[width=0.16\textwidth]{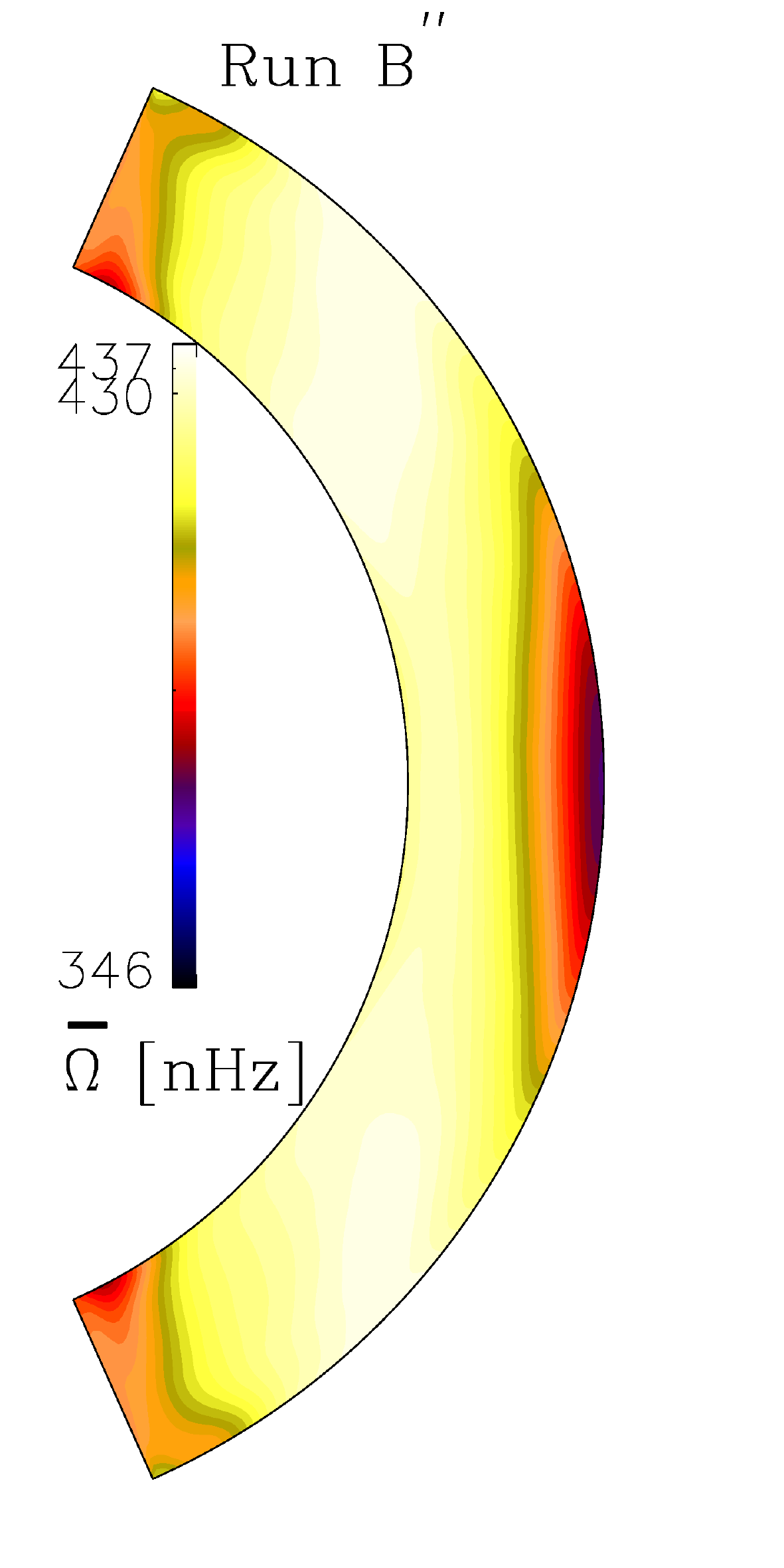}
\caption{Same as Run~B in \Fig{fig:om}, but here the latitudinal extent is different.
For the left panel the latitudinal boundaries are at $\pm84^\circ$, 
whereas for the right one they are at $\pm66^\circ$.
}\label{fig:omlats}
\end{figure}

\begin{figure}[t]
\centering
\includegraphics[width=0.40\textwidth]{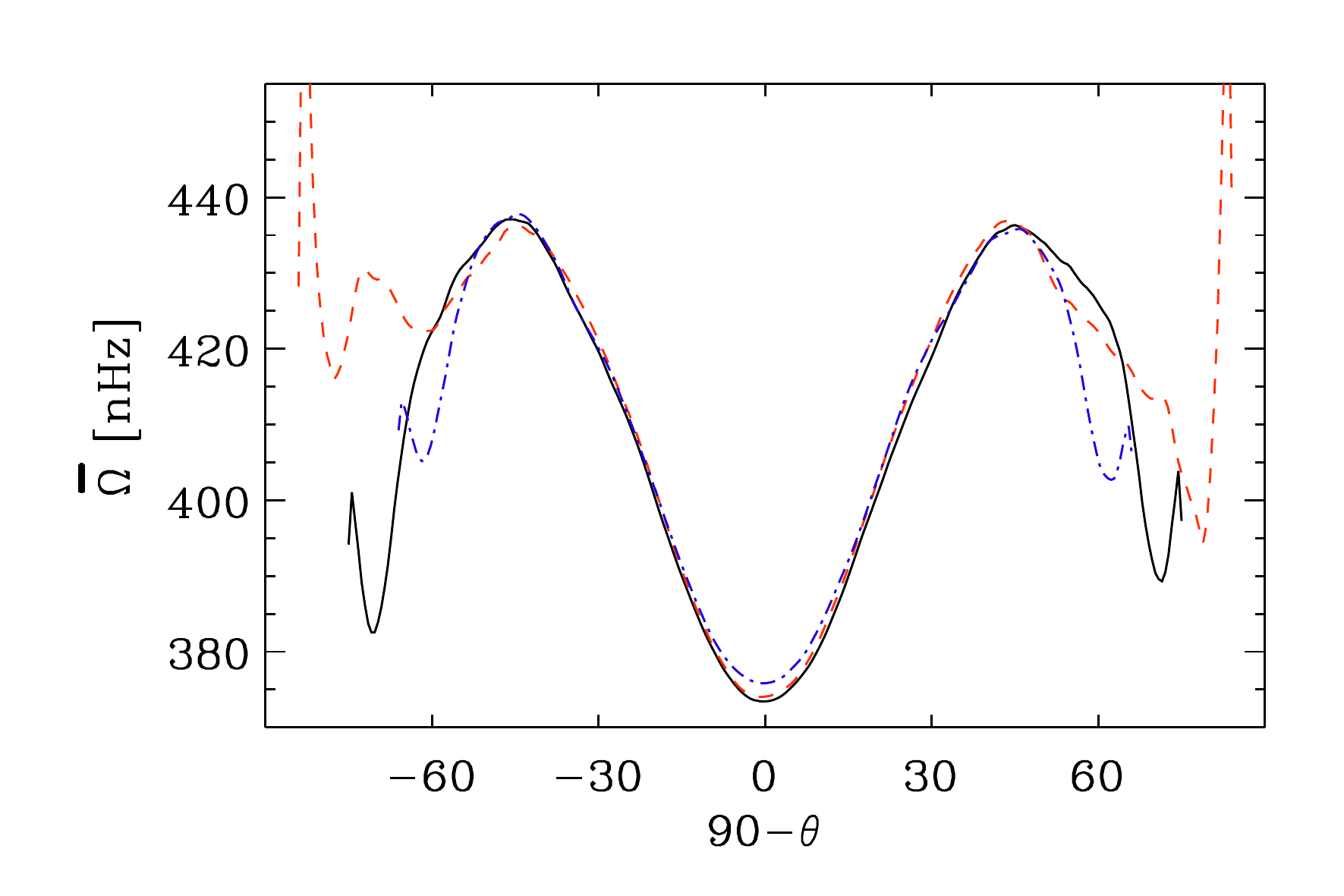}
\caption{Variations of angular velocities $\mean\Omega$ 
at $r = 0.96 R_\odot$ from Runs~B (solid line), 
B$^\prime$ (red dashed), and B$^{\prime\prime}$
(blue dash-dotted).
}\label{fig:omvstheta}
\end{figure}

In \Fig{fig:om} we show the rotation profile
$\mean\Omega=\mean{u}_\phi/r \sin \theta + \Omega_0$ for Runs~A--E.
We see that in Runs~A, B, and BC the equator rotates slower than
the mid- and high latitudes
which is opposite to the rotation profile observed in the Sun. 
However, we find that the regions near the latitudinal boundaries have
slower rotation than mid-latitudes, which was not observed in the
hydrodynamical simulations of \cite{KMB14}.
We have repeated Run~B by increasing and decreasing the extent of 
the latitudinal boundaries.
Runs~B$^\prime$ and B$^{\prime\prime}$ in Table~\ref{tab:runs} correspond to 
these two cases where the latitudinal end points are at $\pm84^\circ$
and $\pm66^\circ$, respectively. In \Fig{fig:omlats}, we show $\mean\Omega$ profiles
for these two runs, whereas in \Fig{fig:omvstheta} we show their latitudinal
variations at $r=0.96R_\odot$. From these two plots we see that
the rotation profiles are very similar in all these runs up to about $\pm50^\circ$
latitudes and the significant departures appear only near the boundaries.
However, the slowly rotating high-latitude branch still exists in all runs.
Therefore this is probably not due to our restricted latitudinal
extent, but may be a real feature which might be caused
by the magnetic fields.

By comparing the rotation profiles from hydrodynamic simulations of \cite{KMB14}
for Runs~A--E, we see that
the pole--equator differential rotation is generally weaker in the magnetic runs,
the reduction being strongest in the runs with a slowly rotating equator.
This will be discussed in Sect.~\ref{sec:mc}, where we
give the ratio of the rotational energy of the hydrodynamic and magnetic
simulations. 
The overall reduction of the pole--equator differential
rotation by magnetic fields agrees with what has been reported before
\citep[e.g.,][]{BMT04,BCRS13,FF14}.
In particular, for slow rotation \cite{FF14} found a switch from AS to SL
differential rotation, which agrees with our results reported below.
By contrast, \cite{BMT04} and \cite {BCRS13} only consider runs
with SL rotation, but in the models by \cite{BMT04} the angular velocity
at high latitudes is not much reduced by magnetic fields.
In the models of \cite{BCRS13}, on the other hand, there is a
lower overshoot layer, where angular velocity is constant and equal
to that at high latitudes when there is a magnetic field.
This is, however, not the case in their non-magnetic runs, so
the magnetic field leads to a strong reduction in their case,
which is similar to ours, but different from what \cite{BMT04} found,
although comparing runs with and without overshoot layers can be
misleading due to different physical effects involved.

\begin{figure}[t]
\centering
\includegraphics[width=0.32\columnwidth]{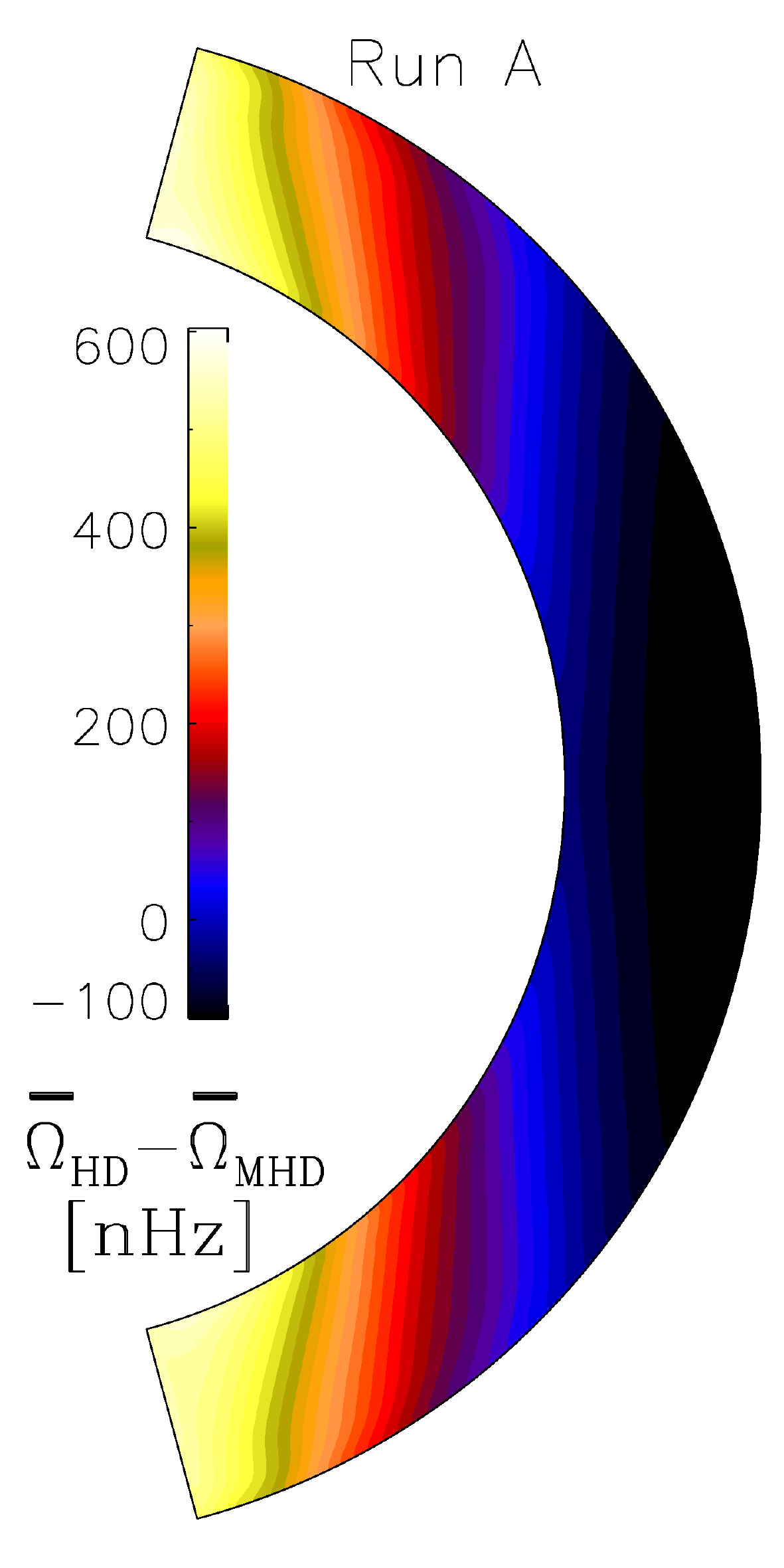}
\includegraphics[width=0.32\columnwidth]{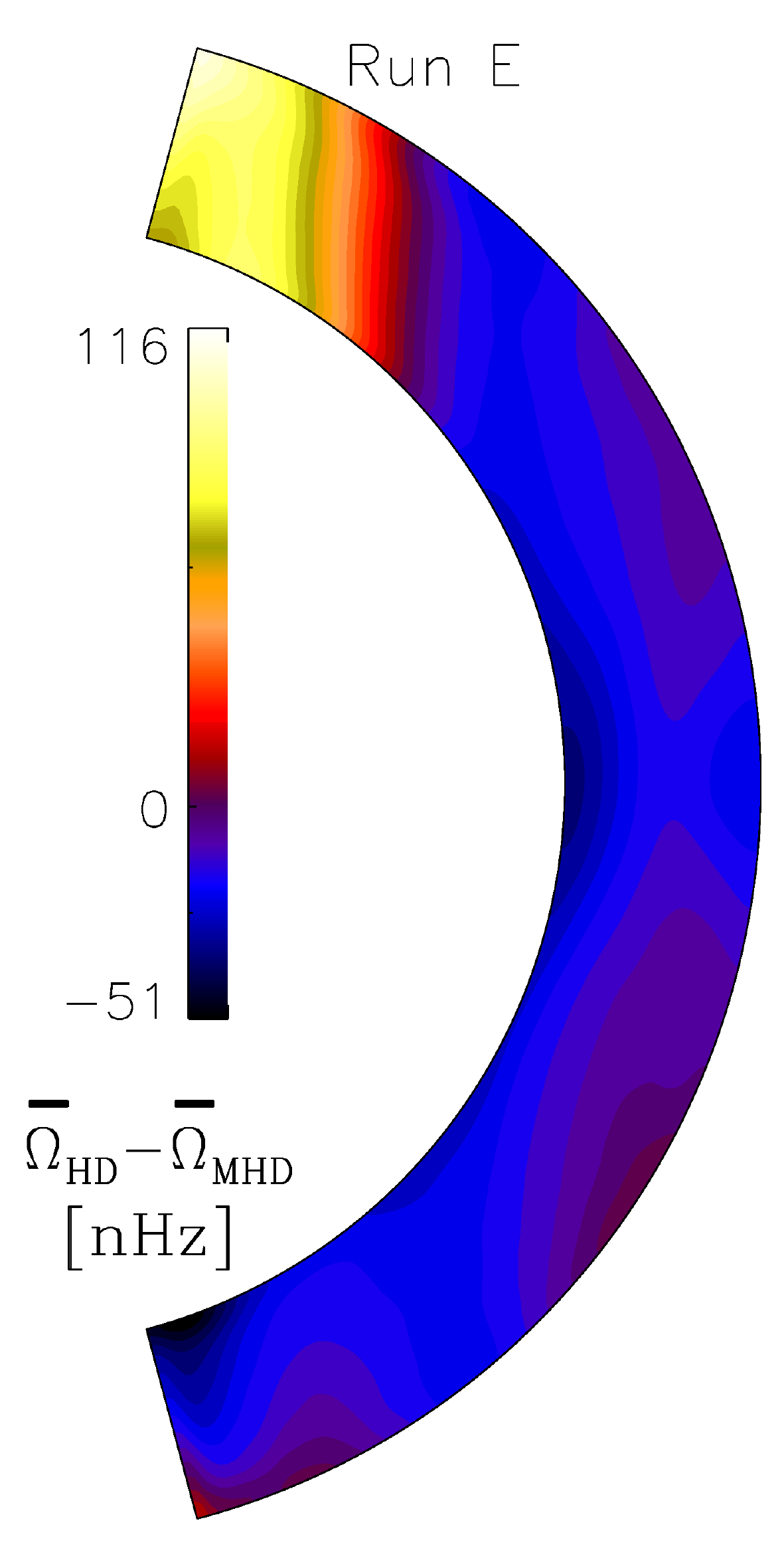}
\caption{Difference of $\mean\Omega$ between the HD and MHD runs.
The left and right panels are for Runs~A and E, respectively.
}\label{fig:Omdiff}
\end{figure}

To illustrate the reduction of differential rotation
in our magnetic runs, we show in \Fig{fig:Omdiff} the differences in
$\mean\Omega$ between the hydrodynamic and magnetic versions of Runs~A
and E. We see that for Run~A the difference is even larger than the
average rotation rate of the Sun.
Possible reasons for the reduction of differential rotation
will be discussed in Sect.~\ref{sec:stress}.

In Runs~C--E, the equator rotates faster than the polar regions,
as is also the case in the Sun.
However, unlike some of the non-magnetic solar-like cases of
\cite{KMB14}, we never
obtain polar vortices or jet-like structures with magnetic fields.
It is also interesting to note that the pole--equator difference in the
rotational velocity is comparable to that of the Sun.
The solar value ($\sim$~430~nHz) is marked in each colorbar of \Fig{fig:om}.

\begin{figure*}[t]
\centering
\includegraphics[width=0.88\columnwidth]{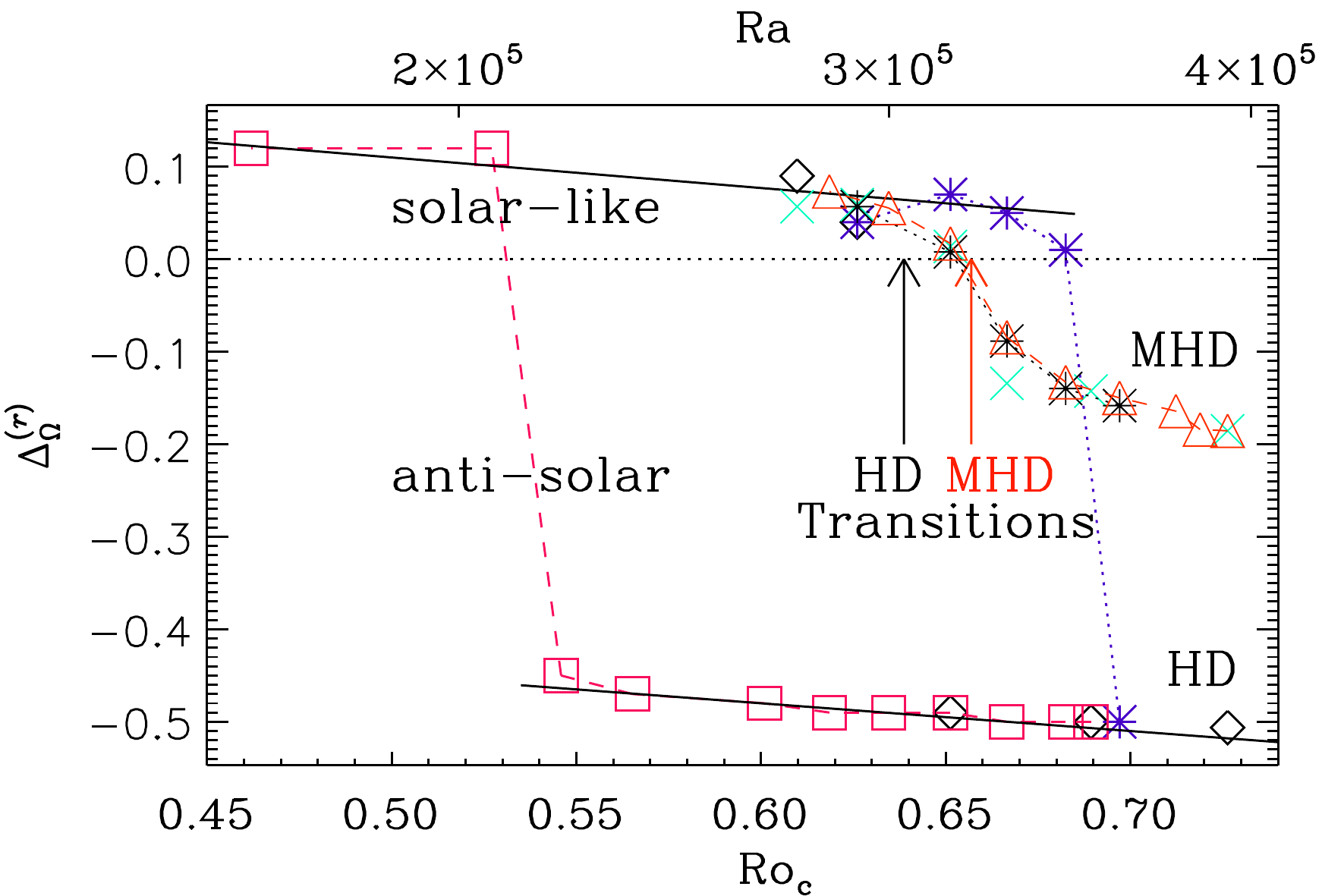}
\includegraphics[width=0.88\columnwidth]{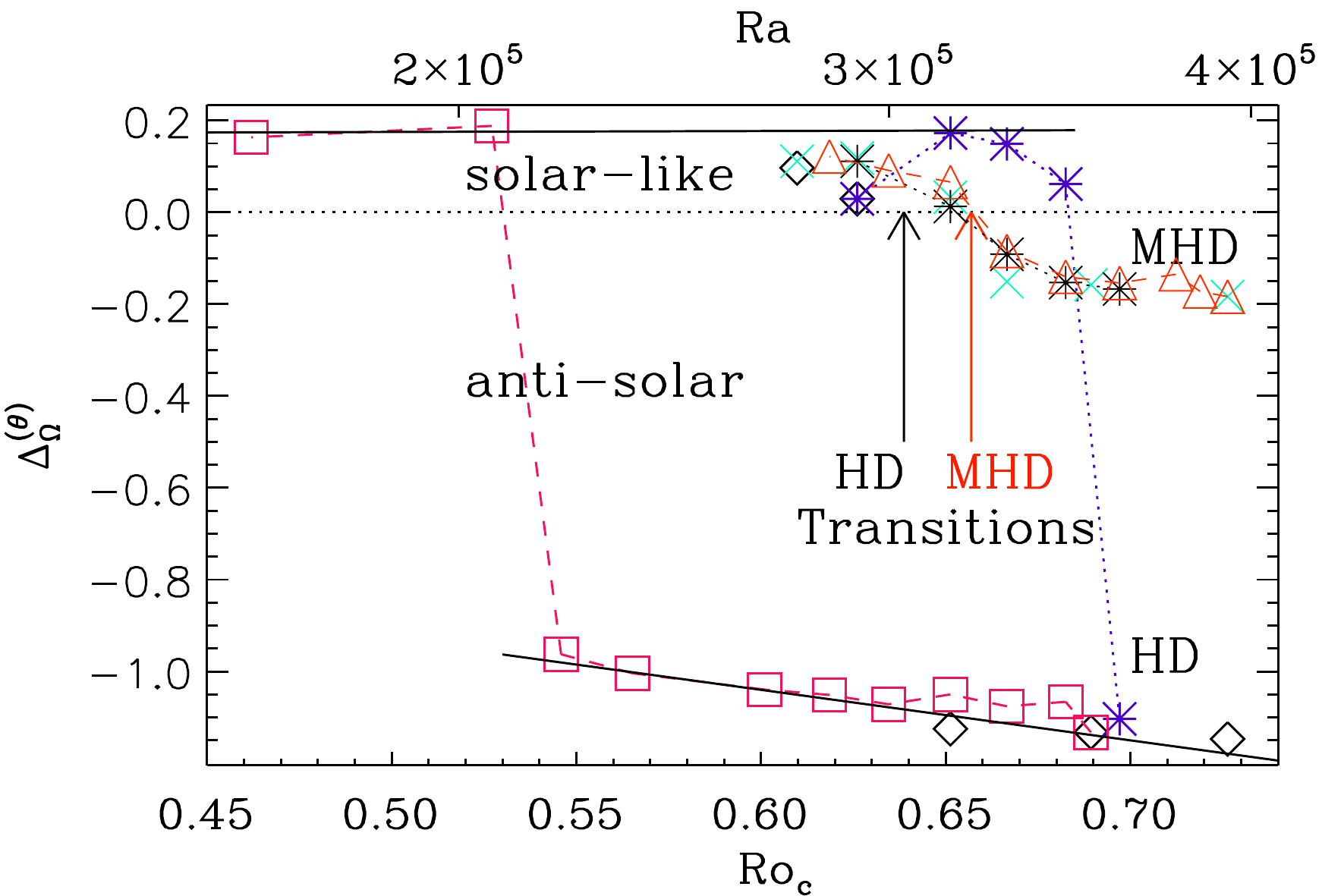}
\caption{Radial (left panel) and latitudinal (right panel) differential 
rotation, defined by \Eq{equ:pDRt}, for Runs~A--E (green crosses), A0-A8 
(red dashed line with triangles), and D0-D4 (black asterisks with dotted line).
The other points are taken from the hydrodynamical simulations (black
diamonds: Runs~A--E, blue dotted line with asterisks: Runs~D0--D4, and
red dashed line with squares: Runs~B0--B10) of \cite{KMB14}.
The horizontal black dotted lines show the zero value.
A red (black) arrow shows the transition point when differential rotation 
in the MHD (HD) simulations changes from AS to SL rotation 
with the decrease of $\Roc$.
}\label{fig:bistab}
\end{figure*}

\subsection{Identifying the SL to AS transition}
\label{sec:SLtoAS}

We measure the relative radial and latitudinal 
differential rotation by the quantities
\begin{eqnarray}
\Delta_\Omega^{(r)}=\frac{\Omega_{\rm eq}-\Omega_{\rm bot}}{\Omega_{\rm eq}},
\quad
\Delta_\Omega^{(\theta)}=\frac{\Omega_{\rm eq}-\Omega_{55}}{\Omega_{\rm eq}},\label{equ:pDRt}
\end{eqnarray}
where $\Omega_{\rm eq}=\mean\Omega(r_1,\pi/2)$ and $\Omega_{\rm
  bot}=\mean\Omega(r_0,\pi/2)$ are the equatorial rotation rates at
the surface and at the base of the convection zone, and $\Omega_{55}
=\onehalf[\mean\Omega(r_1,35^\circ)+\mean\Omega(r_1,145^\circ)]$
is the rotation rate at latitudes $\pm55\degr$ computed as an average
of $\mean{\Omega}$ at $35^\circ$ and $145^\circ$ co-latitudes on the
outer radius.
The arrows in the left panel of \Fig{fig:om} show the positions of these points in the $r$-$\theta$ plane.
The values of $\Delta_\Omega^{(r)}$ and $\Delta_\Omega^{(\theta)}$, listed in Table~\ref{tab:runs}, help us to identify
AS and SL differential rotation.
SL differential rotation implies $\Delta_\Omega^{(r)}>0$ and
$\Delta_\Omega^{(\theta)}>0$.
Following this definition we see that Runs~A, B, and BC are classified
as AS and Runs~C--E as SL. Hence we see that there is a transition from
AS to SL rotation around $\Co \sim 1.4$.
Although this transition has been reported extensively in the literature
\citep{Gi77,RBMD94,KMGBC11,GWA13,GSKM13}, the possibility of a bistability
has only recently been discovered \citep{GYMRW14,KMB14}, i.e., AS and SL
rotation profiles can be obtained for the same input parameters.
By comparing the current results with the hydrodynamical ones, 
we see that the transition happens at a slightly larger value
of $\Roc$, also manifested by the change of the AS hydrodynamical counterpart 
of Run~C, to SL in the MHD regime.
The transition observed in the current dynamo cases is less abrupt than
that of earlier hydrodynamic studies.
Therefore we conclude that the magnetic field helps
to produce SL differential rotation, which has also been found
in the recent anelastic simulations of \cite{FF14}.

In \Fig{fig:bistab} we show differential rotation parameters
$\Delta_\Omega^{(r)}$ and $\Delta_\Omega^{(\theta)}$ computed from
\Eq{equ:pDRt}
for all the runs as functions of $\Roc$ and $\Ra$.
To compare with the hydrodynamic simulations of \cite{KMB14},
we have analyzed the latitudinal differential rotation in
their data with our new definition~(\ref{equ:pDRt}).
The hydrodynamic values shown in Fig.~\ref{fig:bistab}
are now considerably smaller for the AS branch
than the ones reported by \cite{KMB14}.
Note that, had we defined
$\Delta_\Omega^{(\theta)}$ as the difference of
rotation between equator and the endpoints of the domain at
$\theta=\theta_0$ and $\pi-\theta_0$ as in \cite{KMCWB13,KMB14},
instead of $\Omega_{55}$, as we do here, we would have obtained
smaller values of $\Delta_\Omega^{(\theta)}$ because of the slowly
rotating high latitude regions in our magnetic Runs~A--C.
Another way of characterizing the SL or AS 
differential rotation can be done following \cite{KMGBC11}, who approximated 
the surface rotation profile in terms of Gegenbauer polynomials, which is,
$\Omega=\Omega_0\sum_{\ell=1,3,5}\omega_\ell P_\ell^1(\cos\theta)/\sin\theta$.
The sign of $w_3$ indicates whether a rotation is SL or AS.
Following this procedure we obtain the same conclusion for the classification
of the SL and AS differential rotation.

\subsection{Checking for flow bistability}\label{sec:bistab}

Next we study the flow bistability by taking AS and SL cases as
initial conditions.
Firstly, we have performed a set of simulations by starting from the
saturated state of Run A with AS differential rotation and
decreasing $\delta n$ slowly, which corresponds to Runs~A1--A8 in
Table~\ref{tab:runs}. The differential rotation parameters of these runs are
shown as red triangles in Fig.~\ref{fig:bistab}. Secondly, we start
from Run~D with SL differential rotation and increase $\delta n$
slowly to produce Runs~D1--D4.
These are shown as black asterisks in Fig.~\ref{fig:bistab}.
We see that both sets of simulations produce similar results and there is
no evidence for the existence of multiple solutions at the same parameters.
Therefore we conclude that the bistable nature of the differential
rotation, recently discovered by \cite{GYMRW14} and \cite{KMB14},
disappears when dynamically important magnetic fields are allowed
to be generated.
This conclusion is supported by the recent study of \cite{FF14}
who find a stable SL differential rotation 
independent of the history of their convective dynamo simulations.

\subsection{Meridional circulation}\label{sec:mc}

\begin{figure}[t]
\centering
\includegraphics[width=0.494\columnwidth]{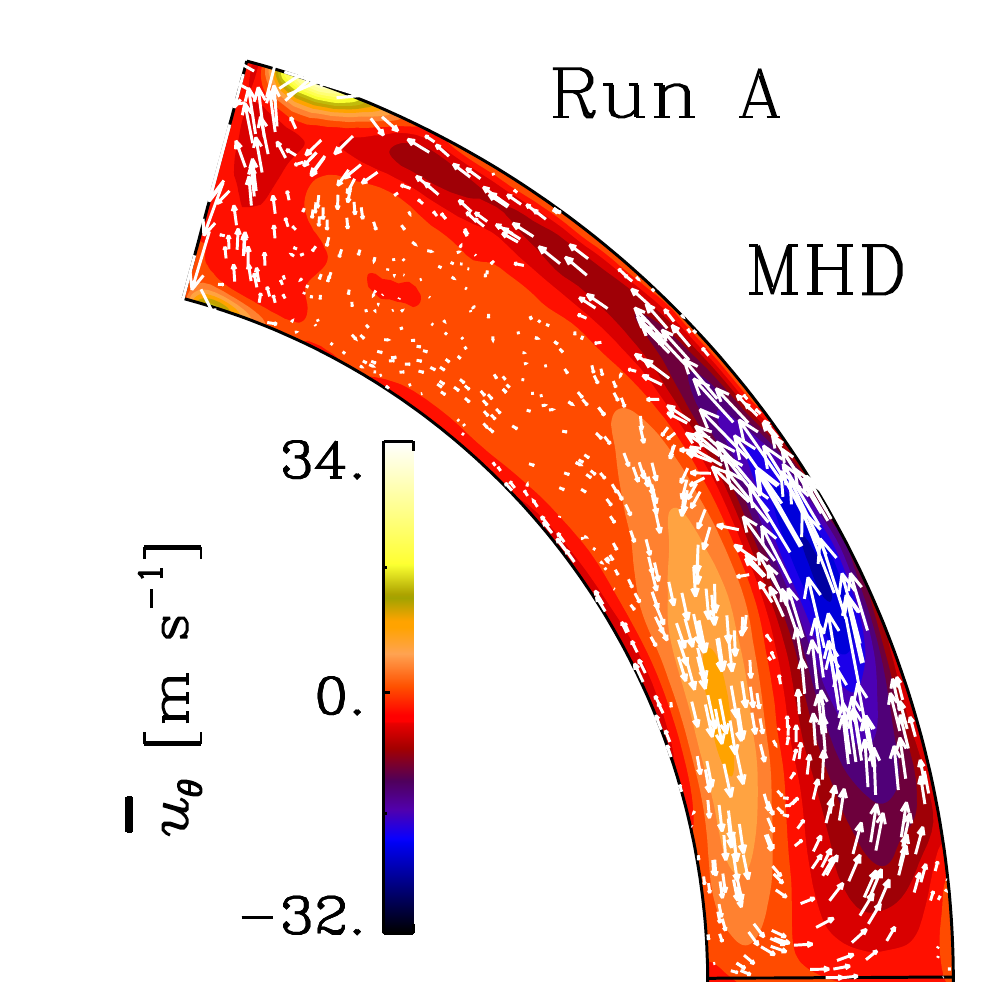}
\includegraphics[width=0.494\columnwidth]{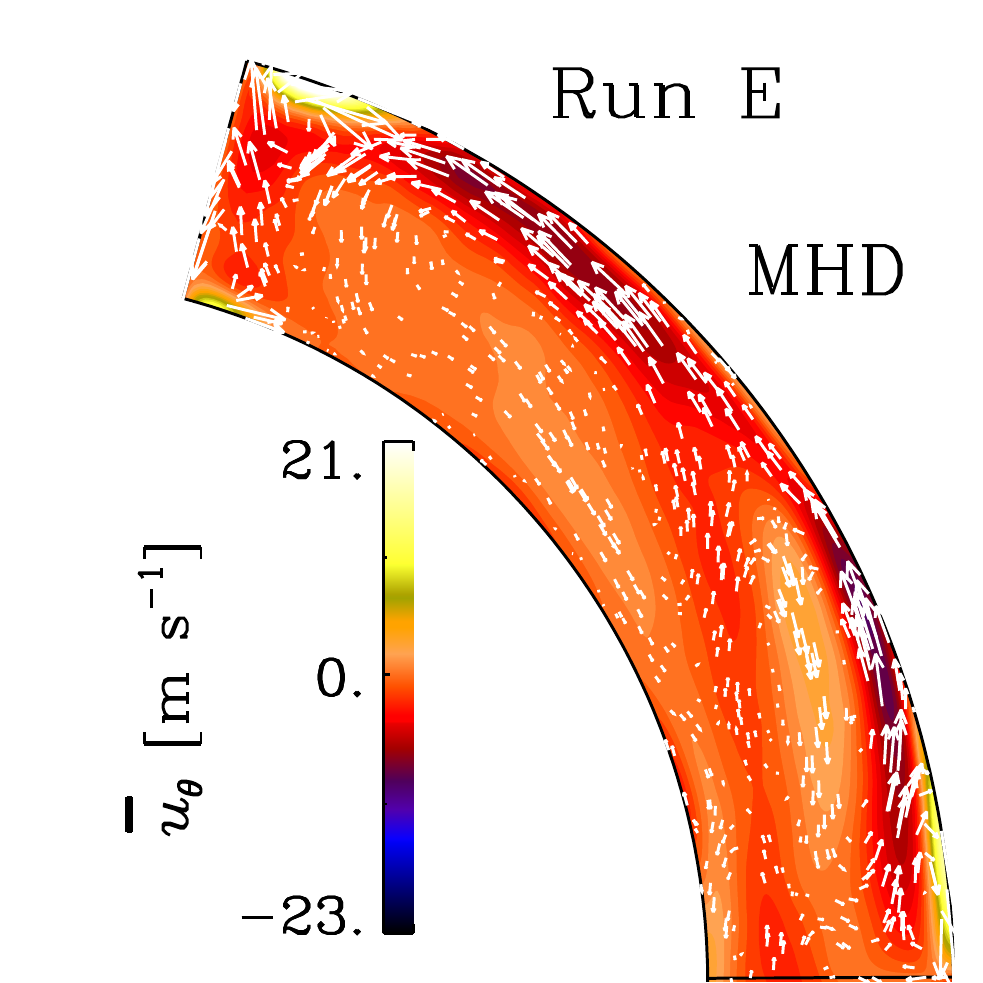}
\includegraphics[width=0.494\columnwidth]{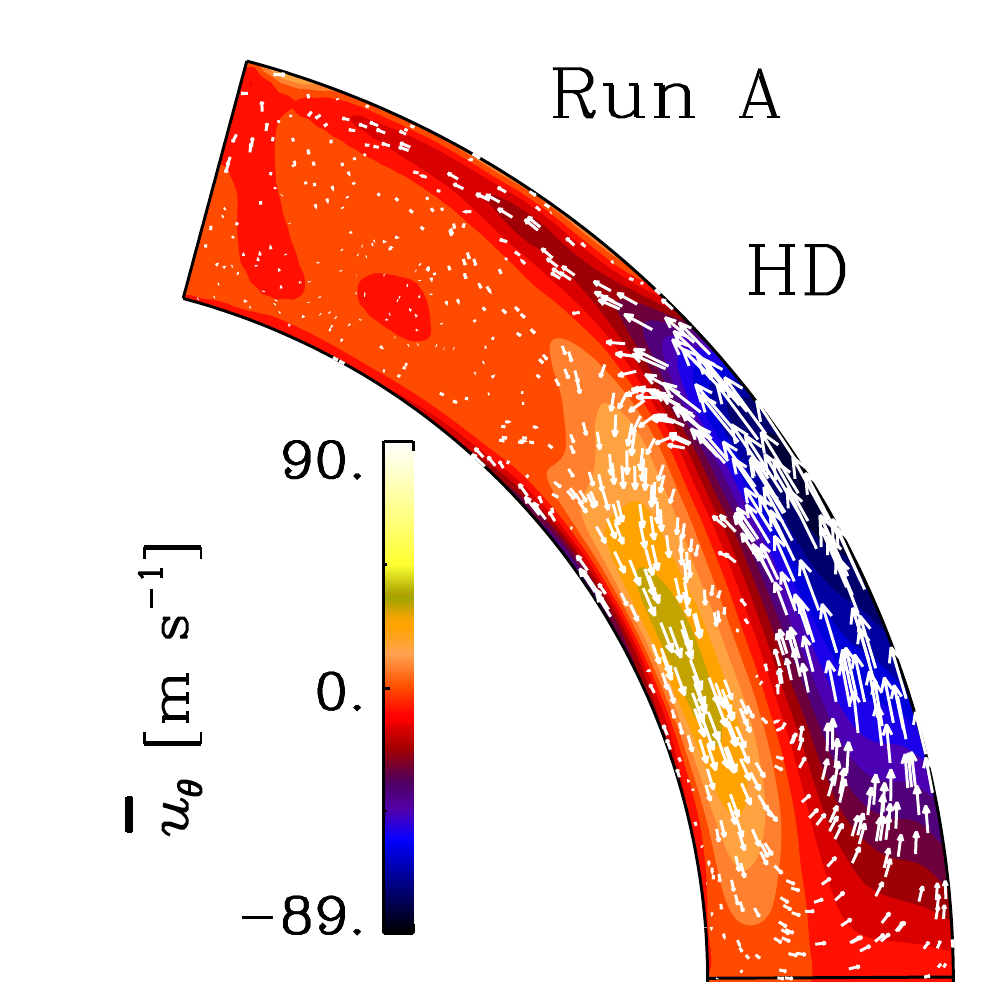}
\includegraphics[width=0.494\columnwidth]{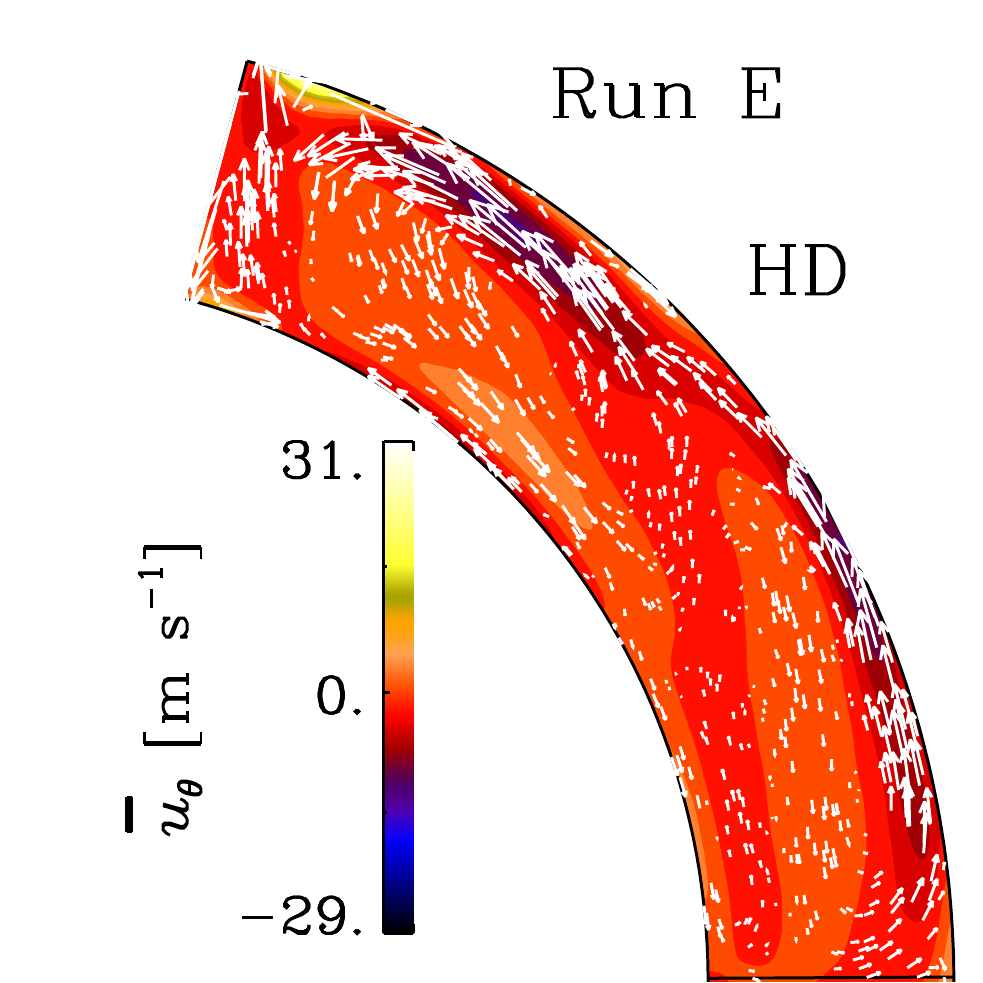}
\caption{Meridional circulation from Runs~A and E. The arrows show the direction of flow 
${\mean{\bm u}}_{\rm m} \equiv (\bar u_r,\bar u_\theta)$ and the background 
color shows $u_\theta$.
Upper (lower) panels are from magnetohydrodynamic (hydrodynamic) simulations.
}\label{fig:mc}
\end{figure}

\begin{table*}[t!]
\centering
\caption[]{Summary of the diagnostic quantities of Runs~A--E.}
      \label{tab:diagnostic}
      \vspace{-0.5cm}
     $$
         \begin{array}{p{0.05\linewidth}ccccccccccrccccccclc}
           \hline
           \noalign{\smallskip}
Run &P_1 {\rm [yr]} (N) &P_2 {\rm [yr]} (N) &P_1 {\rm [yr]} (S) &P_2 {\rm [yr]} (S) &B_{\rm eq} {\rm [G]} & \mean B_{\rm rms} {\rm [G]} &\frac{E_{\rm mag}}{E_{\rm kin}}& \frac{E_{\rm pol}}{E_{\rm kin}} &\frac{E_{\rm tor}}{E_{\rm kin}} & \frac{E_{\rm mer}^{\rm HD}}{E_{\rm mer}^{\rm MHD}} & \frac{E_{\rm rot}^{\rm HD}}{E_{\rm rot}^{\rm MHD}} & {\rm P_{asym}}\\ \hline
A & {\bf 2.96\pm0.22} &4.04\pm0.70 &{\bf 2.85\pm0.25} &4.05\pm0.24 & 3931 & 11412 & 0.119  &   0.066  &   0.052   & 12.1 & ~92.4 & {\rm no} \\ 
B & 5.45\pm0.20 &{\bf 7.88\pm0.57} &{\bf 6.5\pm1.5} &8.7\pm1.4 & 2956 & 11339 & 0.068  &   0.038  &   0.030   & 16.1 & 114.8 & {\rm no} \\ 
BC& 10.6\pm0.33 &- &10.8\pm0.66 &- & 2724 & 11150 & 0.060  &   0.033  &   0.026   &   -  &    -  & {\rm no} \\ 
C & 7.0\pm1.1 &- &- &- & 1789 & 10686 & 0.028  &   0.017  &   0.012   & 37.6 & ~36.9  & {\rm yes}\\ 
D & 6.9\pm1.1 &- &9.4\pm1.3 &- & 3152 &  9220 & 0.120  &   0.064  &   0.053   & ~2.2 & ~~1.2  & {\rm yes}\\ 
E & {\bf 4.49\pm0.38} &2.60\pm0.67 &2.58\pm0.88 &- & 3472 &  8792 & 0.156  &   0.080  &   0.076   & ~1.9 & ~~2.2 & {\rm yes} \\ 
\hline
         \end{array}
     $$
\tablefoot{
Here, $E_{\rm mag}=\brac{\bm{B}^2}/2\mu_0$ is the total magnetic energy, 
$E_{\rm pol}=\brac{(\mean{B}_r^2+\mean{B}_\theta^2)}/2\mu_0$ 
and $E_{\rm tor}=\brac{\mean{B}_\phi^2}/2\mu_0$ are the poloidal and toroidal components 
of the energy of the azimuthally averaged magnetic field.
All quantities are averaged over volume 
and in time over the thermally relaxed state.
Last two columns show the ratios of the meridional 
circulation and rotational energies from hydrodynamic to the magnetic simulations.
\blue{
The first four columns list the candidate periods detected with the
$D^2$ statistics separately for the northern and southern hemispheres.
In a multiperiodic case, the boldface font indicates the most significant period. 
The last column marks cycle period asymmetry between the two hemispheres.
}
}\end{table*}

For all the AS cases (Runs~A, B, and BC) we find single cell
meridional circulation with poleward flow near the surface and
equatorward flow near the bottom of the convection zone. 
This is also the usual assumption in flux transport dynamo
models \citep[e.g.,][]{DC99}, although in such models only the equatorward motion
at the bottom of the convection zone matters \citep{HKC14}.
However, as we go to the SL differential rotation cases, i.e., 
from Run~C to Runs~D and E, the meridional circulation becomes weaker 
and shows multiple cells in radius and latitude,
which has been detected in recent observations \citep{ZBKDH13,STR13,KSJ14}. 
\cite{GSKM13} also find multi-cell meridional circulation for SL differential rotation
and single or two-cell circulation for AS in their hydrodynamic simulations.

Table~\ref{tab:runs} shows that $E_{\rm mer}/E_{\rm kin}$ decreases rapidly
from Runs~A to E (with increasing $\Co$ the energy in the azimuthal component increases).
The upper two panels of \Fig{fig:mc} show the meridional circulation 
for an AS (Run~A) and a SL case (Run~E).
Note that, irrespective of the differential rotation profile, we obtain
a poleward flow
near the surface and its amplitude is in agreement with solar
surface observations \cite[see e.g.,][]{HR10,ZBKDH13}.
The fact that it is poleward both for AS and SL rotation
suggests that the meridional circulation is not just
a consequence of differential rotation.
\blue{
This has been discussed in detail by
\cite{R89}, who has shown that baroclinic forcing can also be an important
driver for the origin of meridional circulation; see \cite{MT09} (Sect.~3),
who have demonstrated how in simulations the convective (and magnetic)
angular momentum flux maintain meridional circulation
in the solar convection zone through the gyroscopic pumping. 
We note that in our simulations we do not have the near-surface shear layer, 
which helps to produce a poleward flow in the upper layers 
through inward angular momentum transport possibly by the downflow 
plumes \citep{KR05,MH11,HRY14b}.
}

It is important to compare these results with
the hydrodynamical counterparts of the same models
shown in the two lower panels. 
We see that the hydrodynamic flow is much stronger, although the overall
pattern is not very different.
In Table~\ref{tab:diagnostic}, we compare the energy ratios
$E_{\rm mer}^{\rm HD}/E_{\rm mer}^{\rm MHD}$ and
$E_{\rm rot}^{\rm HD}/E_{\rm rot}^{\rm MHD}$ of meridional circulation
and rotation respectively with their hydrodynamic counterparts.
The magnetic field clearly suppresses the circulation in the AS cases,
Runs~A--BC, but also in Run~C,
whereas in the other SL cases the effect is small.
Moreover, the flow shows significant temporal variation,
which will be explored later in \S~\ref{sec:modulation}.

\subsection{Magnetic \blue{variability} and butterfly diagrams}\label{sec:bf}

\begin{figure}[t]
\centering
\includegraphics[width=.96\columnwidth]{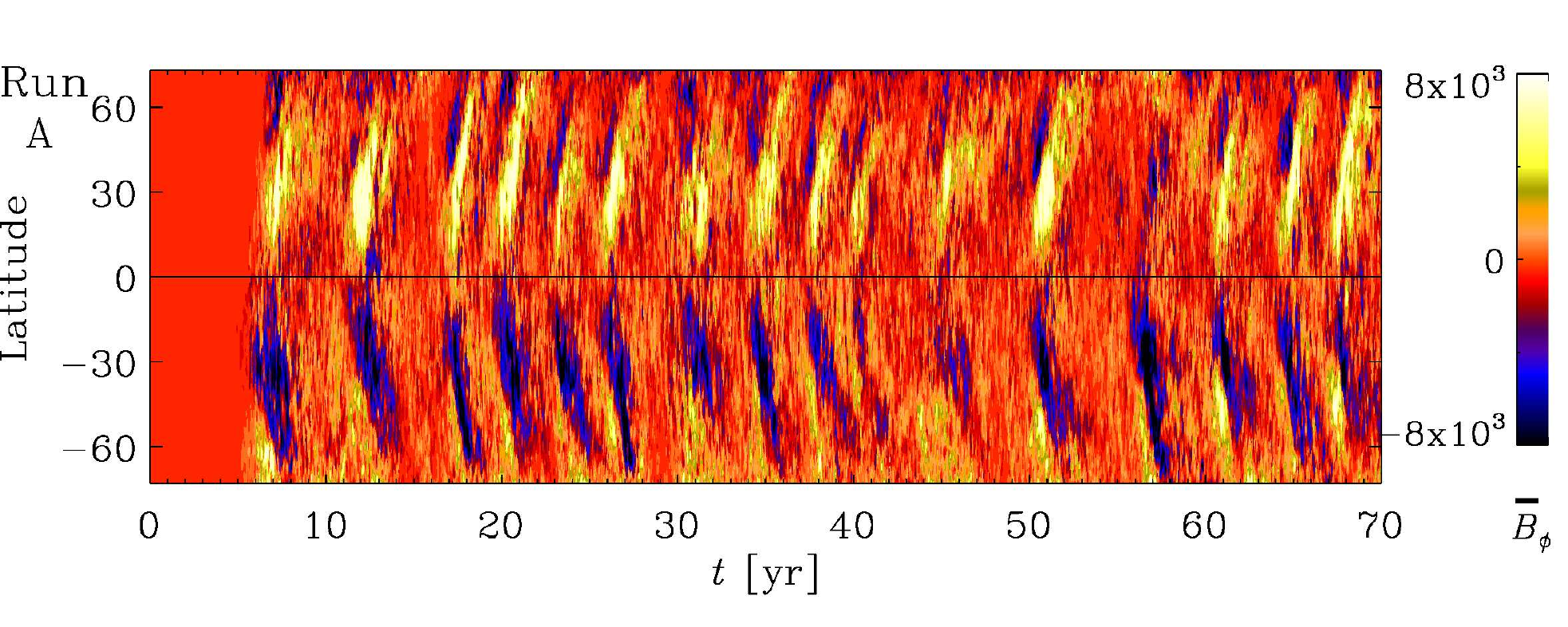}
\includegraphics[width=.96\columnwidth]{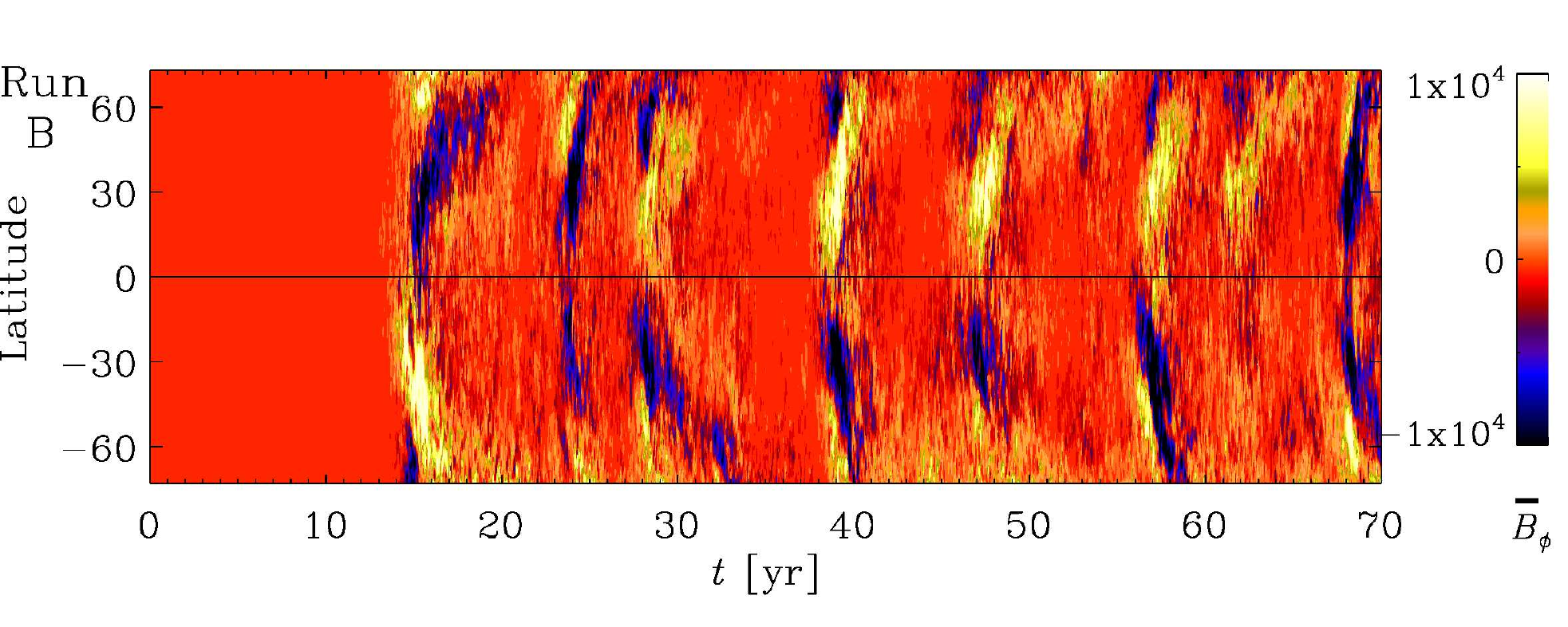}
\includegraphics[width=.96\columnwidth]{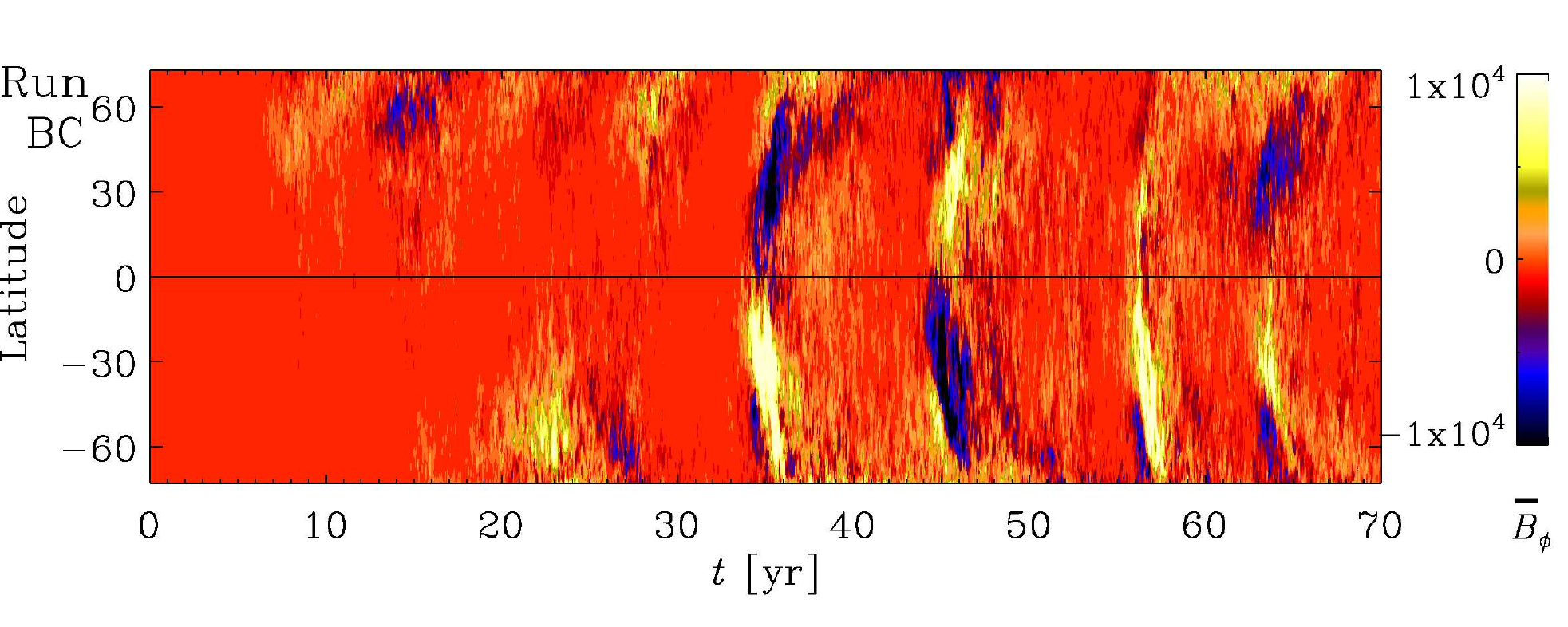}
\includegraphics[width=.96\columnwidth]{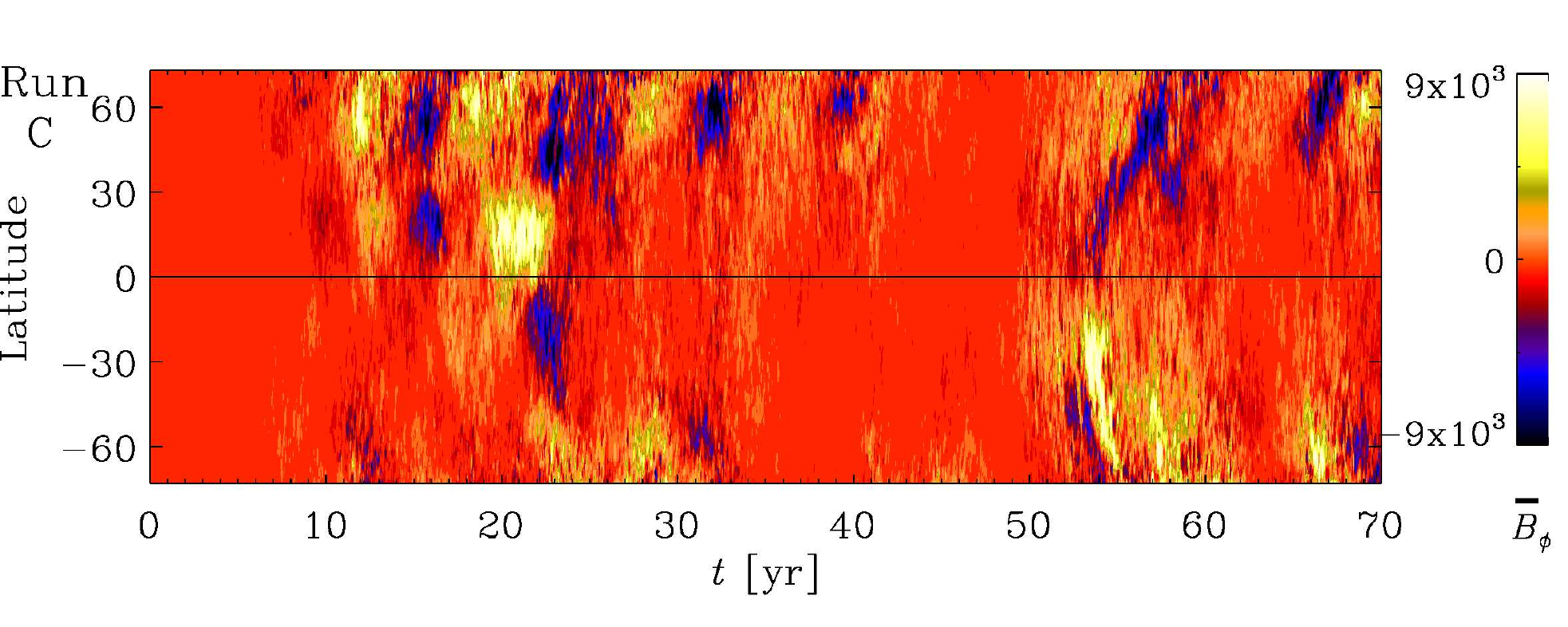}
\includegraphics[width=.96\columnwidth]{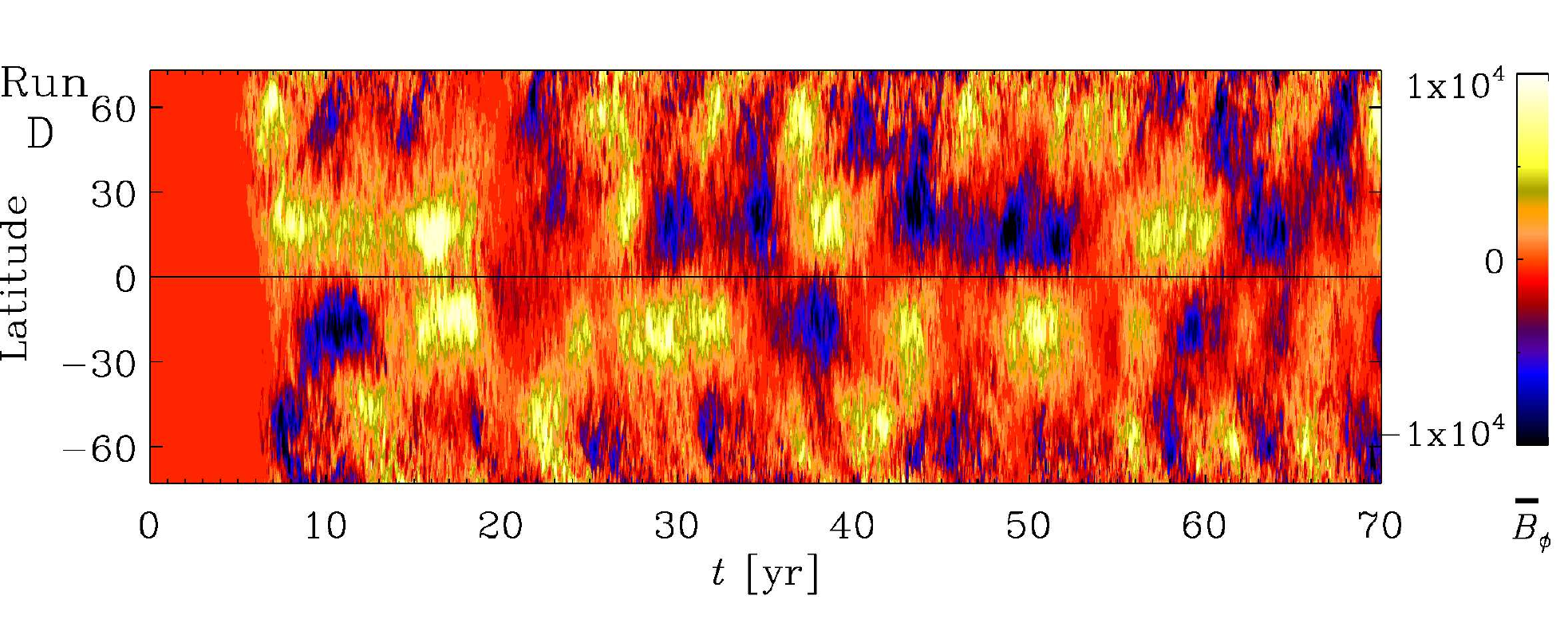}
\includegraphics[width=.96\columnwidth]{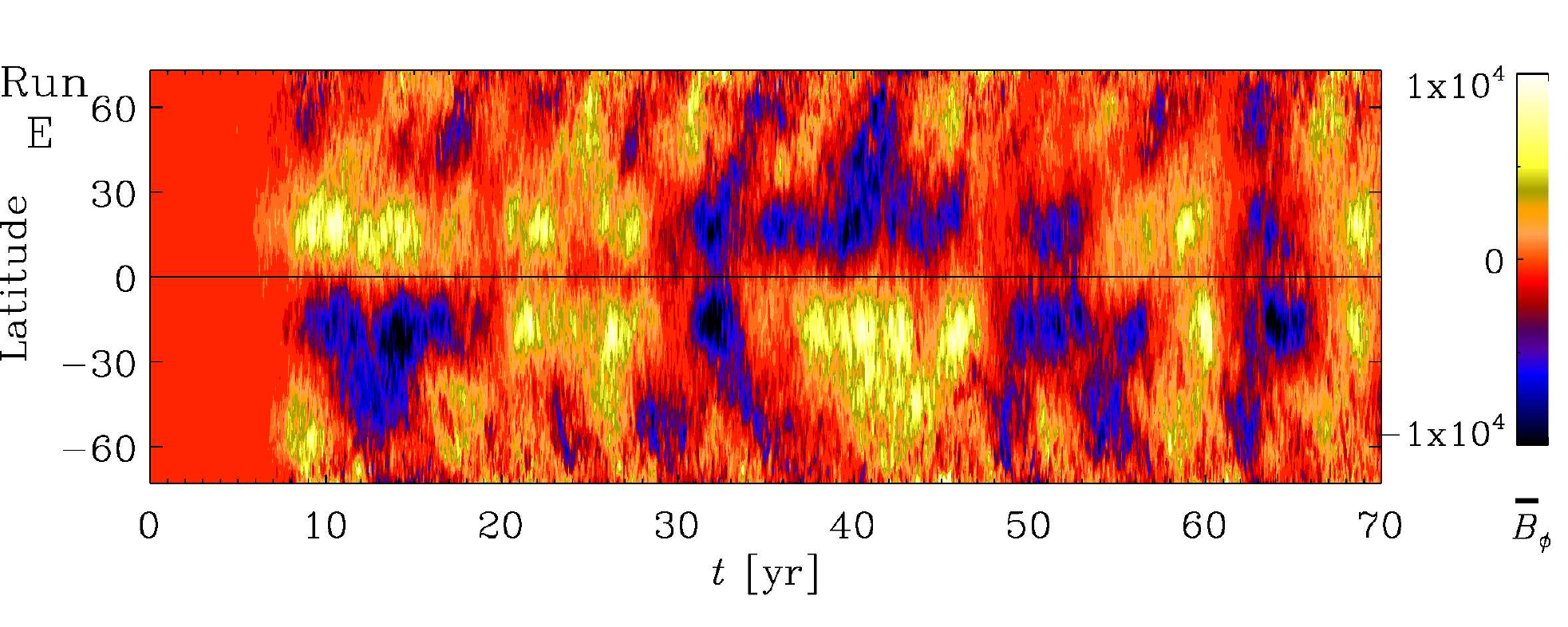}
\caption{Butterfly diagrams: contours of the toroidal field $\mean B_\phi$ (in Gauss) 
at $r = 0.74~R_\odot$ from Runs~A, B, BC, C, D, and E (top to bottom).
}\label{fig:bfbottom}
\end{figure}

\begin{figure}[t]
\centering
\includegraphics[width=.96\columnwidth]{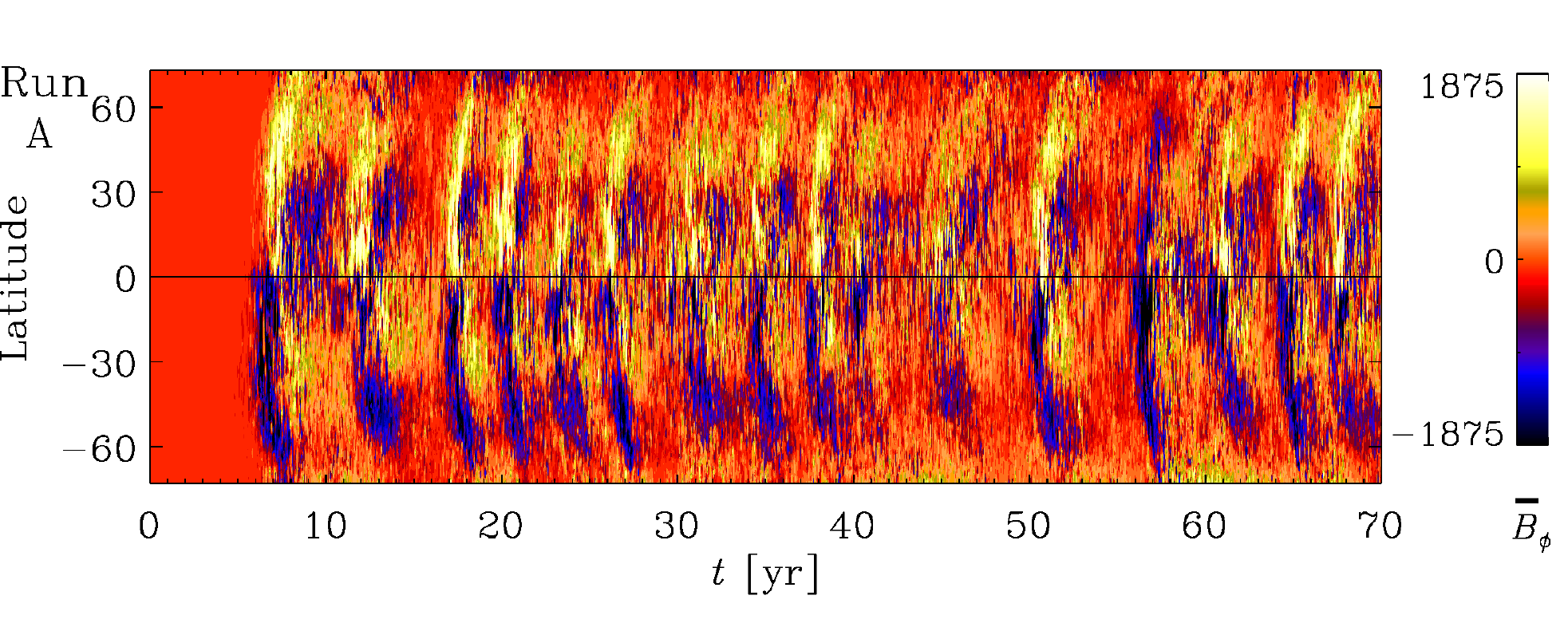}
\includegraphics[width=.96\columnwidth]{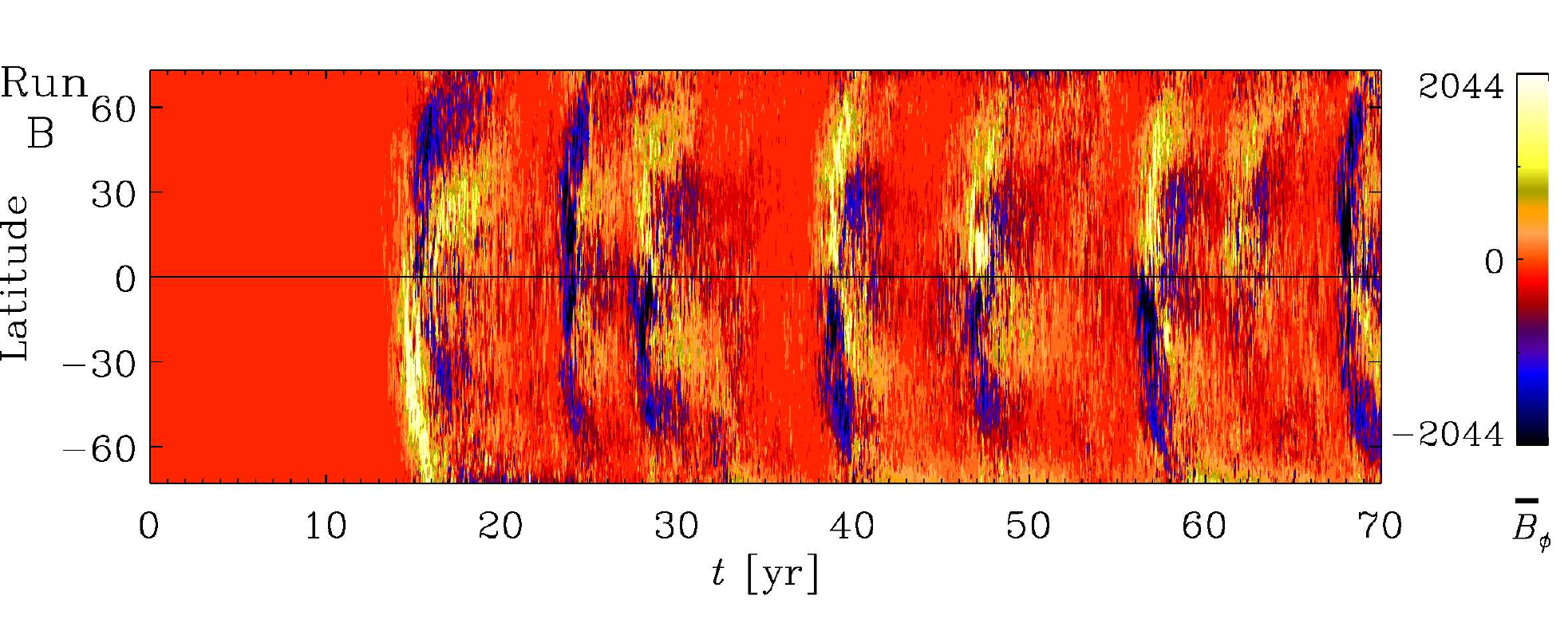}
\includegraphics[width=.96\columnwidth]{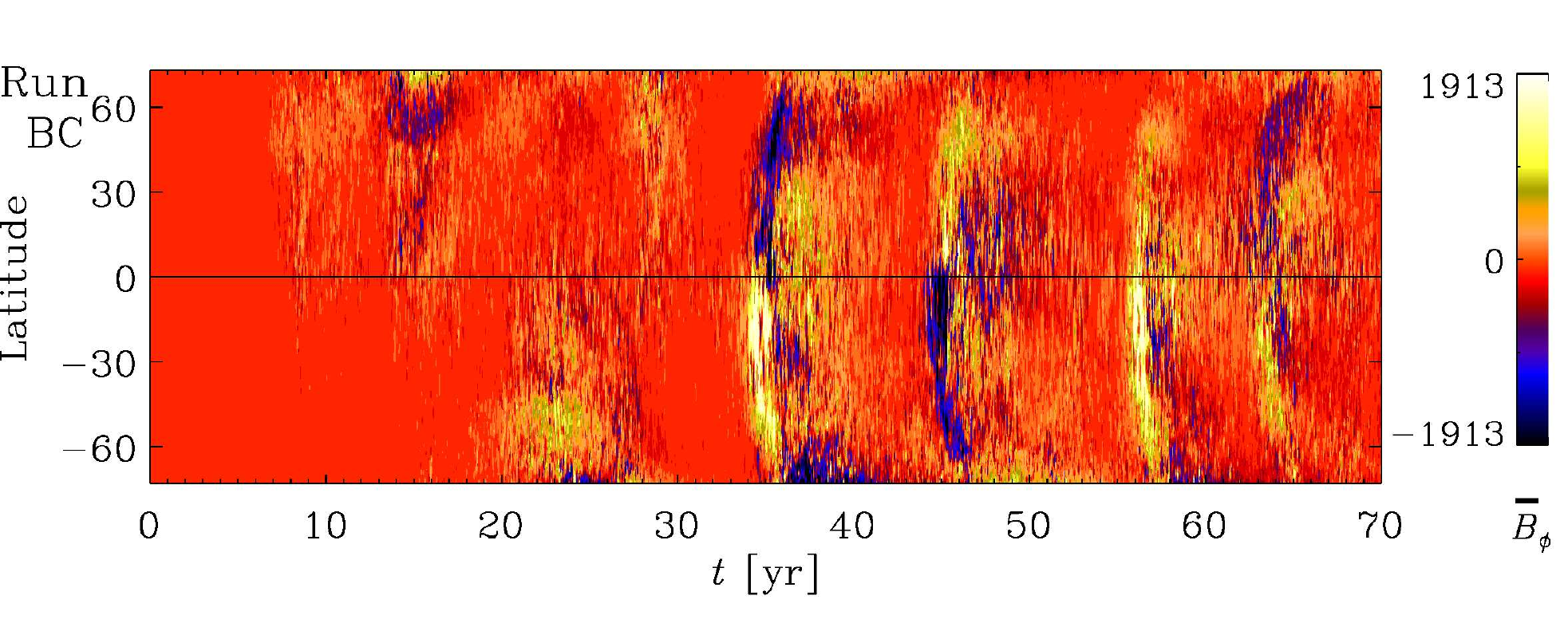}
\includegraphics[width=.96\columnwidth]{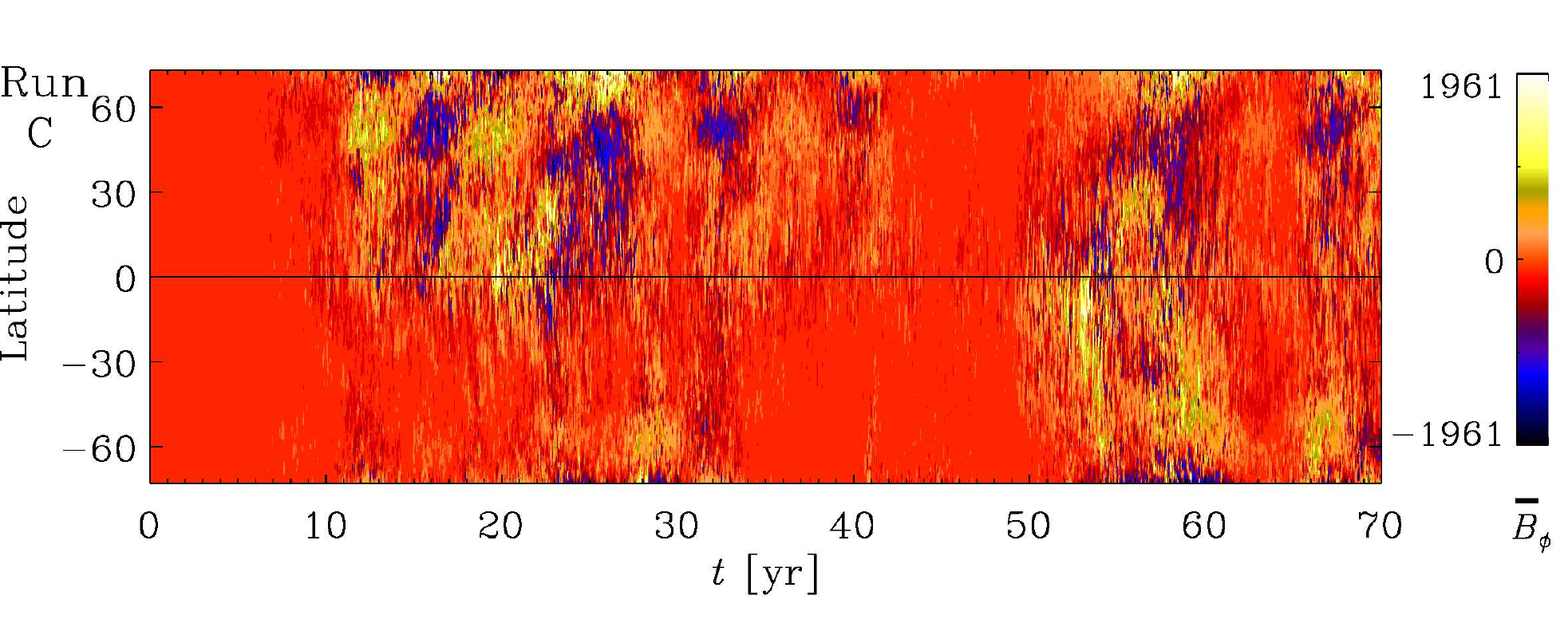}
\includegraphics[width=.96\columnwidth]{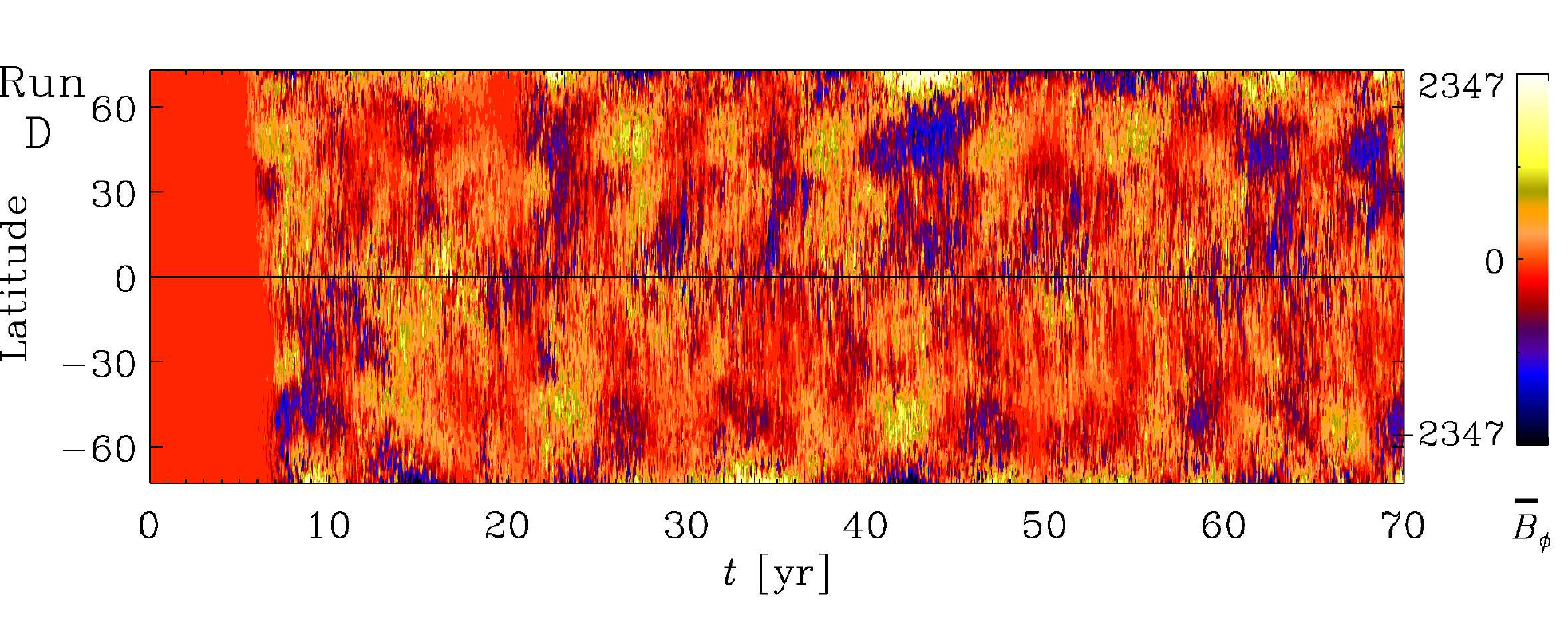}
\includegraphics[width=.96\columnwidth]{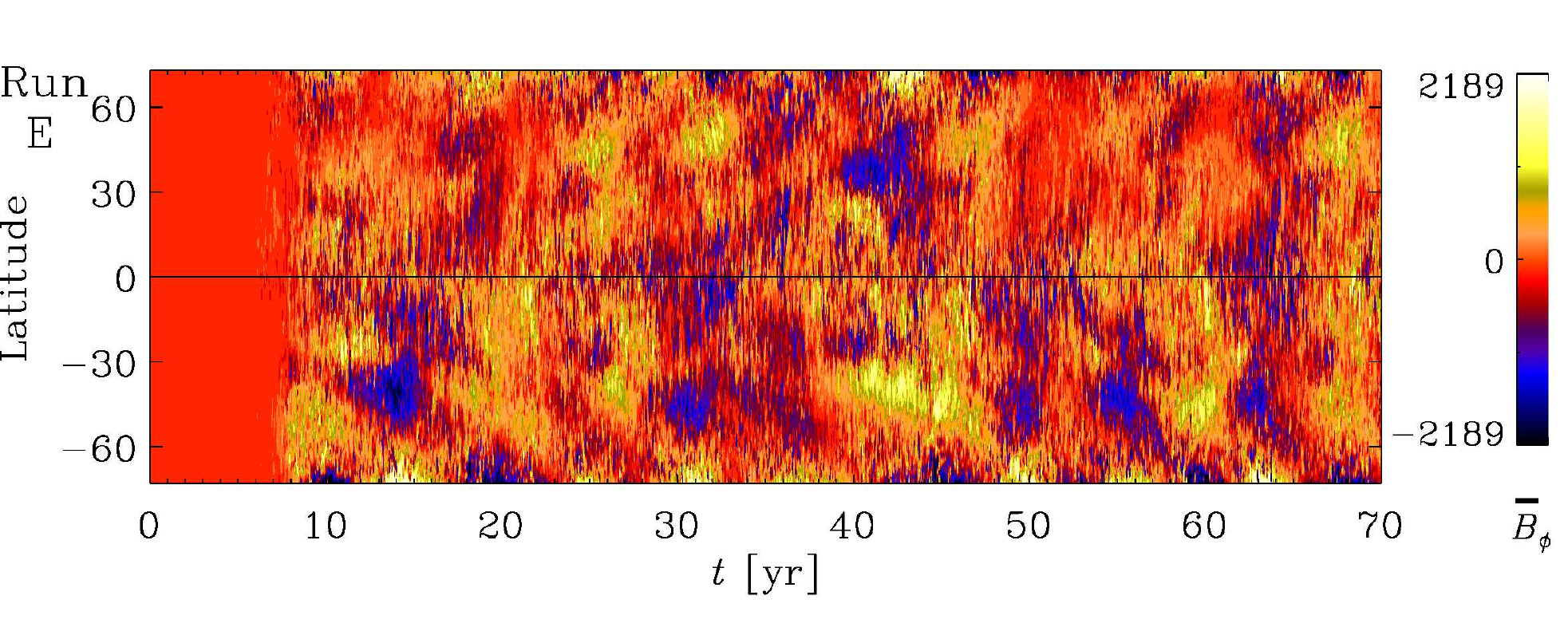}
\caption{Same as \Fig{fig:bfbottom} but here $\mean B_\phi$ is shown at $r = 0.96~R_\odot$.
}\label{fig:bf}
\end{figure}

The large-scale spatio-temporal organization of the magnetic field can be
seen from a time--latitude or butterfly diagram of $\mean B_\phi$,
for example.
The degree of radial coherence can be judged by looking at different depths.
We show such butterfly diagrams for Runs~A--E both
at $r = 0.74~R_\odot$ (Fig.~\ref{fig:bfbottom}) and
at $r = 0.96~R_\odot$ (Fig.~\ref{fig:bf}),
where $\mean B_\phi$ is given in Gauss.
Here we only show the first 70~years of each simulation,
although in some cases we ran for longer times (see below).
We see that Runs~A, B, and BC, which produce AS differential rotation,
show prominent \blue{activity cycles}, but no clear polarity reversals. 
Therefore, these \blue{activity} cycles are different from the Hale polarity
cycle of the Sun.
\blue{
However, we want to stress here that
polarity reversals are verified only for the Sun, whereas for stellar cycles,
most commonly detected either from photometry or from Ca H \& K lines with 
spectroscopy, such information is not retrievable, and therefore the
reported variability might equally well be related to variations in the
magnetic field strength as to polarity reversals.
}

For Runs~A and B, the \blue{activity} cycles are associated with
a weak poleward propagation at high latitudes.
Another aspect we notice is that, as we go from Run~A to Run~BC,
the \blue{activity} cycles appear at a later time.
This is surprising given that the rotational influence
on the turbulence actually increases.
For Run~C, with SL rotation, the variations are rather irregular
and show periods of very weak mean fields.
The variations in Runs~D and E are irregular and a different dynamo mode
appears to be excited in these simulations, which is also
characterized by a weaker influence on the differential rotation than
in the AS cases.
By comparing Figs.~\ref{fig:bfbottom} and \ref{fig:bf}, we also see that 
$\mean B_\phi$ is stronger near the bottom of the
convection zone, which could be caused by downward pumping of the
mean magnetic field \citep{NBJRRST92,BJNRST96,TBCT01}.

Comparing with earlier work, we note that
in several recent global dynamo simulations
\citep{GCS10,RCGBS11,BMBBT11,KMB12,KMCWB13,CKMB14}, regular cycles
develop, resulting in butterfly diagrams with polarity reversals and
sometimes even equatorward migration of toroidal field at low latitudes.
However all these simulations are for larger Coriolis numbers
\citep[for example, $\Co\ge5$ in][]{CKMB14}, in which regime more coherent fields have been found to be favored
\citep{BBBMT10,BMBBT11}.

\subsection{Diagnostic stellar activity diagrams}\label{sec:Diagnostic}

The nature of stellar cycles can be characterized by the ratio of
cycle frequency $\omega_{\rm cyc}=2\pi/P$ to the rotation rate
$\Omega_0$,
where $P$ is an estimate of the cycle period.
There is a tendency
for stars of different characteristics to group at different positions
in a ``diagnostic'' diagram of $\omega_{\rm cyc}/\Omega_0$ versus $\Co$
\citep{BST98,SB99}.
The more rapidly rotating stars of \cite{KMCWB13} were found to be
located in three groups with increasing slope in two of them and
decreasing slope in one.

\begin{figure*}[t]
\centering
\includegraphics[width=1.0\textwidth]{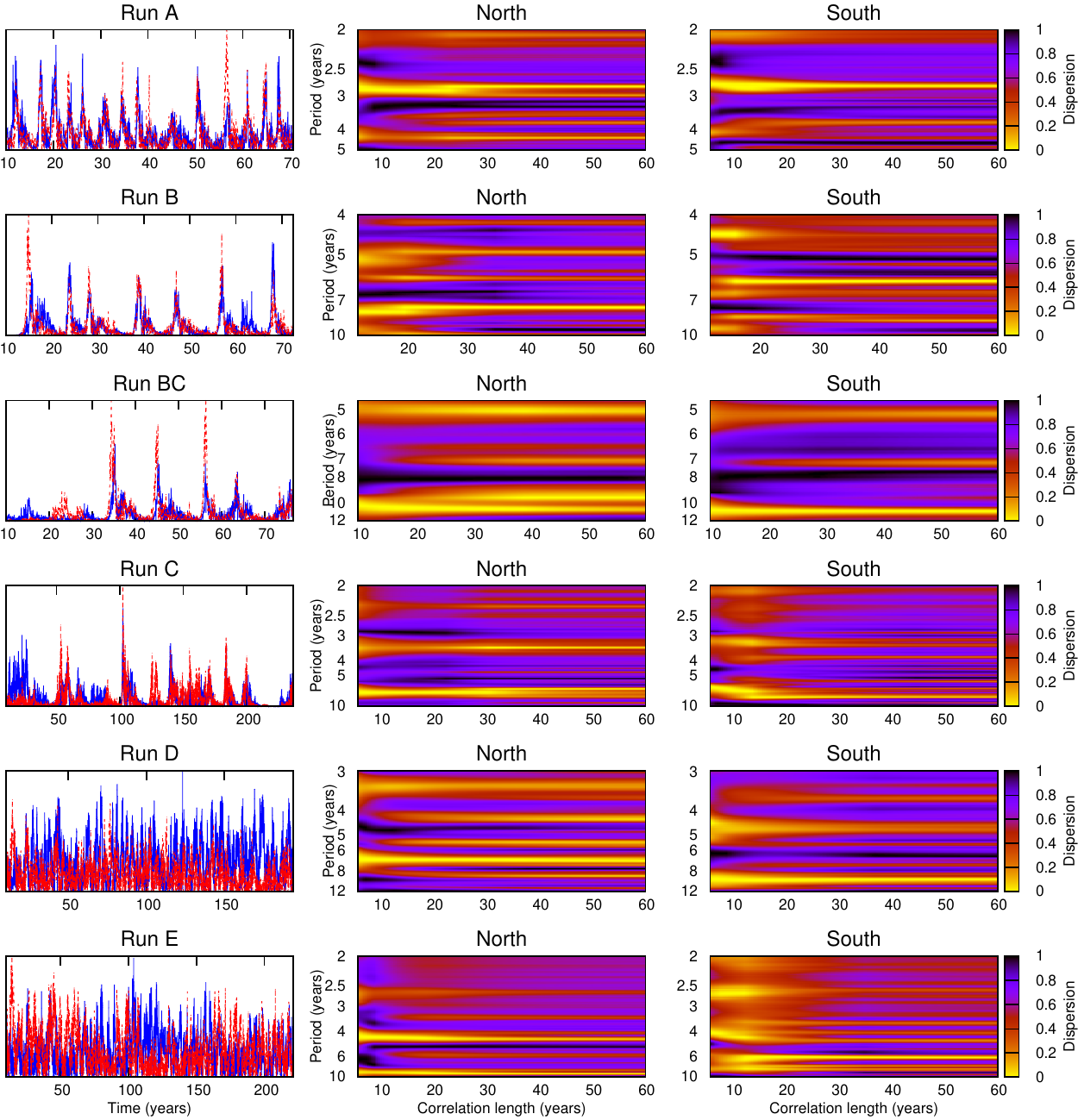}
\caption{Phase dispersion analysis results; from top to bottom, Runs~A--E,
in the middle column for the northern hemisphere, on the right for the southern 
hemisphere. On the left column we show the original time series 
analyzed (blue/red for north/south).
The time series are normalized by their means, so values are not shown.
}\label{fig:D2_all}\end{figure*}

In the present work, the values of $\Co$ are much smaller, so it is
important to repeat such an analysis for the more slowly rotating stars
of the present paper.
\blue{
However, even by visual inspection of the butterfly diagrams, it is evident that
these variations are not strictly harmonic, and therefore Fourier transform is
not useful for the analysis. Instead, we use the phase dispersion method 
\citep{pelt1983,lindborg2013}. It is based on the statistics 
\begin{equation}
D^2(P) = \frac{1}{2\sigma^2}\frac{{\sum\limits_{i = 1}^{N - 1} {\sum\limits_{j = i + 1}^N {g(t_i } } ,t_j ,P,\Delta t)[f(t_i ) - f(t_j )]^2 }}{{\sum\limits_{i = 1}^{N - 1} {\sum\limits_{j = i + 1}^N {g(t_i } } ,t_j ,P,\Delta t)}},
\end{equation}
where $f(t_i),i = 1,\dots,N$ is the input time series, $\sigma^2$ is its variance, $g(t_i,t_j,P,\Delta t)$ is the selection function,
which in the general case is significantly different from zero only when
\begin{eqnarray}
t_j  - t_i  &\approx& kP,k =  \pm 1, \pm 2, \ldots {\rm \ \ \ and}\\
\left| {t_j  - t_i } \right| &\le& \Delta t.
\end{eqnarray}  
In the latter condition, $\Delta t$ is the so-called {\it correlation length}.
In our case the time series is evenly sampled, in which case
a modification to the general top-hat selection function
is needed to prevent artefacts: in this study we choose $g$ as the
product of two Gaussians: one with a half-width-at-half-maximum (HWHM)
of $\Delta t$ and the other with an HWHM of a preselected phase
separation limit, for which we adopt 0.1.
For the particular case when $\Delta t$ is longer than the full data span,
the $D^2(P)$ statistics is essentially a slight reformulation
of the well known Stellingwerf statistics \citep{stellingwerf}.
As the correlation length is made shorter, we match nearby cycles in
a progressively narrower region, and consequently estimate a certain
mean period, which needs not to be coherent for the full time span.
After having continued Runs~C, D and E for a longer time, we 
apply this statistics to the time series of $\mean B_\phi^2$
at $r = 0.96~R_\odot$, averaged over $10^\circ$ to $50^\circ$ latitude,
separately for north and south.
This is shown in the left column of \Fig{fig:D2_all}, where we now
depict the full length of the simulation, excluding however the initial
exponential growth phase of the dynamo.}

\blue{
To determine the possible average period of the cycle in the time
series we proceed as follows: first we calculate the $D^2$ statistics
for preselected period and correlation length ranges.
Then looking at the plot we detect the correlation lengths
for which there exist only distinct flat minima around certain periods (this must be the case for the leftmost correlation length if the lower bound has been chosen correctly). After eliminating doubles or halves of the actual periods we obtain a set of candidate periods. The period with the 
strongest minimum is selected as the best candidate, others being possible modulating periods.
}

\blue{
In Fig.~\ref{fig:D2_all}, the results for the phase dispersion analysis
are depicted; the periods with error estimates for all the runs can be
found from Table~\ref{tab:diagnostic}.
The error estimates were calculated by building bootstrap resamples for
the differences between pairs of data points with approximately the same
time separation.
For Run~A, a stable period of slightly less than three years is prominent especially for the southern hemisphere, which is manifested by the fact that the 
minimum does not split even if the correlation length is increased. For the northern hemisphere, a period with roughly the same value is obtained, but 
it is less coherent and splits at correlation lengths of roughly 25 years. A secondary period of roughly 4 years is also present in both hemispheres.
The hemispheres appear less synchronized for Run~B: again, two significant
periods are detected for both hemispheres, the most prominent ones being 8
and 7 years, respectively (albeit with large error estimates).
Again, the dominant periods persist even when the correlation length
is increased.
For Run~BC, a stable period of about 11 years is found in both hemispheres.
In the southern hemisphere, it is more persistent as the correlation length
is increased.
For Run~C, only the northern hemisphere shows a prominent period
around 7 years, while the splitting of the minima starts already for
the correlation length of 10 years in the south.
Run~D shows prominent periods around 7 and 9 years for the northern
and southern hemispheres, respectively.
Although splitting occurs especially in the north, we regard these periods
as significant.
Finally in the case of Run~E, multiperiodicity was detected for northern
hemisphere with stronger period around 4.5 years and a weaker one around
2.5 years, while the latter period was detected also for the southern
hemisphere.
}

\begin{figure}[t]
\centering
\includegraphics[width=0.94\columnwidth]{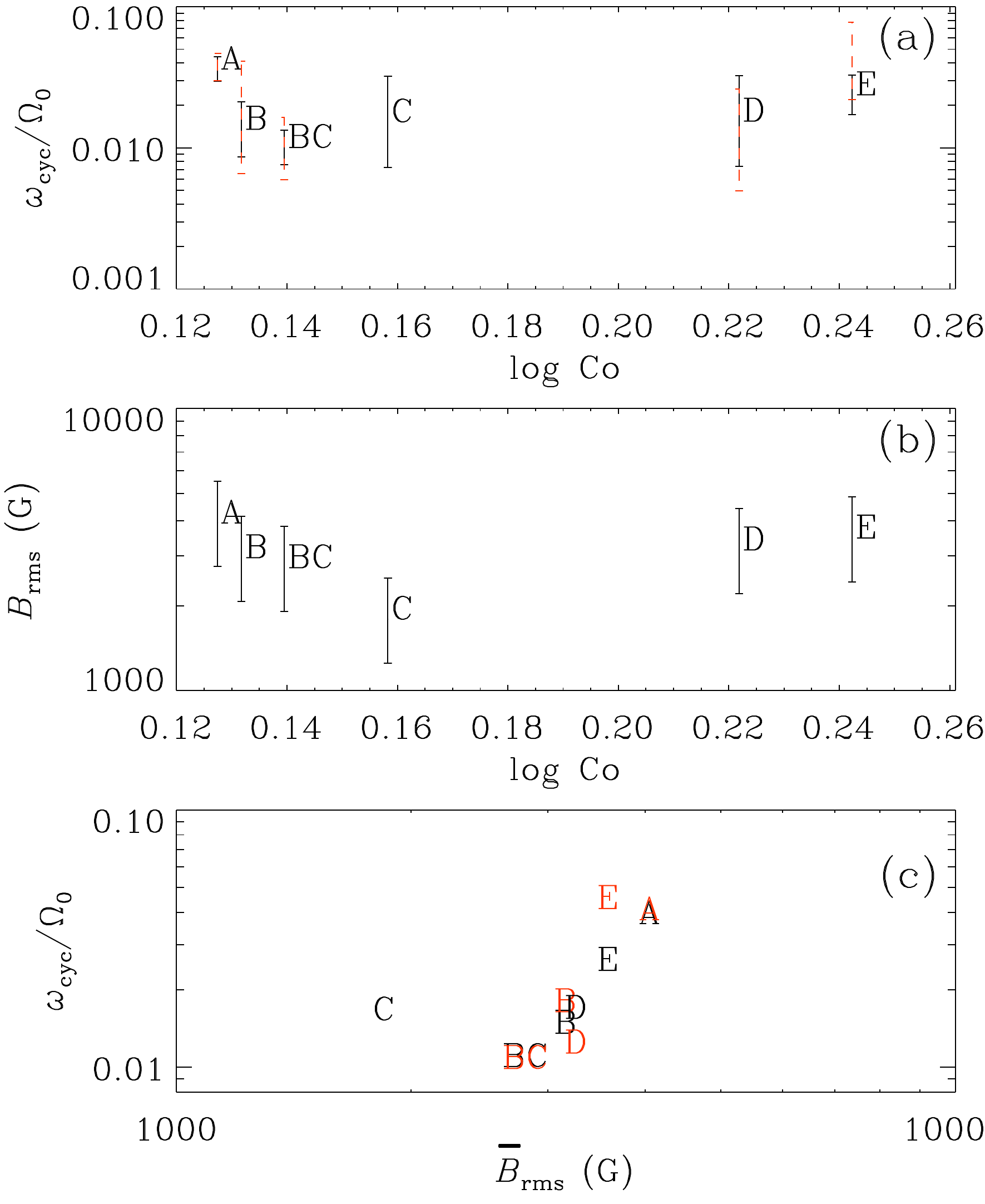}
\caption{Diagnostic diagrams showing
(a) $\omega_{\rm {cyc}}/\Omega_0$,
(b) the strength of the large-scale magnetic field
over the whole convection zone measured by
$\mean{B}_{\rm rms} = \brac{\brac{B_r}_{\phi}^2 + \brac{B_{\theta}}_{\phi}^2+\brac{B_\phi}_{\phi}^2}_{r\theta t}^{1/2}$
vs.\ log $\Co$, and (c) $\omega_{\rm {cyc}}/\Omega_0$
vs.\ $\mean{B}_{\rm rms}$.
The red color shows the values computed for the southern hemisphere.
}\label{fig:pocycom}
\end{figure}

The plot of $\omega_{\rm cyc}/\Omega_0$ versus $\Co$
is shown in \Fig{fig:pocycom}(a), where we see that stars with AS
and SL rotation are located in two different positions in such a diagram.
We know that the Sun lies on the upper left branch
in such a diagram for real stars \citep{BST98}, but in
\Fig{fig:pocycom}(a) this branch corresponds to AS rotation, which
obviously disagrees with the observations.
However, it is plausible that there are stars with AS rotation that have
simply not yet been observed.
Our simulations therefore suggest a possible prediction in that such
stars might occupy a possibly separate branch further to the left of the
solar branch, which already corresponds to the domain of inactive stars.
We thus expect there to be either a separate branch or a part of the
branch with inactive stars like the Sun, having however AS rotation.

In \Fig{fig:pocycom}(b) we show the mean magnetic field. 
Here the errors of $\mean{B}_{\rm rms}$ are computed as the largest
departure of the mean from any one third of the full time series.
We see that for the AS differential rotation runs (Runs~A, B, and BC)
the magnetic field decreases with rotation rate, whereas for the SL
rotation cases (Runs~C, D, and E) it increases.

In \Fig{fig:pocycom}(c) we plot the cycle frequency ratio
against the mean magnetic field. 
Following the interpretation of \cite{BST98}, the cycle frequency ratio
is essentially a measure of the $\alpha$ effect in a mean-field dynamo,
so the increasing trend in the cycle frequency ratio suggests that
the $\alpha$ effect increases with magnetic field strength,
which is referred to as antiquenching.
This interpretation hinges on some ill-known assumptions, for example
the turbulent transport in this model is assumed to depend only on the
largest scale of the mean magnetic field and of course the rather
unconventional assumption of antiquenching itself.
Antiquenching of both $\alpha$ and turbulent diffusivity has actually
been detected in simulations \citep{CMRB11}, and may be possible more
easily in a sphere than in a Cartesian layer, but we have at present no
further indication that this interpretation is applicable to our model.

\subsection{Magnetic modulation of the flow}\label{sec:modulation}

\begin{figure}[t]
\centering
\includegraphics[width=1.05\columnwidth]{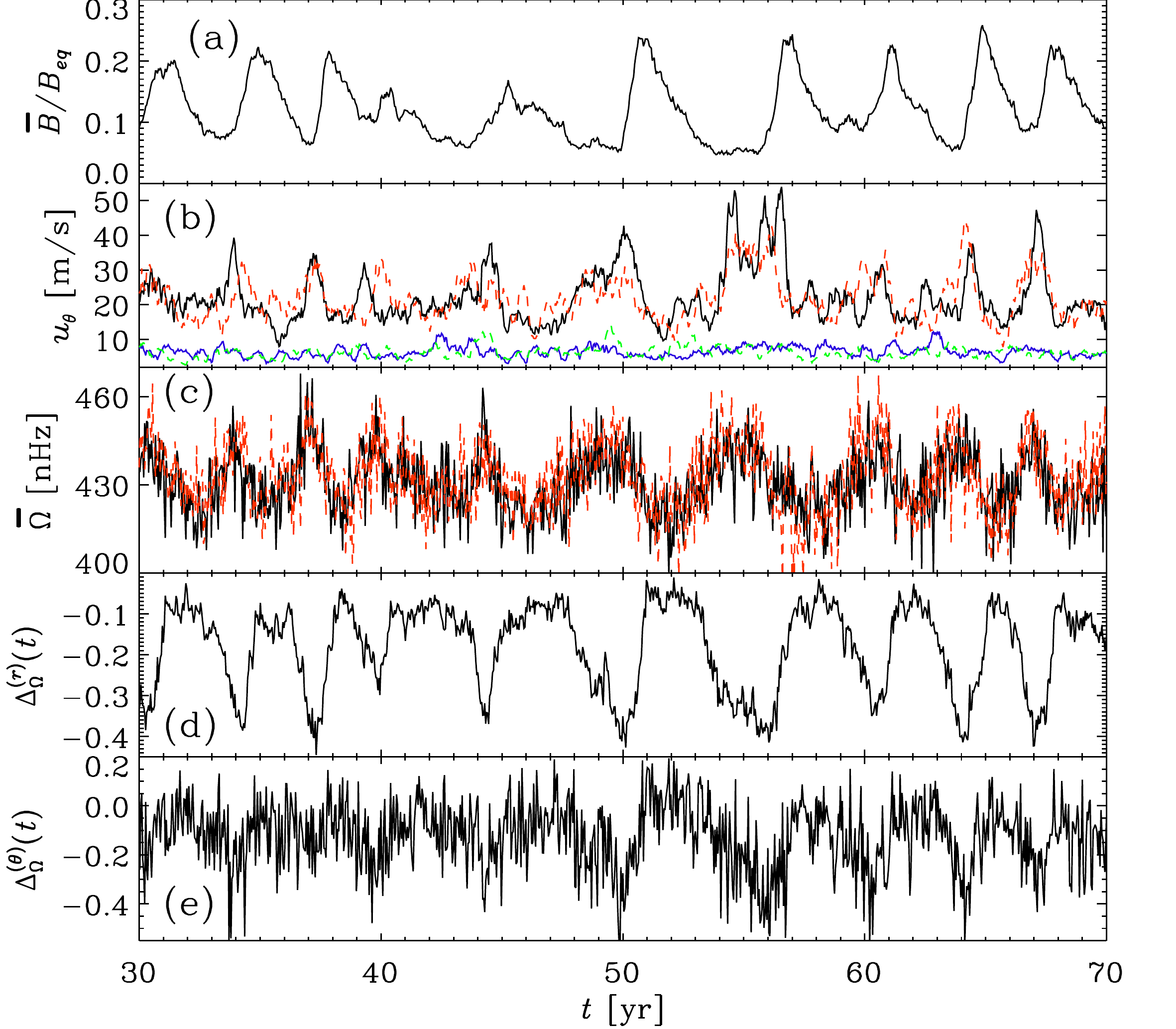}
\caption{From Run A: (a) The large-scale magnetic field over the whole convection zone
$\mean{B}= \brac{\brac{B_r}_{\phi}^2 + \brac{B_{\theta}}_{\phi}^2+\brac{B_\phi}_{\phi}^2}_{r\theta}^{1/2}$
normalized by $B_{\rm eq}$.
(b) The latitudinal component of meridional circulation $u_\theta(r,\pm32^\circ)$
(smoothed over 5 months) 
at $r\approx0.95R_\odot$ (black and red) and $r\approx0.73R_\odot$ 
(blue and green), 
(c) azimuthally averaged angular velocity $\mean\Omega(0.95R_\odot,\pm32^\circ)$,
\blue{
(d) $\mean\Omega(r,0^\circ)$ at $r=0.73R_\odot$ (red dashed) and $r=0.95R_\odot$ (black).
}
The dashed (solid) line corresponds to the southern (northern) hemisphere.
(e) Radial shear $\Delta_\Omega^{(r)}$, 
and (f) latitudinal shear $\Delta_\Omega^{(\theta)}$,
defined in \Eq{equ:pDRt}, as functions of time.
}\label{fig:flucA}
\end{figure}

We have seen that some of our runs show clear \blue{activity} cycles.
Therefore we expect to see a corresponding modulation of the flow.
In \Fig{fig:flucA}, we show for Run~A the temporal variation of the mean large-scale magnetic 
field ($\mean B$) normalized by $B_{\rm eq}$,
the latitudinal component of the meridional circulation $u_\theta(r,\pm32^\circ)$
at $r\approx0.95R_\odot$ and $r\approx0.73R_\odot$,
the mean rotation rate $\mean\Omega(0.95R_\odot,\pm32^\circ)$,
as well as the latitudinal and 
radial differential rotation $\Delta_\Omega^{(r)}$ and
$\Delta_\Omega^{(\theta)}$, defined in \Eq{equ:pDRt}. 
We see that the meridional circulation varies with the magnetic 
\blue{field}, becoming weaker during maximum and stronger during minimum, 
the overall temporal variation being about $50\%$ in this case.
\blue{(The linear correlation coefficient between $\mean B$ and $u_\theta(0.95R_\odot,\pm32^\circ) 
\approx -0.36, -0.38$.)}
This kind of weak anti-correlation between the activity cycle and the meridional flow 
has been found in solar observations \citep{CD01,HR10} and is believed to arise 
at least in part from the Lorentz force of the dynamo-generated 
magnetic fields \cite[see e.g.,][]{Re06,KC12,PCB12}. 
\blue{ 
The meridional circulation at the bottom is also weakly correlated with the activity cycle
(correlation coefficients between $\mean B$ and $u_\theta(0.73R_\odot,\pm32^\circ) \approx -0.22, -0.47$).
We see that $\mean\Omega(0.95R_\odot,\pm32^\circ)$  (\Fig{fig:flucA}(c)) also shows a weak anti-correlation with
the magnetic variations
(having correlation coefficient $\approx-0.25$).
The strong magnetic fields during 
maxima change $\mean\Omega$ by a few per cent ($\approx 6\%$).
However $\mean\Omega(0.95R_\odot,0^\circ)$ (\Fig{fig:flucA}(d)) shows positive correlation
(correlation coefficient $\approx 0.36$) and the overall variation is larger ($\approx 12\%$).
Due to this variation of $\mean\Omega$ at the equator, the values of
$\Delta_\Omega^{(r)}$ and $\Delta_\Omega^{(\theta)}$ (\Fig{fig:flucA}e-f)
show a positive correlation with the magnetic field (correlation
coefficients 0.36, 0.21) with the overall variation being $\sim 75\%$
and $166\%$, respectively.
}

\begin{figure}[t]
\centering
\includegraphics[width=1.05\columnwidth]{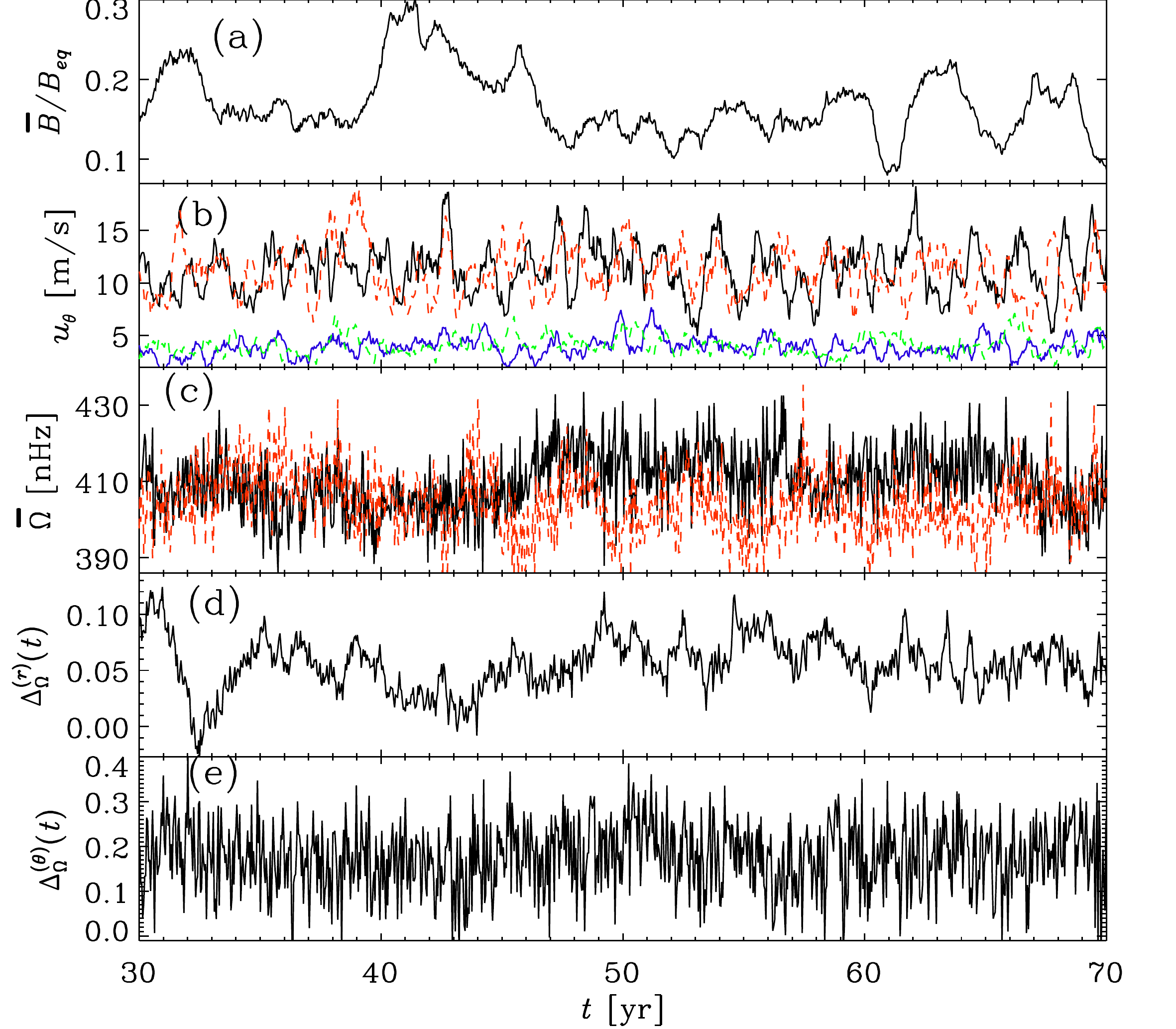}
\caption{Same as \Fig{fig:flucA}, but from Run~E which produces SL differential rotation.
}
\label{fig:flucB}
\end{figure}

We have seen that Runs~B, BC, and C produce clear \blue{activity cycles}
similar to Run~A and in all these runs we do see a \blue{corresponding}
variation in the flow.
However, for Runs~C, D, and E, which produce SL differential
rotation, the magnetic \blue{variations} are not so regular.
In \Fig{fig:flucB}, we show the temporal variations for the SL
differential rotation case Run~E.
We see that the large-scale magnetic field does not have a regular \blue{cycle}.
The meridional circulation does not appear to show a close correlation
with the magnetic field (\blue{with correlation coefficients being about
$-0.1$ and $-0.3$ for surface and bottom meridional circulation, respectively}).
However the significant variations (up to 
$45\%$) exist. Such irregular variation in
meridional circulation is found to be crucial in modeling
many aspects of the solar cycle in flux transport dynamo model \citep{KC11,KC13}.
The situation is similar in the case of differential rotation and rotational shear: the
early part of the time series ($t=30\ldots50$~yr) show an
anti-correlation with the
magnetic field strength, but at later times this correlation is not
so obvious \blue{(the overall linear correlation coefficients are
$-0.32$, $-0.22$, $-0.43$, $0.28$, $-0.55$ and $0.19$ 
for $\mean\Omega(0.95R_\odot,\pm32^\circ)$, $\mean\Omega(0.95R_\odot,0^\circ)$, $\mean\Omega(0.73R_\odot,0^\circ)$, $\Delta_\Omega^{(r)}$, and $\Delta_\Omega^{(\theta)}$, respectively).}
The variation in differential rotation is about $4\%$, whereas for 
the radial and latitudinal shear it is about $50\%$ and $60\%$, respectively.

Significant variations observed in the large-scale flows in all the simulations
motivate us to measure the Lorentz force.
The Lorentz force can change the flow by acting in two ways,
through large-scale and small-scale magnetic fields.
Firstly, it can act directly on the large-scale flow
which is known as `macro-feedback' (caused by the mean Lorentz force)
and has been applied in several mean-field models \citep[e.g.,][]{Sch79,BMT92}.
Secondly, it can affect the large-scale flow by affecting the convective 
motions and the best example of this is the magnetic quenching of the $\Lambda$-effect, which
is known as `micro-feedback' \citep[for an application, see][]{KAR99}.

\begin{figure}[t]
\centering
\includegraphics[width=0.490\columnwidth]{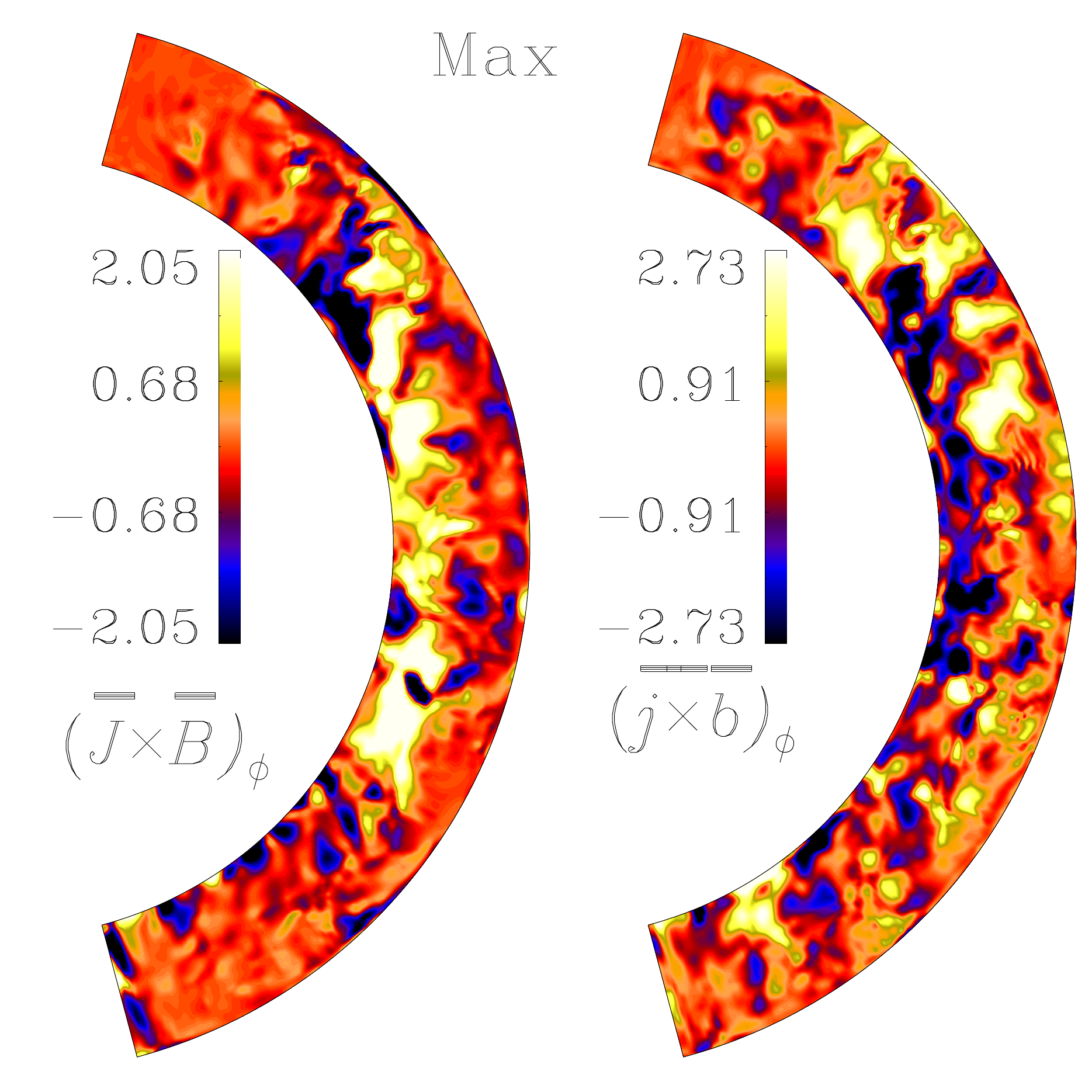}
\includegraphics[width=0.490\columnwidth]{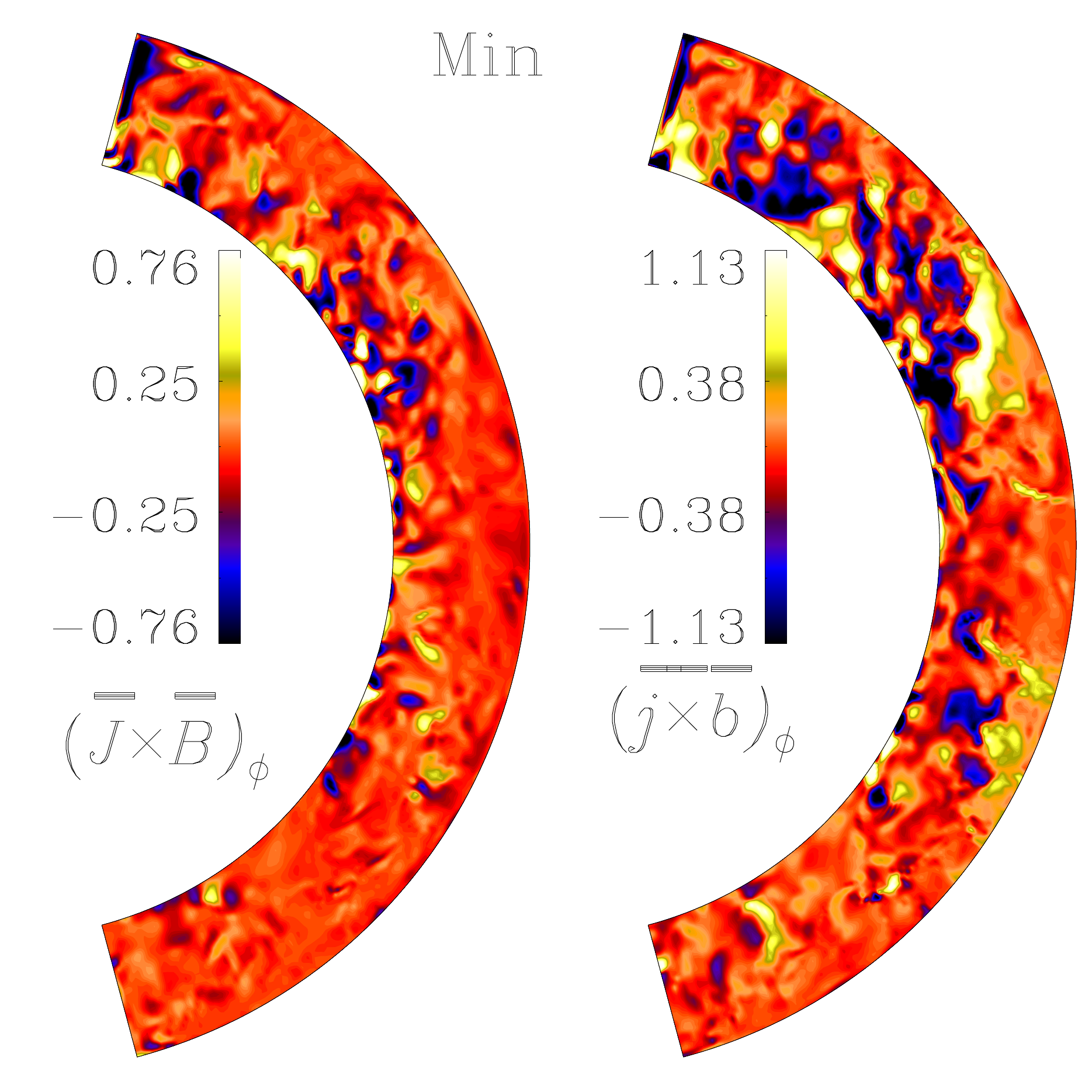}
\caption{From Run~A: Contributions of the $\phi$ component of the large-scale 
Lorentz force $\meanv{J} \times \meanv{B}$ and small-scale Lorentz force 
$\mean{{\bm j}\times{\bm b}}$ during a magnetic maximum (left two panels)
and minimum (right two panels).
Values are given in units of $10^{-9}$~N~m$^{-3}$.
}\label{fig:lorforce}
\end{figure}

To get an idea of these effects we measure the $\phi$ component of the Lorentz force
(which appears in the zonal momentum equation)
from the large-scale magnetic field $(\meanv{J}\times
\meanv{B})_\phi$, and the small-scale contribution $(\mean{{\bm j}
  \times {\bm b}})_\phi$ which are shown in \Fig{fig:lorforce}
both during magnetic maximum (left two panels) and minimum (right two panels) from Run~A.
We see that the small-scale Lorentz force, which enters into the
total stress and thus the $\Lambda$ effect, is typically stronger 
than the Lorentz force of the large-scale field and both have significant
temporal variations, becoming weaker during magnetic minimum.
Clearly both contributions are important, which was already
emphasized by \cite{BCRS13}, who discussed in detail the differences
in the integrated stresses for hydrodynamic and magnetic models in
the case of SL rotation.
Our results also show that both small-scale and large-scale contributions
to the Lorentz force are responsible
for producing temporal variations in the large-scale flows.
\cite{BCRS13} also emphasized the importance of magnetic fields in providing
coupling to the radiative interior which is absent in their hydrodynamic
models, but also in our magnetic models which lack the presence of a lower
overshoot layer.
Finally, we note that, while the Lorentz force from the mean field shows,
during magnetic maximum, a systematic variation with distance from the axis,
there are variations on similarly small scales both from the mean and
fluctuating fields.
This property is related to poor scale separation and may also
apply to real stars.

\section{Turbulent angular momentum transport}\label{sec:stress}

Numerical simulations have been used on various occasions to study
angular momentum transport in both hydrodynamic and magnetic cases,
but they usually focus on the regime of SL rotation \citep{BT02,BMT04,BCRS13}.
Furthermore, the contributions to the stress are often integrated over
latitude or radius.
\blue{In such representations, the stresses from opposite signs tend to cancel.}

\begin{figure*}[t]
\centering
\includegraphics[width=\textwidth]{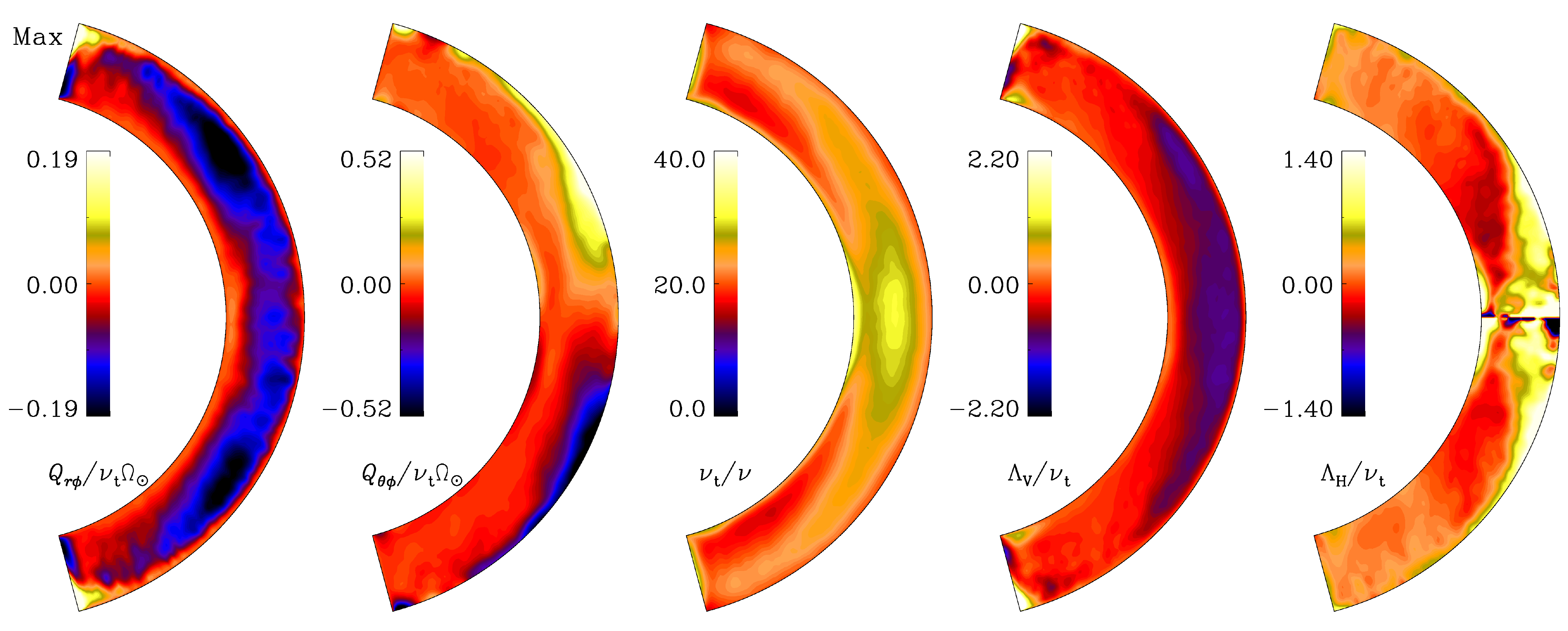}
\includegraphics[width=\textwidth]{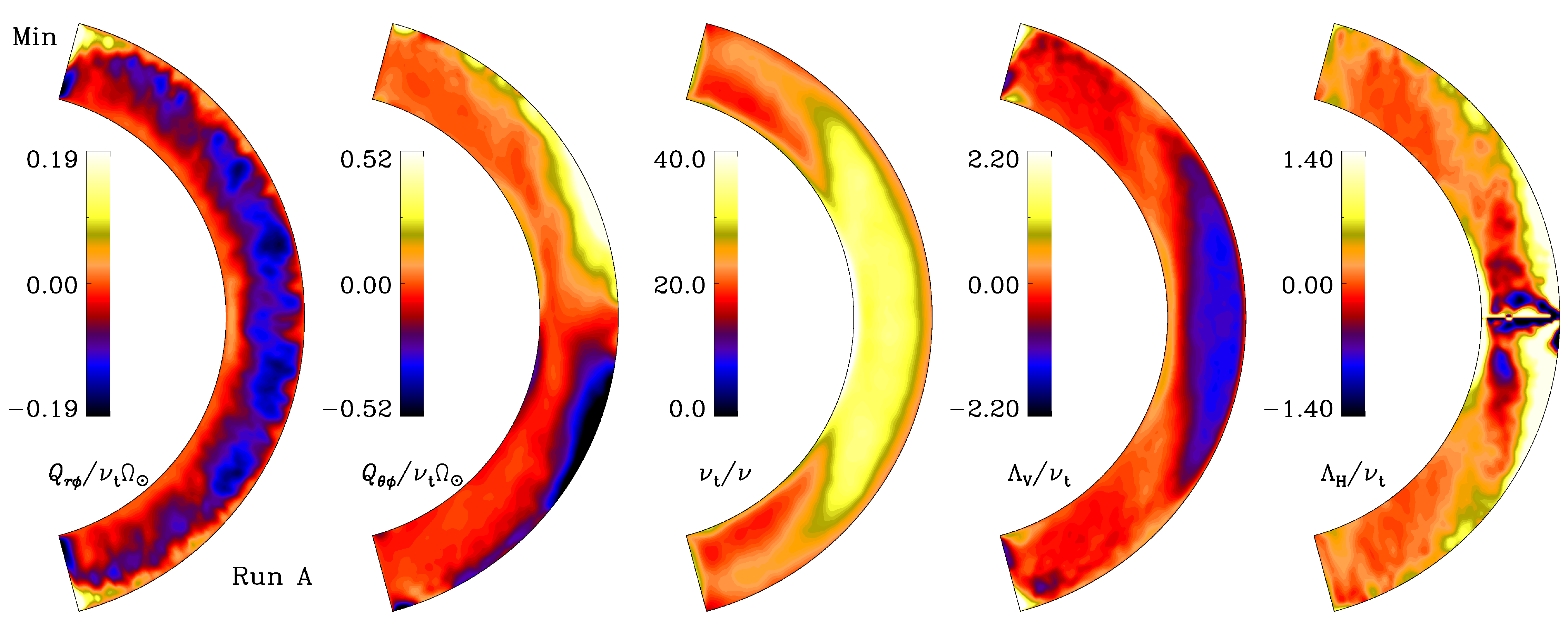}
\caption{Normalized profiles of $\qrp$, $\qtp$, $\nut$,
$\Lambda_{\rm V}$, and $\Lambda_{\rm H}$ for Run~A. 
The top and bottom panels show data averaged over four maxima and minima, 
respectively.
}\label{fig:lambA}
\end{figure*}

To make contact with mean-field theory invoking the $\Lambda$ effect
\citep{R80,R89}, it is useful to look at the profiles without integration
over latitude or radius and to separate between diffusive and nondiffusive
contributions.
Simulations have confirmed many aspects of $\Lambda$ effect in
mean field theory \citep{PTBNS93,RBMD94}.
The aim here is to interpret the differences between hydrodynamic and
magnetic cases in the AS and SL regimes in terms of corresponding changes
in the underlying $\Lambda$ effect.
For this purpose we first compute the contributions of the Reynolds stress
$\qij=\mean{u_i' u_j'}$, and the Maxwell stress
$\Mij=(\rho \mu_0)^{-1}\mean{B_i' B_j'}$ to the angular momentum balance in
the convection zone. Here, primes denote fluctuating quantities
which are calculated by subtracting the longitudinal mean from the
original quantity, e.g., $u_i' = u_i - \mean u_i$.
The radial and latitudinal angular momentum transports are determined by the
off-diagonal components of $\qij$ and $\Mij$, namely $\qrp$, $\qtp$,
$\mrp$ and $\mtp$, respectively.
In the mean-field theory of hydrodynamics, the Reynolds stress
contributions to angular momentum transport
are approximated in terms of the turbulent viscosity $\nut$ and the 
$\Lambda$-effect \citep{R80,R89}:
\begin{eqnarray}
\qrp &=& \mean{u_r' u_\phi'} \equiv \LamV \sin\theta\mean\Omega - \nut r \sin \theta \frac{\pd \mean\Omega}{\pd r}, \label{equ:Rrp} \\
\qtp &=& \mean{u_\theta' u_\phi'} \equiv \LamH \cos\theta\mean\Omega - \nut \sin \theta \frac{\pd \mean\Omega}{\pd \theta}. \label{equ:Rtp}
\end{eqnarray}
The coefficients $\LamV$ and $\LamH$ are the vertical and horizontal
$\Lambda$-effects, which are non-diffusive contributions to the
Reynolds stress that arise from the interaction of anisotropic
turbulence and rotation \citep{KR95}.
As in earlier work \citep{KMB14},
we adopt the mixing length formula to estimate $\nut$ 
\begin{equation}
\nut=\onethird \urms \alpha_{\rm MLT} H_p,
\end{equation}
where $\alpha_{\rm MLT}=1.7$ and $H_p (r) =-(\pd \ln p/\pd r)^{-1}$.
By computing $\urms=\urms(r,\theta)$ using $\phi$ averages, we get the
profile of $\nut$ in the meridional plane.

\begin{figure*}[t]
\centering
\includegraphics[width=\textwidth]{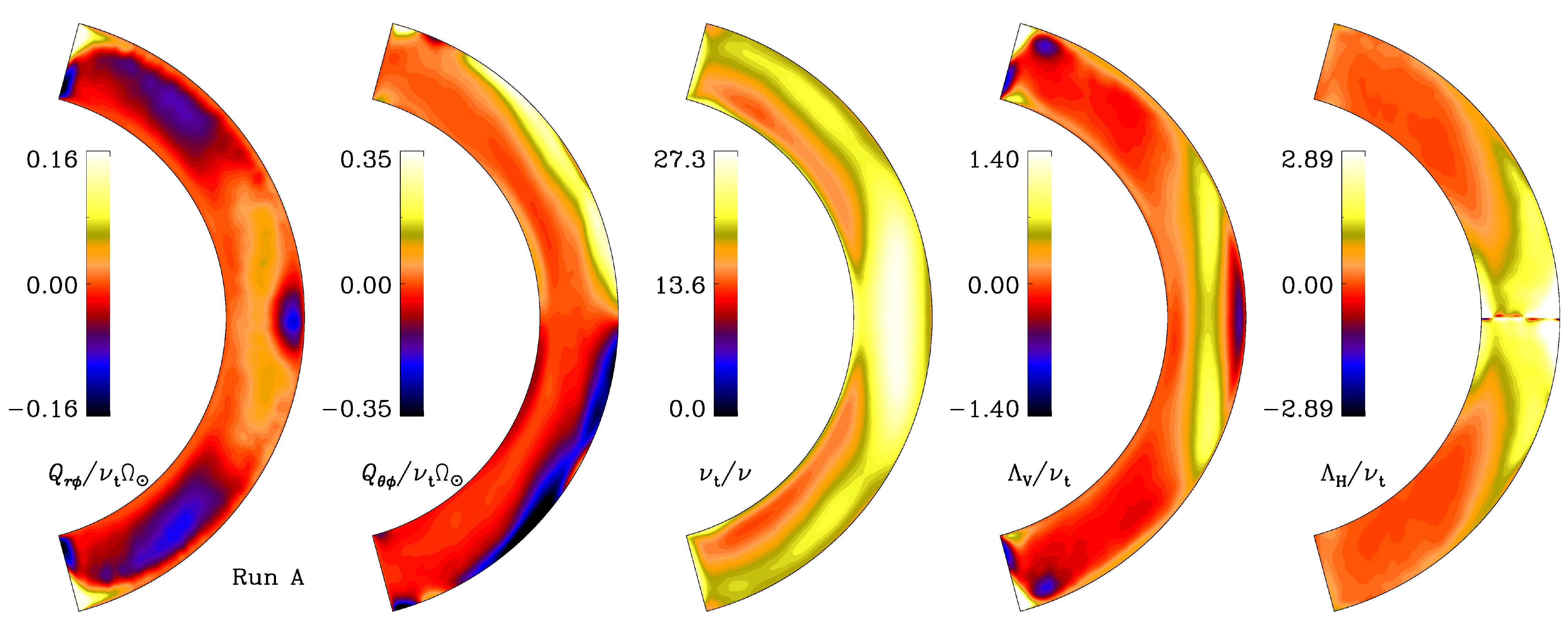}
\caption{Similar to \Fig{fig:lambA}, but for Run~E and time-averaged
over the last few maxima and minima.
}\label{fig:lambE}
\end{figure*}

The two leftmost panels of \Fig{fig:lambA} show $\qrp$ and $\qtp$
normalized by $\nut \Omega_0$, and the third panel shows $\nut$
normalized by the microphysical viscosity $\nu$ from Run~A.
We see that $\qrp$ is negative in most of the convection zone
which implies inward transport of angular momentum.
This is expected given the fact that the differential rotation in
this case is AS.
The negative $\qrp$ is in agreement with \cite{RBMD94} and \cite{KMB14}.
However for Run~E, which produces SL differential rotation, $\qrp$ is positive
at low latitudes.
On the other hand, the latitudinal stress $\qtp$ is positive (negative)
in the northern (southern) hemisphere,
which implies equatorward angular momentum transport.
This is true for both Runs~A and E (Figs.~\ref{fig:lambA} and \ref{fig:lambE}).
We note that the recent observation of $\qtp$ \citep{Hat13} finds the same sign.
 
The magnetic field is expected to change the total (Reynolds and Maxwell)
stress over the \blue{activity cycle}.
In fact, magnetic quenching of the Reynolds stress is known
to have important consequences in mean-field models,
particularly in connection with explaining the origin
of the torsional oscillation of the Sun or 
grand minima \cite[e.g.,][]{KPMT99,KAR99}. 
Therefore, to see the variations over the \blue{activity cycle}, we show
$\qrp$ and $\qtp$ both during maximum and minimum phases. 
Note that these are not computed from one snapshot, but from
averages over four maxima or minima phases.
By comparing top and bottom panels of \Fig{fig:lambA} we find that $\qrp$ is slightly stronger 
during magnetic maximum, although $\qtp$ is weaker.
\blue{
Even if we ignore the fact that our model is very far from that of
\citet{BCRS13}, a quantitative comparison of our results is not straightforward.
Moreover, \citet{BCRS13} show the temporal variations of fluxes integrated 
across spherical shells; see their definition of $I_r$ in Eq.~(22) and Figure~7.
During magnetic maximum, our $\qrp$ becomes stronger, whereas $I_r$ in \citet{BCRS13}
becomes weaker and its overall variation is higher than ours.
The temporal variations of Reynolds stresses show spatial coherence over all depths as seen in \citet{BCRS13}.
}

Next, we see in the middle panels of \Fig{fig:lambA} that $\nut$ is significantly weaker during 
maximum due to magnetic quenching.
The normalized value is around 40, which is similar to the value of $\Rey$.
We note that $\nut$ decreases towards high latitudes,
but increases towards deeper regions of the convection zone.

\begin{figure*}[t]
\centering
\includegraphics[width=0.32\textwidth]{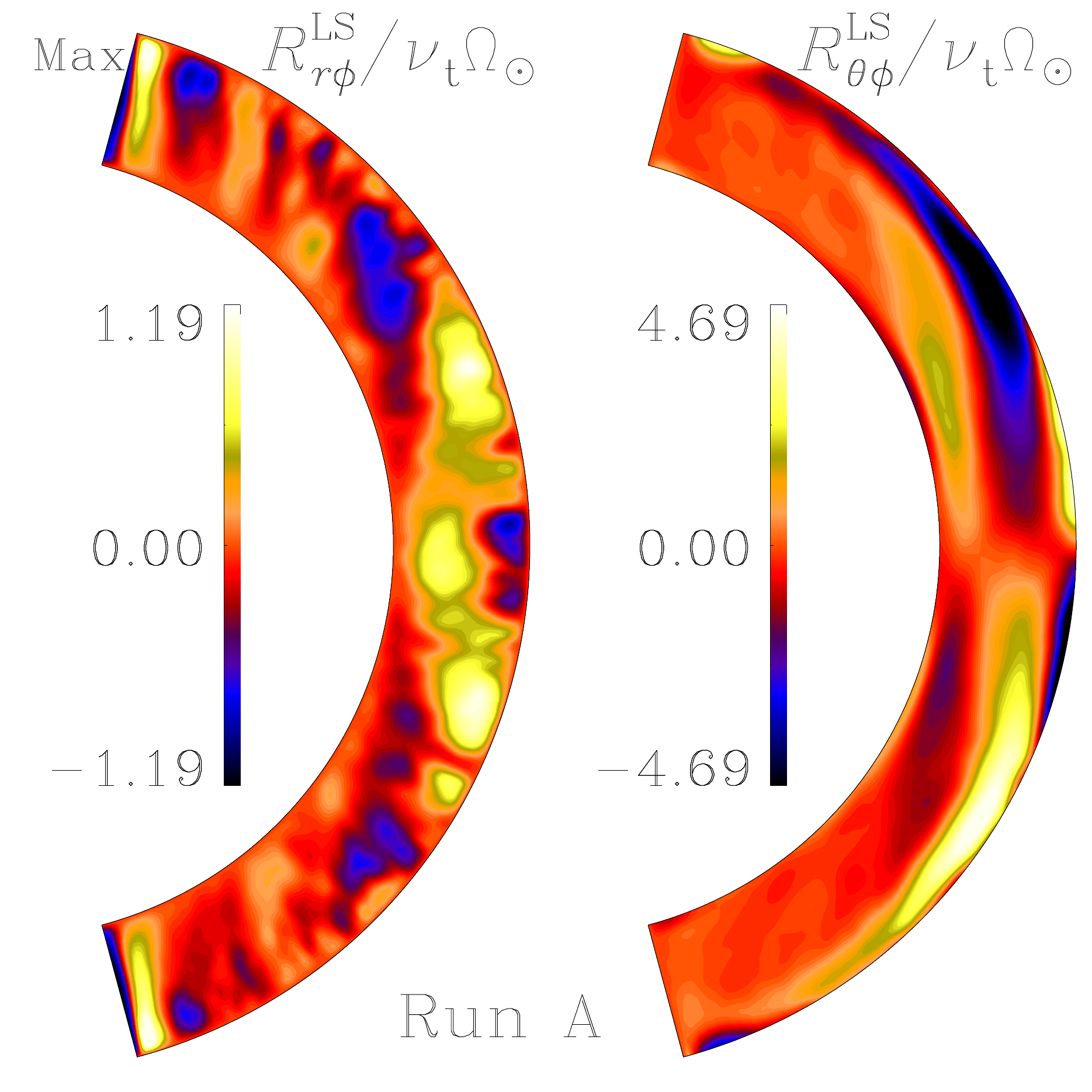}
\includegraphics[width=0.32\textwidth]{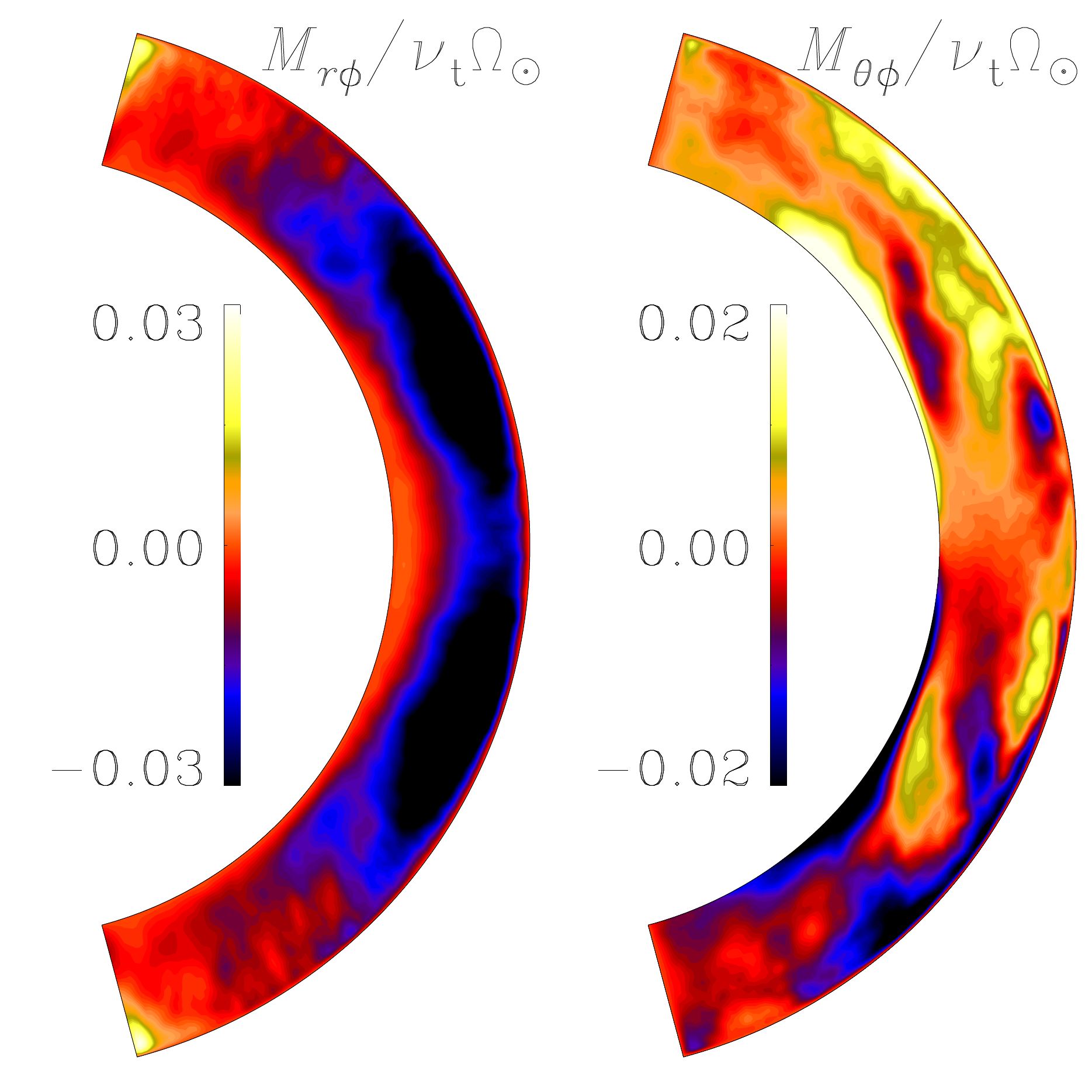}
\includegraphics[width=0.32\textwidth]{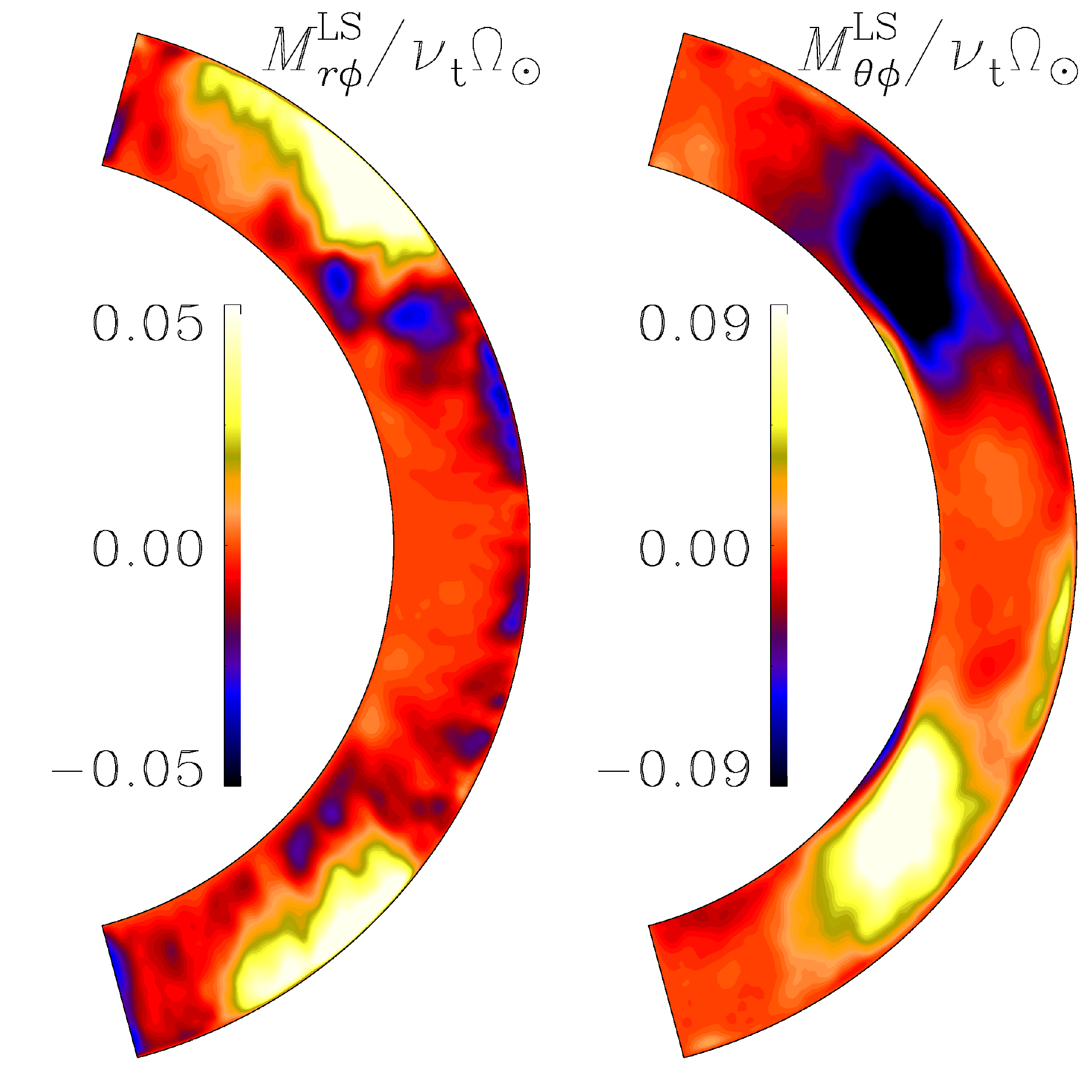}
\includegraphics[width=0.32\textwidth]{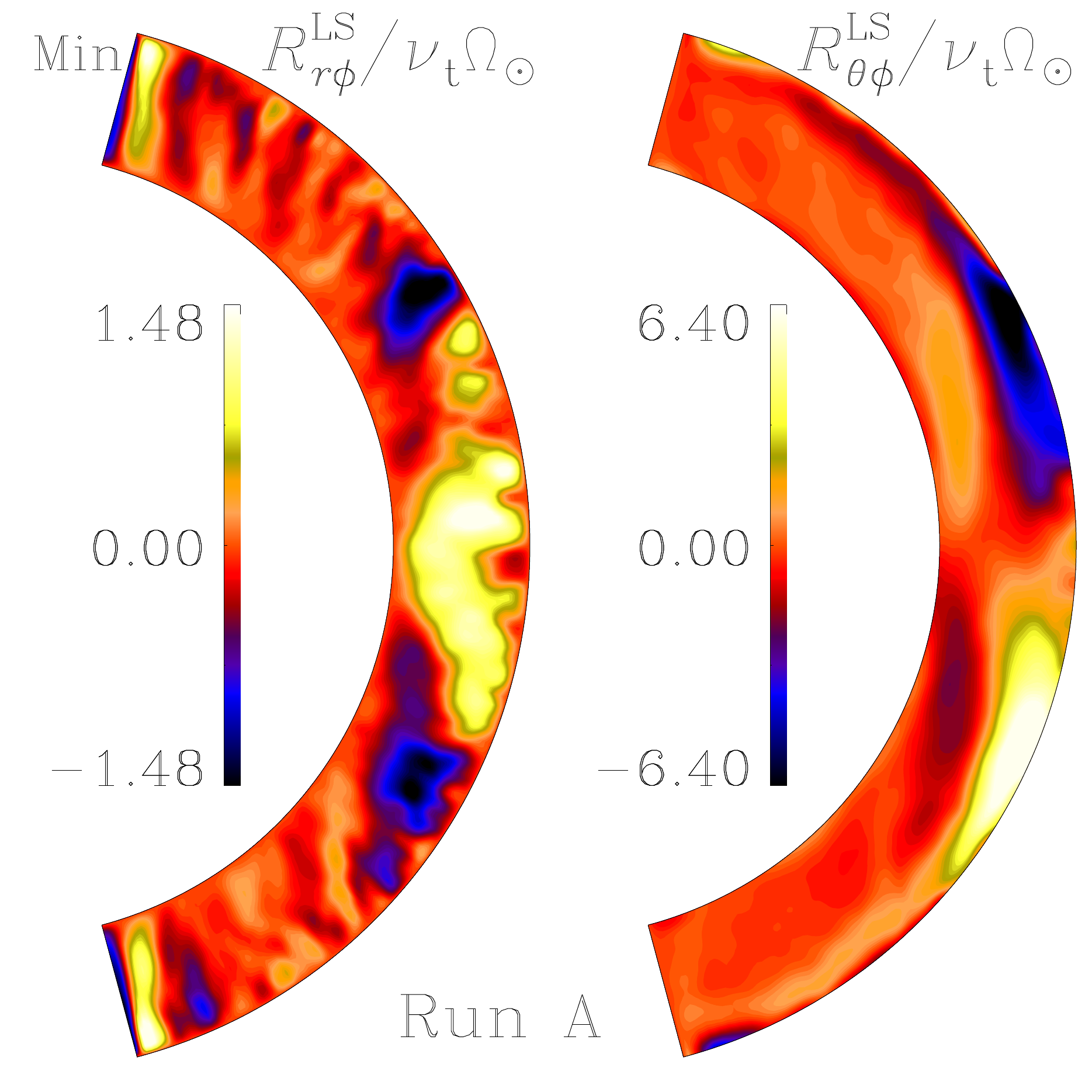}
\includegraphics[width=0.32\textwidth]{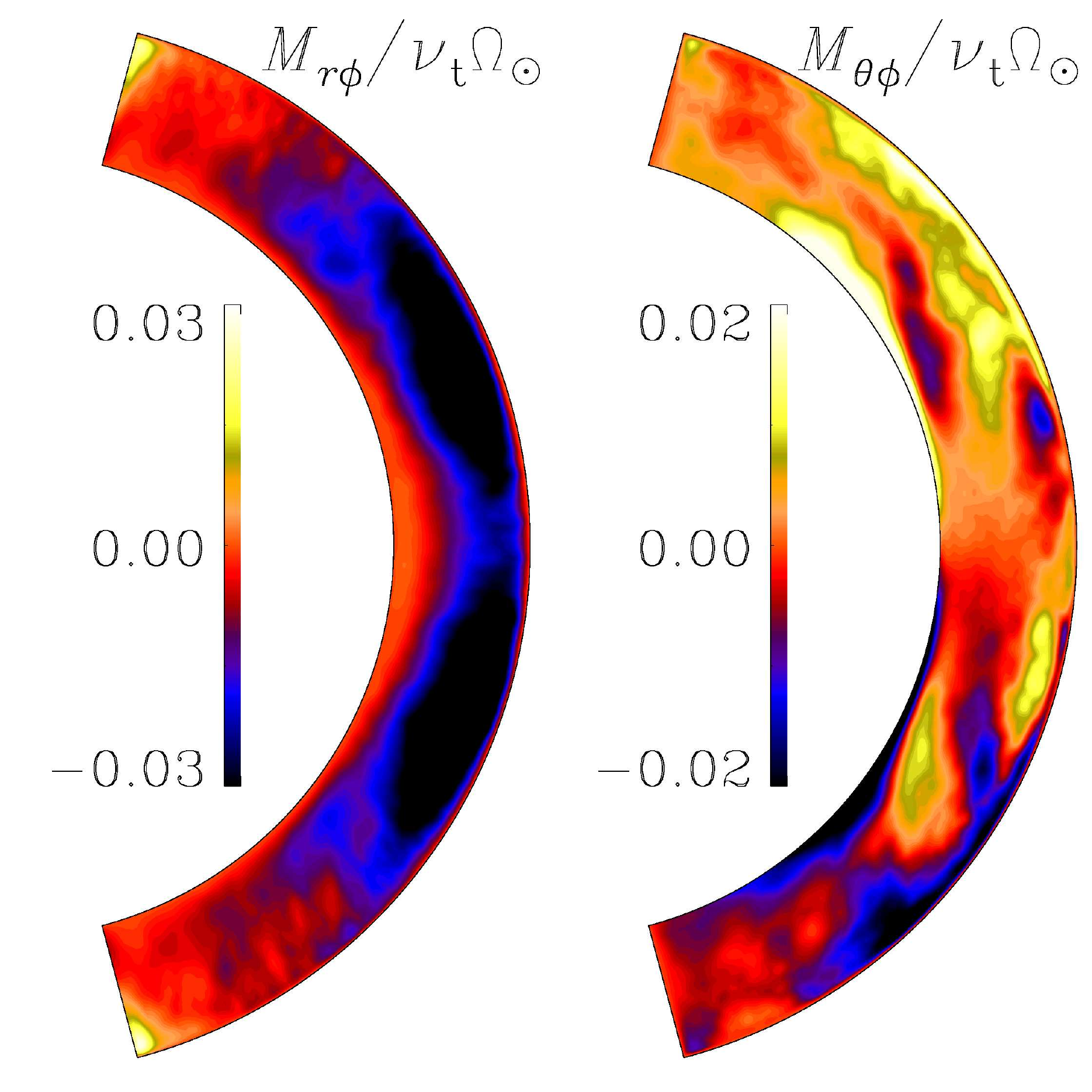}
\includegraphics[width=0.32\textwidth]{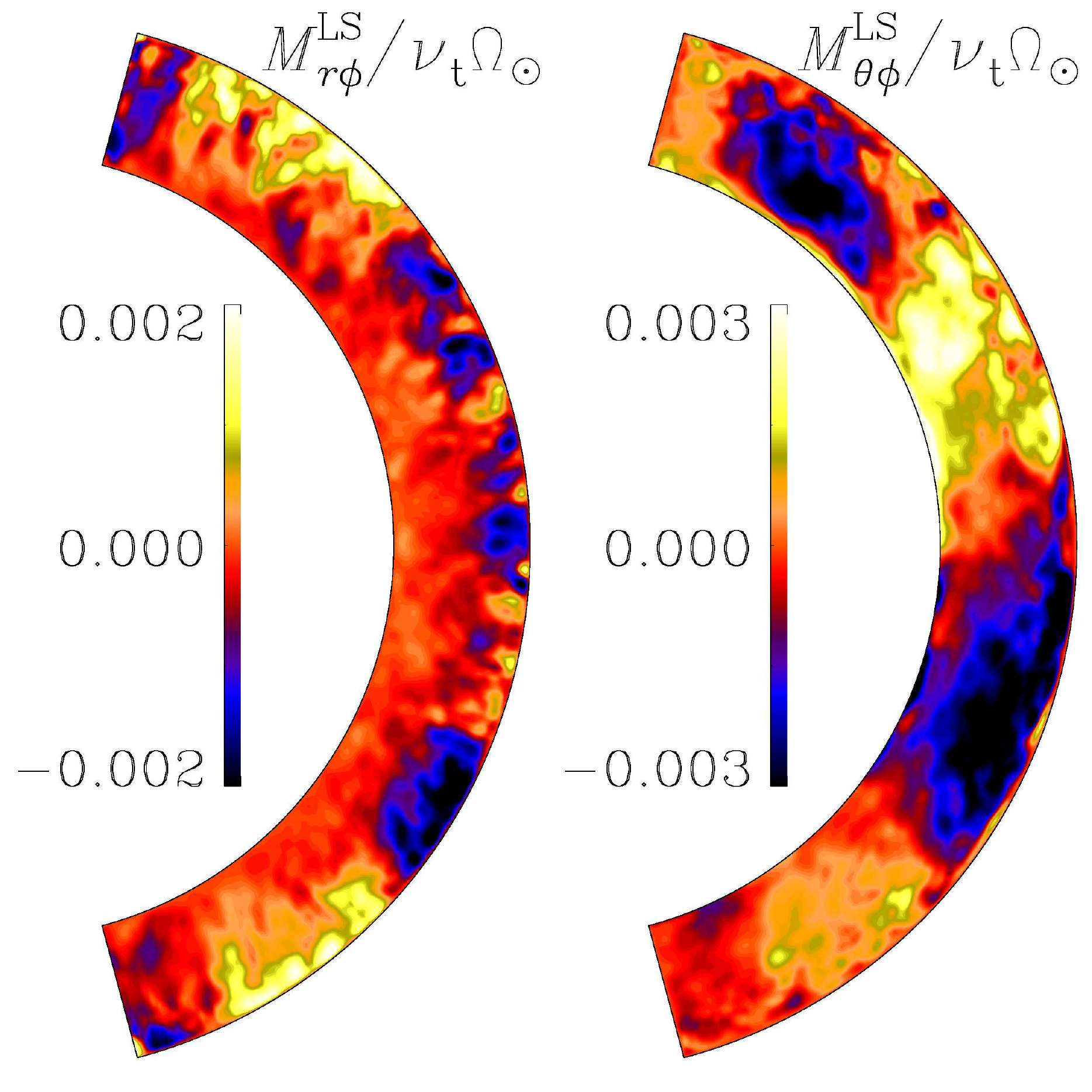}
\caption{
From Run~A: $R_{r\phi}^{\rm LS}=\mean{u}_r(\mean{u}_\phi+\Omega_0 r \sin\theta)$, 
$R_{\theta\phi}^{\rm LS}=\mean{u}_\theta(\mean{u}_\phi+\Omega_0 r \sin\theta)$ (left two panels), 
$\mrp=\mean{ B'_r B'_\phi}/\rho\mu_0$, $\mtp=\mean{B'_\theta B'_\phi}/\rho\mu_0$ (middle two),
$M_{r\phi}^{\rm LS}=\mean B_r \mean B_\phi/\rho\mu_0$ 
and $M_{\theta\phi}^{\rm LS}=\mean B_\theta \mean B_\phi/\rho\mu_0$ (right two), 
normalized by $\nut\Omega_\odot$. 
Upper and lower panels are during magnetic maximum and minimum, respectively.
}\label{fig:amfA}
\end{figure*}

\begin{figure*}[t]
\centering
\includegraphics[width=0.32\textwidth]{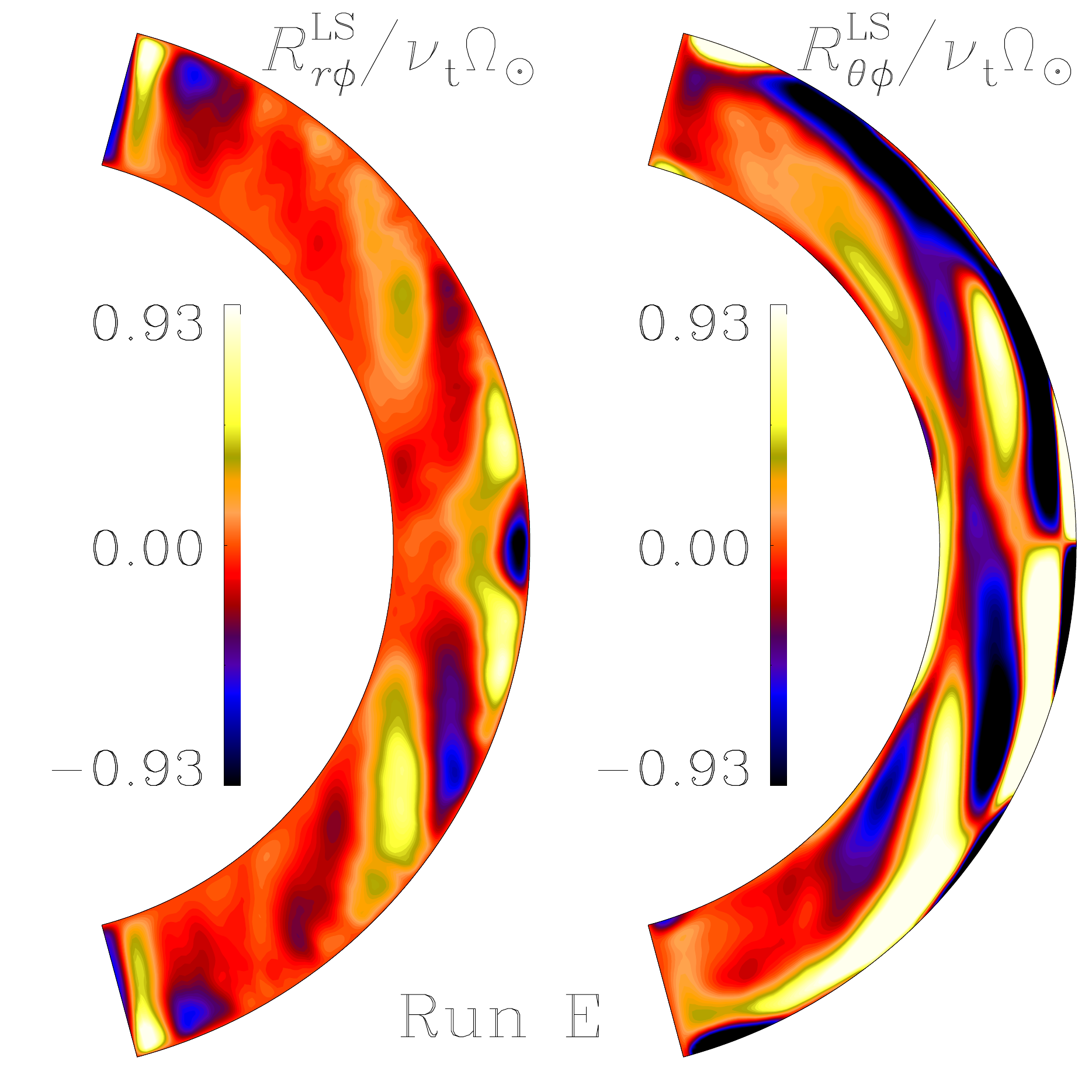}
\includegraphics[width=0.32\textwidth]{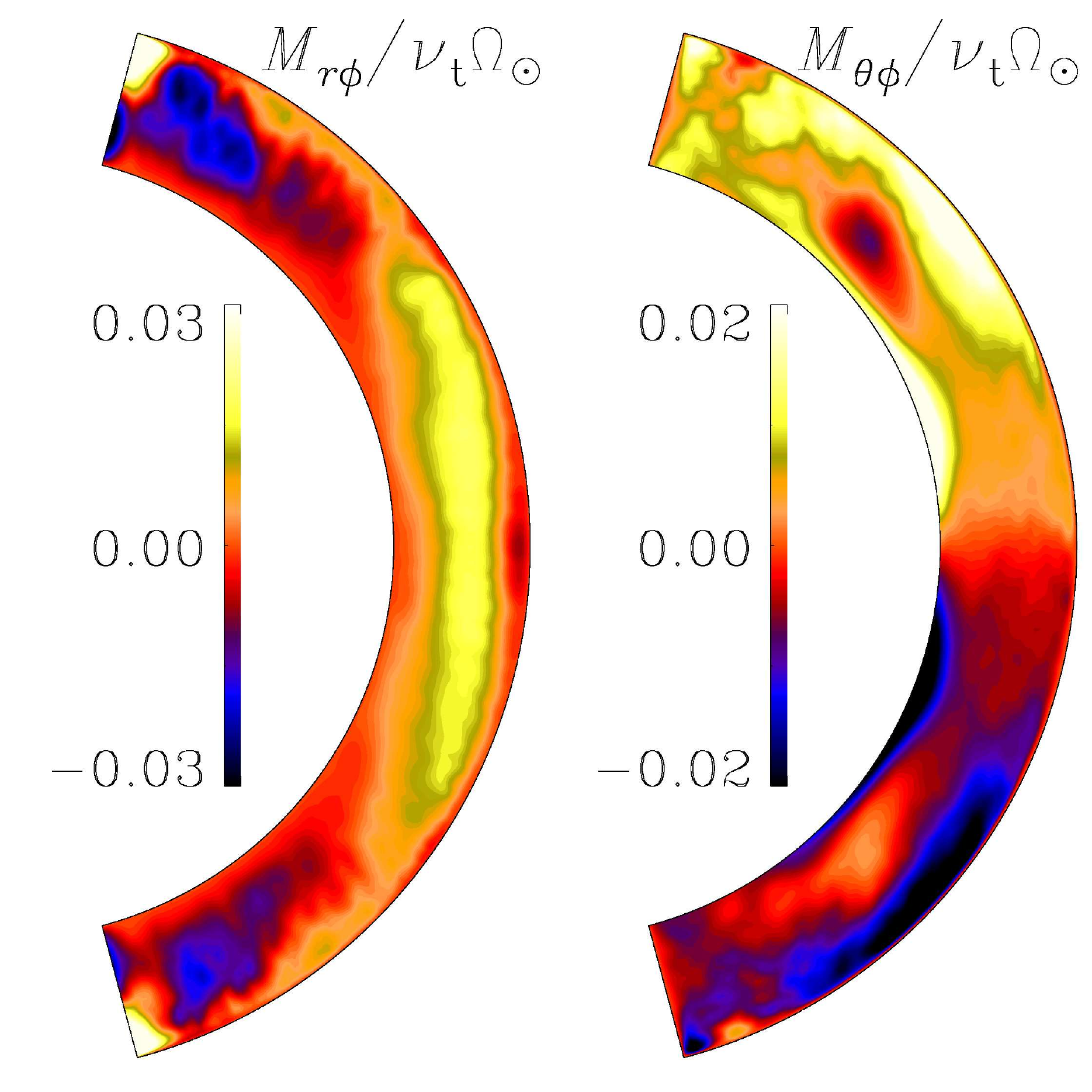}
\includegraphics[width=0.32\textwidth]{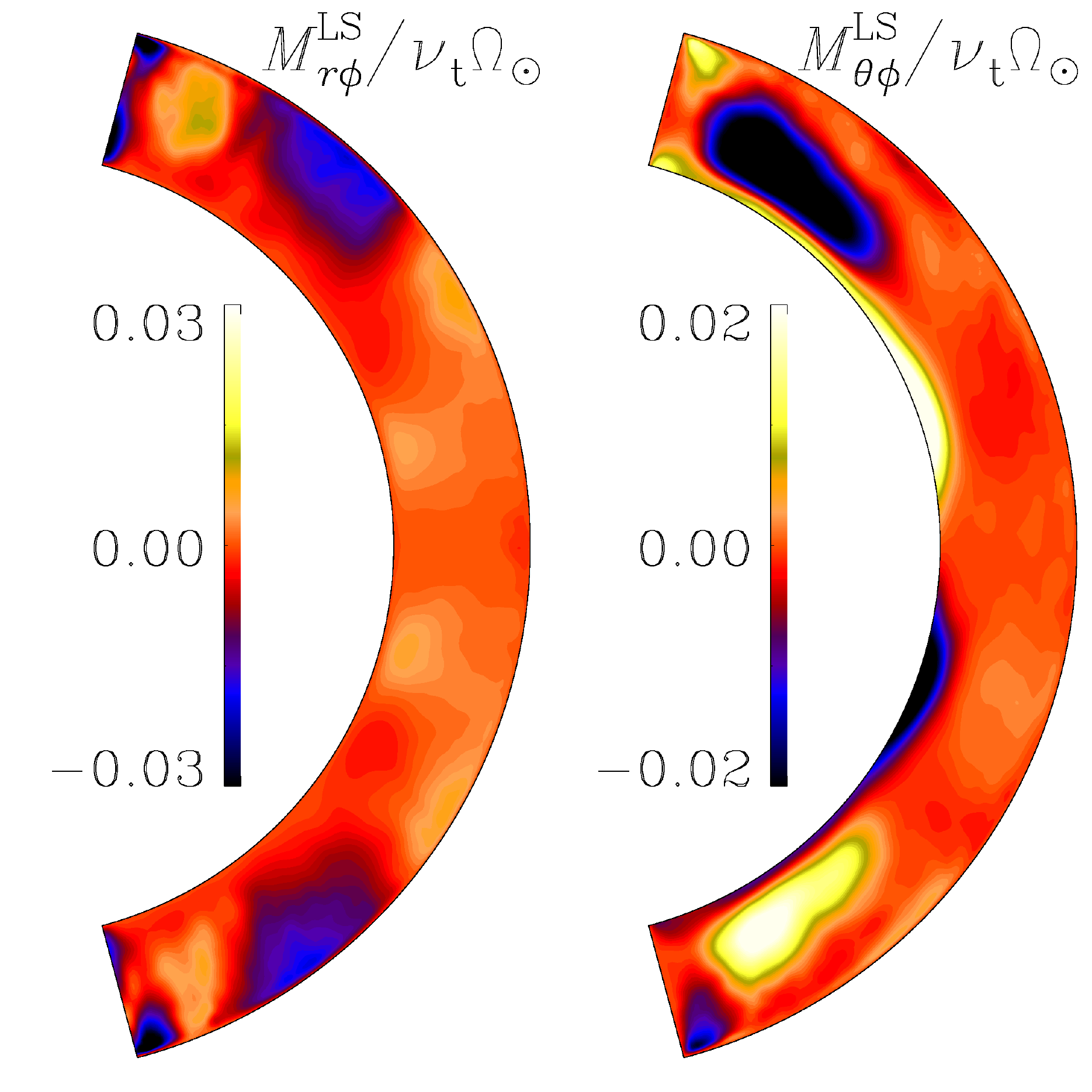}
\caption{
Same as \Fig{fig:amfA} but from Run~E and averaged over 
last few maxima and minima.
}\label{fig:amfE}
\end{figure*}

After solving \Eqs{equ:Rrp}{equ:Rtp} for $\LamV$ and $\LamH$
by using the computed values of $\qrp$, $\qtp$, and $\nut$,
we find $\LamV$ and $\LamH$
again during magnetic maximum and minimum.
These are shown in the last two panels of \Fig{fig:lambA}.
We see that $\LamV$ is negative in most of the convection zone,
which is in agreement with the
findings from the first-order smoothing approximation \citep{KR95,KR05},
and with earlier numerical studies
\citep[e.g.,][]{PTBNS93,KKT04,REZ05,KB08,KMB14}.
However, for Run~E with SL rotation, $\LamV$ is positive at low latitudes
and does not change much from maximum to minimum, which is why we show
in \Fig{fig:lambE} averages over all time.
Another important property is that $\LamV$ is of the same order as $\nut$,
which is consistent with earlier findings \citep{KBKSN10}.
We find that for both Runs~A and E, $\LamH$ is positive
which is again in agreement with the analytical results.
By comparing the present results with the hydrodynamic counterparts
of \cite{KMB14}, we find that the $\Lambda$-effect, is weaker.
This could be a reason for getting strongly suppressed differential rotation,
compared to the hydrodynamic simulations,
particularly in AS cases (Runs~A--BC).
Furthermore we see a significant
\blue{cycle-related} variation in the $\Lambda$-effect.
In AS cases, both $\LamV$ and $\LamH$ are
much smaller during maximum (see last two columns in \Fig{fig:lambA}). 
This could be one of the reasons for significant temporal variations in the large-scale flow.

Next we compute the radial and latitudinal components to the angular momentum transport
by the meridional circulation: $R_{r\phi}^{\rm LS}=\mean{u}_r(\mean{u}_\phi+\Omega_0 r \sin\theta)$, and 
$R_{\theta\phi}^{\rm LS}=\mean{u}_\theta(\mean{u}_\phi+\Omega_0 r \sin\theta)$. We show these in \Fig{fig:amfA}
for Run~A both during magnetic maximum and minimum.
Again, they are normalized by $\nut\Omega_\odot$
using the $\nut$ computed earlier.
We see that these stresses, particularly the latitudinal component, 
are the most dominating ones for transporting angular momentum. This is
because in this run we have strong meridional circulation, 
mostly poleward near surface and equatorward near the bottom. 
We also observe temporal modulations of transport due to meridional
circulation, which become stronger during magnetic minimum.
Therefore this temporal variation could lead to a variation in the differential rotation.
In fact, \citet{BCRS13} found significant temporal variation in the
angular momentum flux from meridional circulation
and suggested that it could be the primary source of the torsional oscillations.
\blue{
The temporal variations of $R_{r\phi}^{\rm LS}$ are not well-correlated spatially,
which is also seen in \citet{BCRS13} (see Fig.~7B).
}
For Run~E, which is shown in \Fig{fig:amfE},
we find smaller values of the stresses from the mean flows
than in Run~A, which is expected because
of a much weaker meridional circulation in this case.

In the middle four panels of \Fig{fig:amfA} we show the radial and latitudinal 
components of the Maxwell stress: 
$\mrp =\mean{ B'_r B'_\phi}/\rho\mu_0$ and $\mtp=\mean{B'_\theta B'_\phi}/\rho\mu_0$, 
respectively from Run~A, both during magnetic maximum and minimum.
We see that these are an order of magnitude smaller
than $\qrp$ and $\qtp$ and they have the same signs as $\qrp$ and $\qtp$,
respectively.
During magnetic maximum, both $\mrp$ and $\mtp$
(top middle panels of \Fig{fig:amfA}) 
are significantly larger than during magnetic minimum (lower middle panels).
Therefore the Maxwell stresses
may be another source of the temporal variation in the differential rotation.
\blue{
We recall that \citet{BCRS13} also find large variations of the 
Maxwell stress in their simulation with highly correlated spatial variations.
}
In Run~E, shown in middle panels of \Fig{fig:amfE}, we again find similar 
values of $\mrp$ and $\mtp$ and acting in the opposite to $\qrp$ and $\qtp$, 
respectively.

The Maxwell stress discussed above contributes to the redistribution of
angular momentum by fluctuating (non-axisymmetric) magnetic fields.
The mean axisymmetric field also contributes to the angular momentum balance 
\cite[see e.g.,][]{BMT04,BCRS13}.
This mean contribution is the result of correlations 
of the mean toroidal field with the radial and latitudinal 
components of the mean magnetic fields which are 
$M_{r\phi}^{\rm LS}=\mean B_r \mean B_\phi/\rho\mu_0$ and 
$M_{\theta\phi}^{\rm LS} =\mean B_\theta \mean B_\phi/\rho\mu_0$.
The two rightmost columns of panels of \Fig{fig:amfA} show these two
terms from Run~A,
both during magnetic maximum and minimum.
For Run~E, they are shown in the two rightmost panels of \Fig{fig:amfE}. 
Again we see that both components of the stress from the mean field
are much smaller than the Reynolds stresses 
$\qrp$ and $\qtp$ and comparable to the fluctuating contributions,
$\mrp$ and $\mtp$.
Importantly, they become much weaker during magnetic minimum.
\blue{
The variation in the large-scale magnetic tension is much larger than the variations in the other stresses
and also much larger than what \citet{BCRS13} found.
They concluded that the variation in the large-scale magnetic 
tension is not responsible for the torsional oscillations.
By comparing the top and bottom rows of the four rightmost panels of
\Fig{fig:amfA} we see that the temporal coherence of the spatial 
structures is not preserved during the activity cycle,
which was also observed by \citet{BCRS13}.
However, we must bear in mind that
here we are comparing our Run~A, which produces AS differential rotation,
with a SL case in \citet{BCRS13}. In our SL cases, there are
no prominent cycles and the temporal variations in the stresses are 
less pronounced than in Run~A.
}

\section{Conclusion}\label{sec:con}

Motivated by the recently discovered bistability of stellar
differential rotation in hydrodynamic spherical convection
simulations \citep{GYMRW14,KMB14}, we have performed several sets of magnetohydrodynamic simulations.
Except for the allowance of dynamo-generated magnetic fields,
our models are essentially the same as those of \cite{KMB14}.
By taking different radiative conductivities,
the convective velocities and hence the rotational influence
are varied in the simulations.
Runs~A, B, and BC ($\Co=1.34$, 1.35, and 1.38) produce AS differential rotation,
whereas Runs~C, D, and E ($\Co=1.44$, 1.67, and 1.75) produce SL
differential rotation.
When we take an AS (SL) rotation profile as initial
condition and perform a set of simulations by increasing (decreasing)
the radiative conductivity slowly, we find similar states of
differential rotation as in simulations which were started from scratch,
i.e., with an initially rigid rotation profile.
Therefore the bistable states of differential rotation
seem to disappear in the MHD simulations,
and we only find mono-stable solutions.

Besides the disappearance of the bistable differential rotation in MHD
simulations, we find several other new results.
(i) The abrupt transition from AS to SL rotation in hydrodynamic
simulations now seems to become more gradual and this transition happens
at slightly larger values of $\Roc$ ($\approx 0.66$) and thus smaller
values of $\Co$ ($\approx 1.4$).
This means that the magnetic field helps to produce SL differential rotation,
which is also in agreement with the recent study of \cite{FF14}.
(ii) The polar vortices or jet-like structures that were observed in
previous hydrodynamic simulations \citep[e.g.,][]{HA07,KMB14} are now absent.
(iii) Both differential rotation and meridional circulation are now
strongly suppressed in comparison to the hydrodynamic values and lie
within the observed range.
The suppression of these large-scale flows in the magnetic runs, 
particularly in AS differential rotation cases, could be a consequence of 
a reduction of the $\Lambda$ effect and the presence of the Lorentz force
of the magnetic field acting on the flow \citep{MP75}. 
(iv) The large-scale flows show significant time variation
as a consequence of the magnetic variations.
All cases with AS differential rotation (Runs~A--BC) show clear \blue{activity}
cycles and their large-scale flows have corresponding variations.
Run~A, which is more solar-like in terms of its
highest convective flux and lowest $\Co$ value, shows $\approx6\%$ 
variation in $\mean\Omega(r,\theta)$ about its mean.
However the variation in meridional circulation is as large as $60\%$
and in large-scale shear, $\Delta_\Omega^{(r)}$ and $\Delta_\Omega^{(\theta)}$, the variations are about $75\%$ and $160\%$, respectively.
All runs which produce SL differential rotation (Runs~C--E) also show
some magnetic variations, although they are not as regular and prominent
as for AS differential rotation.
In these runs we also see a detectable temporal variation in the large-scale flows
which is primarily caused by the variations of the large-scale (axisymmetric) magnetic tension, 
the Maxwell stresses, and the stress from the mean flow.
For Run~E, which has the smallest convective flux and SL differential rotation,
the variation in $\mean\Omega(r,\theta)$ is about $4\%$, whereas 
in both meridional circulation and large-scale shear the variation is 
about $50\%$.

Many authors simulate the large-scale flows in stellar convection
zones using hydrodynamical models assuming that the magnetic field does
not have a significant effect
\citep[to mention just a few]{BT02,BBT07,GWA13,GYMRW14,GSKM13,KMB14}. 
It is difficult to quantify the effects of dynamo-generated magnetic fields
on flows in the Sun from simulations as all the existing models are still far from the real Sun.
However, from observations \citep{CD01,HR10,ABC08} we do see significant
variations in both meridional circulation and rotational shear
as well as a small variation in differential rotation in the form of torsional oscillations,
which are believed to be (at least partially) coming from cyclic variations of the magnetic fields.
The present study now suggests that magnetic fields cannot be neglected
in simulating the large-scale flows in solar convection zone.

\begin{acknowledgements}
We thank an anonymous referee and J\"orn Warnecke for a careful reading 
the paper and for suggestions, which improved the presentation.
BBK wishes to thank the University of Helsinki for hospitality 
during the initiation of this work.
Financial support from the
Academy of Finland grants No.\ 136189, 140970, 272786 (PJK)
and 272157 to the ReSoLVE Centre of Excellence (MJK),
as well as the Swedish Research Council grants 621-2011-5076 and 2012-5797,
the Research Council of Norway under the FRINATEK grant 231444,
and the European Research Council under the AstroDyn Research Project
227952 are acknowledged as well as the HPC-Europa2 project, funded by
the European Commission - DG Research in the Seventh Framework
Programme under grant agreement No.\ 228398.
The computations have been carried out at the National Supercomputer
Centres in Link\"oping and Ume{\aa} and the Center for Parallel
Computers at the Royal Institute of Technology in Sweden, the Nordic
High Performance Computing Center in Iceland, and the supercomputers
hosted by CSC -- IT Center for Science in Espoo, Finland.
\end{acknowledgements}

\bibliographystyle{aa}
\bibliography{paper}

\begin{thebibliography}{91}
\expandafter\ifx\csname natexlab\endcsname\relax\def\natexlab#1{#1}\fi

\bibitem[{{Antia} {et~al.}(2008){Antia}, {Basu}, \& {Chitre}}]{ABC08}
{Antia}, H.~M., {Basu}, S., \& {Chitre}, S.~M. 2008, \apj, 681, 680

\bibitem[{{Augustson} {et~al.}(2013){Augustson}, {Brun}, {Miesch}, \&
  {Toomre}}]{ABMT13}
{Augustson}, K., {Brun}, A.~S., {Miesch}, M.~S., \& {Toomre}, J. 2013,
  arXiv:1310.8417

\bibitem[{{Ballot} {et~al.}(2007){Ballot}, {Brun}, \&
  {Turck-Chi{\`e}ze}}]{BBT07}
{Ballot}, J., {Brun}, A.~S., \& {Turck-Chi{\`e}ze}, S. 2007, \apj, 669, 1190

\bibitem[{{Beaudoin} {et~al.}(2013){Beaudoin}, {Charbonneau}, {Racine}, \&
  {Smolarkiewicz}}]{BCRS13}
{Beaudoin}, P., {Charbonneau}, P., {Racine}, E., \& {Smolarkiewicz}, P.~K.
  2013, \solphys, 282, 335

\bibitem[{{Brandenburg} {et~al.}(2005){Brandenburg}, {Chan}, {Nordlund}, \&
  {Stein}}]{BCNS05}
{Brandenburg}, A., {Chan}, K.~L., {Nordlund}, {\AA}., \& {Stein}, R.~F. 2005,
  AN, 326, 681

\bibitem[{{Brandenburg} {et~al.}(1996){Brandenburg}, {Jennings}, {Nordlund},
  {Rieutord}, {Stein}, \& {Tuominen}}]{BJNRST96}
{Brandenburg}, A., {Jennings}, R.~L., {Nordlund}, {\AA}., {et~al.} 1996,
  Journal of Fluid Mechanics, 306, 325

\bibitem[{{Brandenburg} {et~al.}(1992){Brandenburg}, {Moss}, \&
  {Tuominen}}]{BMT92}
{Brandenburg}, A., {Moss}, D., \& {Tuominen}, I. 1992, \aap, 265, 328

\bibitem[{{Brandenburg} {et~al.}(1998){Brandenburg}, {Saar}, \&
  {Turpin}}]{BST98}
{Brandenburg}, A., {Saar}, S.~H., \& {Turpin}, C.~R. 1998, \apjl, 498, L51

\bibitem[{{Brown} {et~al.}(2010){Brown}, {Browning}, {Brun}, {Miesch}, \&
  {Toomre}}]{BBBMT10}
{Brown}, B.~P., {Browning}, M.~K., {Brun}, A.~S., {Miesch}, M.~S., \& {Toomre},
  J. 2010, \apj, 711, 424

\bibitem[{{Brown} {et~al.}(2011){Brown}, {Miesch}, {Browning}, {Brun}, \&
  {Toomre}}]{BMBBT11}
{Brown}, B.~P., {Miesch}, M.~S., {Browning}, M.~K., {Brun}, A.~S., \& {Toomre},
  J. 2011, \apj, 731, 69

\bibitem[{{Brown} {et~al.}(1989){Brown}, {Christensen-Dalsgaard},
  {Dziembowski}, {Goode}, {Gough}, \& {Morrow}}]{Brownea89}
{Brown}, T.~M., {Christensen-Dalsgaard}, J., {Dziembowski}, W.~A., {et~al.}
  1989, \apj, 343, 526

\bibitem[{{Brun} {et~al.}(2004){Brun}, {Miesch}, \& {Toomre}}]{BMT04}
{Brun}, A.~S., {Miesch}, M.~S., \& {Toomre}, J. 2004, \apj, 614, 1073

\bibitem[{{Brun} \& {Palacios}(2009)}]{BP09}
{Brun}, A.~S. \& {Palacios}, A. 2009, \apj, 702, 1078

\bibitem[{{Brun} \& {Toomre}(2002)}]{BT02}
{Brun}, A.~S. \& {Toomre}, J. 2002, \apj, 570, 865

\bibitem[{{Chan}(2010)}]{Chan10}
{Chan}, K.~L. 2010, in IAU Symposium, Vol. 264, IAU Symposium, ed. A.~G.
  {Kosovichev}, A.~H. {Andrei}, \& J.-P. {Rozelot}, 219--221

\bibitem[{{Chatterjee} {et~al.}(2011){Chatterjee}, {Mitra}, {Rheinhardt}, \&
  {Brandenburg}}]{CMRB11}
{Chatterjee}, P., {Mitra}, D., {Rheinhardt}, M., \& {Brandenburg}, A. 2011,
  \aap, 534, A46

\bibitem[{{Chou} \& {Dai}(2001)}]{CD01}
{Chou}, D.-Y. \& {Dai}, D.-C. 2001, \apjl, 559, L175

\bibitem[{{Choudhuri} {et~al.}(1995){Choudhuri}, {Sch\"ussler}, \&
  {Dikpati}}]{CSD95}
{Choudhuri}, A.~R., {Sch\"ussler}, M., \& {Dikpati}, M. 1995, \aap, 303, L29

\bibitem[{{Cole} {et~al.}(2014){Cole}, {K{\"a}pyl{\"a}}, {Mantere}, \&
  {Brandenburg}}]{CKMB14}
{Cole}, E., {K{\"a}pyl{\"a}}, P.~J., {Mantere}, M.~J., \& {Brandenburg}, A.
  2014, ApJL, 780, L22

\bibitem[{{Dikpati} \& {Charbonneau}(1999)}]{DC99}
{Dikpati}, M. \& {Charbonneau}, P. 1999, \apj, 518, 508

\bibitem[{{Fan} \& {Fang}(2014)}]{FF14}
{Fan}, Y. \& {Fang}, F. 2014, \apj, 789, 35

\bibitem[{{Gastine} {et~al.}(2013){Gastine}, {Wicht}, \& {Aurnou}}]{GWA13}
{Gastine}, T., {Wicht}, J., \& {Aurnou}, J.~M. 2013, Icarus, 225, 156

\bibitem[{{Gastine} {et~al.}(2014){Gastine}, {Yadav}, {Morin}, {Reiners}, \&
  {Wicht}}]{GYMRW14}
{Gastine}, T., {Yadav}, R.~K., {Morin}, J., {Reiners}, A., \& {Wicht}, J. 2014,
  \mnras, 438, L76

\bibitem[{{Ghizaru} {et~al.}(2010){Ghizaru}, {Charbonneau}, \&
  {Smolarkiewicz}}]{GCS10}
{Ghizaru}, M., {Charbonneau}, P., \& {Smolarkiewicz}, P.~K. 2010, \apjl, 715,
  L133

\bibitem[{{Gilman}(1977)}]{Gi77}
{Gilman}, P.~A. 1977, Geophys. Astrophys. Fluid Dynam., 8, 93

\bibitem[{{Gilman}(1983)}]{Gi83}
{Gilman}, P.~A. 1983, \apjs, 53, 243

\bibitem[{{Guerrero} {et~al.}(2013){Guerrero}, {Smolarkiewicz}, {Kosovichev},
  \& {Mansour}}]{GSKM13}
{Guerrero}, G., {Smolarkiewicz}, P.~K., {Kosovichev}, A.~G., \& {Mansour},
  N.~N. 2013, \apj, 779, 176

\bibitem[{{Hathaway} \& {Rightmire}(2010)}]{HR10}
{Hathaway}, D.~H. \& {Rightmire}, L. 2010, Science, 327, 1350

\bibitem[{{Hathaway} {et~al.}(2013){Hathaway}, {Upton}, \& {Colegrove}}]{Hat13}
{Hathaway}, D.~H., {Upton}, L., \& {Colegrove}, O. 2013, Science, 342, 1217

\bibitem[{{Hazra} {et~al.}(2014){Hazra}, {Karak}, \& {Choudhuri}}]{HKC14}
{Hazra}, G., {Karak}, B.~B., \& {Choudhuri}, A.~R. 2014, \apj, 782, 93

\bibitem[{{Heimpel} \& {Aurnou}(2007)}]{HA07}
{Heimpel}, M. \& {Aurnou}, J. 2007, Icarus, 187, 540

\bibitem[{{Hotta} {et~al.}(2014){Hotta}, {Rempel}, \& {Yokoyama}}]{HRY14b}
{Hotta}, H., {Rempel}, M., \& {Yokoyama}, T. 2014, arXiv:1410.7093

\bibitem[{{Hotta} \& {Yokoyama}(2011)}]{HY11}
{Hotta}, H. \& {Yokoyama}, T. 2011, \apj, 740, 12

\bibitem[{{K{\"a}pyl{\"a}} \& {Brandenburg}(2008)}]{KB08}
{K{\"a}pyl{\"a}}, P.~J. \& {Brandenburg}, A. 2008, \aap, 488, 9

\bibitem[{{K{\"a}pyl{\"a}} {et~al.}(2010{\natexlab{a}}){K{\"a}pyl{\"a}},
  {Brandenburg}, {Korpi}, {Snellman}, \& {Narayan}}]{KBKSN10}
{K{\"a}pyl{\"a}}, P.~J., {Brandenburg}, A., {Korpi}, M.~J., {Snellman}, J.~E.,
  \& {Narayan}, R. 2010{\natexlab{a}}, \apj, 719, 67

\bibitem[{{K{\"a}pyl{\"a}} {et~al.}(2014){K{\"a}pyl{\"a}}, {K{\"a}pyl{\"a}}, \&
  {Brandenburg}}]{KMB14}
{K{\"a}pyl{\"a}}, P.~J., {K{\"a}pyl{\"a}}, M.~J., \& {Brandenburg}, A. 2014,
  \aap, 570, A43

\bibitem[{{K{\"a}pyl{\"a}} {et~al.}(2010{\natexlab{b}}){K{\"a}pyl{\"a}},
  {Korpi}, {Brandenburg}, {Mitra}, \& {Tavakol}}]{KKBMT10}
{K{\"a}pyl{\"a}}, P.~J., {Korpi}, M.~J., {Brandenburg}, A., {Mitra}, D., \&
  {Tavakol}, R. 2010{\natexlab{b}}, Astron. Nachr., 331, 73

\bibitem[{{K{\"a}pyl{\"a}} {et~al.}(2004){K{\"a}pyl{\"a}}, {Korpi}, \&
  {Tuominen}}]{KKT04}
{K{\"a}pyl{\"a}}, P.~J., {Korpi}, M.~J., \& {Tuominen}, I. 2004, \aap, 422, 793

\bibitem[{{K{\"a}pyl{\"a}} {et~al.}(2006){K{\"a}pyl{\"a}}, {Korpi}, \&
  {Tuominen}}]{KKT06}
{K{\"a}pyl{\"a}}, P.~J., {Korpi}, M.~J., \& {Tuominen}, I. 2006, Astron.
  Nachr., 327, 884

\bibitem[{{K{\"a}pyl{\"a}} {et~al.}(2011{\natexlab{a}}){K{\"a}pyl{\"a}},
  {Mantere}, \& {Brandenburg}}]{KMB11}
{K{\"a}pyl{\"a}}, P.~J., {Mantere}, M.~J., \& {Brandenburg}, A.
  2011{\natexlab{a}}, Astron. Nachr., 332, 883

\bibitem[{{K{\"a}pyl{\"a}} {et~al.}(2012){K{\"a}pyl{\"a}}, {Mantere}, \&
  {Brandenburg}}]{KMB12}
{K{\"a}pyl{\"a}}, P.~J., {Mantere}, M.~J., \& {Brandenburg}, A. 2012, \apjl,
  755, L22

\bibitem[{{K{\"a}pyl{\"a}} {et~al.}(2013){K{\"a}pyl{\"a}}, {Mantere}, {Cole},
  {Warnecke}, \& {Brandenburg}}]{KMCWB13}
{K{\"a}pyl{\"a}}, P.~J., {Mantere}, M.~J., {Cole}, E., {Warnecke}, J., \&
  {Brandenburg}, A. 2013, \apj, 778, 41

\bibitem[{{K{\"a}pyl{\"a}} {et~al.}(2011{\natexlab{b}}){K{\"a}pyl{\"a}},
  {Mantere}, {Guerrero}, {Brandenburg}, \& {Chatterjee}}]{KMGBC11}
{K{\"a}pyl{\"a}}, P.~J., {Mantere}, M.~J., {Guerrero}, G., {Brandenburg}, A.,
  \& {Chatterjee}, P. 2011{\natexlab{b}}, \aap, 531, A162

\bibitem[{{Karak}(2010)}]{Ka10}
{Karak}, B.~B. 2010, \apj, 724, 1021

\bibitem[{{Karak} \& {Choudhuri}(2011)}]{KC11}
{Karak}, B.~B. \& {Choudhuri}, A.~R. 2011, \mnras, 410, 1503

\bibitem[{{Karak} \& {Choudhuri}(2012)}]{KC12}
{Karak}, B.~B. \& {Choudhuri}, A.~R. 2012, \solphys, 278, 137

\bibitem[{{Karak} \& {Choudhuri}(2013)}]{KC13}
{Karak}, B.~B. \& {Choudhuri}, A.~R. 2013, Research in Astronomy and
  Astrophysics, 13, 1339

\bibitem[{{Karak} {et~al.}(2014){Karak}, {Kitchatinov}, \& {Choudhuri}}]{KKC14}
{Karak}, B.~B., {Kitchatinov}, L.~L., \& {Choudhuri}, A.~R. 2014, \apj, 791, 59

\bibitem[{{Kholikov} {et~al.}(2014){Kholikov}, {Serebryanskiy}, \&
  {Jackiewicz}}]{KSJ14}
{Kholikov}, S., {Serebryanskiy}, A., \& {Jackiewicz}, J. 2014, \apj, 784, 145

\bibitem[{{Kitchatinov} \& {Olemskoy}(2011)}]{KO11}
{Kitchatinov}, L.~L. \& {Olemskoy}, S.~V. 2011, \mnras, 411, 1059

\bibitem[{{Kitchatinov} {et~al.}(1999){Kitchatinov}, {Pipin}, {Makarov}, \&
  {Tlatov}}]{KPMT99}
{Kitchatinov}, L.~L., {Pipin}, V.~V., {Makarov}, V.~I., \& {Tlatov}, A.~G.
  1999, \solphys, 189, 227

\bibitem[{{Kitchatinov} \& {R\"udiger}(1995)}]{KR95}
{Kitchatinov}, L.~L. \& {R\"udiger}, G. 1995, \aap, 299, 446

\bibitem[{{Kitchatinov} \& {R{\"u}diger}(2004)}]{KR04}
{Kitchatinov}, L.~L. \& {R{\"u}diger}, G. 2004, Astron. Nachr., 325, 496

\bibitem[{{Kitchatinov} \& {R\"udiger}(2005)}]{KR05}
{Kitchatinov}, L.~L. \& {R\"udiger}, G. 2005, Astron. Nachr., 326, 379

\bibitem[{{Kitchatinov} {et~al.}(1994){Kitchatinov}, {R\"udiger}, \&
  {K\"uker}}]{KRK94}
{Kitchatinov}, L.~L., {R\"udiger}, G., \& {K\"uker}, M. 1994, \aap, 292, 125

\bibitem[{{Kovari} {et~al.}(2014){Kovari}, {Kriskovics}, {K{\"u}nstler},
  {Carroll}, {Strassmeier}, {Vida}, {Olah}, {Bartus}, \& {Weber}}]{KKKea14}
{Kovari}, Z., {Kriskovics}, L., {K{\"u}nstler}, A., {et~al.} 2014,
  arXiv:1411.1774

\bibitem[{{K{\"u}ker} {et~al.}(1999){K{\"u}ker}, {Arlt}, \&
  {R{\"u}diger}}]{KAR99}
{K{\"u}ker}, M., {Arlt}, R., \& {R{\"u}diger}, G. 1999, \aap, 343, 977

\bibitem[{{K{\"u}ker} \& {R{\"u}diger}(2011)}]{KR11}
{K{\"u}ker}, M. \& {R{\"u}diger}, G. 2011, Astron. Nachr., 332, 933

\bibitem[{{Lindborg} {et~al.}(2013){Lindborg}, {Mantere}, {Olspert}, {Pelt},
  {Hackman}, {Henry}, {Jetsu}, \& {Strassmeier}}]{lindborg2013}
{Lindborg}, M., {Mantere}, M.~J., {Olspert}, N., {et~al.} 2013, \aap, 559, A97

\bibitem[{{Malkus} \& {Proctor}(1975)}]{MP75}
{Malkus}, W.~V.~R. \& {Proctor}, M.~R.~E. 1975, J. Fluid Mech., 67, 417

\bibitem[{{Matt} {et~al.}(2011){Matt}, {Do Cao}, {Brown}, \& {Brun}}]{MCBB11}
{Matt}, S.~P., {Do Cao}, O., {Brown}, B.~P., \& {Brun}, A.~S. 2011, Astron.
  Nachr., 332, 897

\bibitem[{{Miesch} {et~al.}(2006){Miesch}, {Brun}, \& {Toomre}}]{MBT06}
{Miesch}, M.~S., {Brun}, A.~S., \& {Toomre}, J. 2006, \apj, 641, 618

\bibitem[{{Miesch} \& {Dikpati}(2014)}]{MD14}
{Miesch}, M.~S. \& {Dikpati}, M. 2014, \apjl, 785, L8

\bibitem[{{Miesch} \& {Hindman}(2011)}]{MH11}
{Miesch}, M.~S. \& {Hindman}, B.~W. 2011, \apj, 743, 79

\bibitem[{{Miesch} \& {Toomre}(2009)}]{MT09}
{Miesch}, M.~S. \& {Toomre}, J. 2009, Ann. Rev. Fluid Mech., 41, 317

\bibitem[{{Nelson} {et~al.}(2013){Nelson}, {Brown}, {Brun}, {Miesch}, \&
  {Toomre}}]{NBBMT13}
{Nelson}, N.~J., {Brown}, B.~P., {Brun}, A.~S., {Miesch}, M.~S., \& {Toomre},
  J. 2013, \apj, 762, 73

\bibitem[{{Nordlund} {et~al.}(1992){Nordlund}, {Brandenburg}, {Jennings},
  {Rieutord}, {Ruokolainen}, {Stein}, \& {Tuominen}}]{NBJRRST92}
{Nordlund}, A., {Brandenburg}, A., {Jennings}, R.~L., {et~al.} 1992, \apj, 392,
  647

\bibitem[{{Passos} {et~al.}(2012){Passos}, {Charbonneau}, \&
  {Beaudoin}}]{PCB12}
{Passos}, D., {Charbonneau}, P., \& {Beaudoin}, P. 2012, \solphys, 279, 1

\bibitem[{{Pelt}(1983)}]{pelt1983}
{Pelt}, J. 1983, in ESA Special Publication, Vol. 201, Statistical Methods in
  Astronomy, ed. E.~J. {Rolfe}, 37--42

\bibitem[{{Pipin} \& {Kosovichev}(2011)}]{PK11}
{Pipin}, V.~V. \& {Kosovichev}, A.~G. 2011, \apjl, 727, L45

\bibitem[{{Pulkkinen} {et~al.}(1993){Pulkkinen}, {Tuominen}, {Brandenburg},
  {Nordlund}, \& {Stein}}]{PTBNS93}
{Pulkkinen}, P., {Tuominen}, I., {Brandenburg}, A., {Nordlund}, A., \& {Stein},
  R.~F. 1993, \aap, 267, 265

\bibitem[{{Racine} {et~al.}(2011){Racine}, {Charbonneau}, {Ghizaru}, {Bouchat},
  \& {Smolarkiewicz}}]{RCGBS11}
{Racine}, {\'E}., {Charbonneau}, P., {Ghizaru}, M., {Bouchat}, A., \&
  {Smolarkiewicz}, P.~K. 2011, \apj, 735, 46

\bibitem[{{Rempel}(2005)}]{Re05}
{Rempel}, M. 2005, \apj, 622, 1320

\bibitem[{{Rempel}(2006)}]{Re06}
{Rempel}, M. 2006, \apj, 647, 662

\bibitem[{{Rieutord} {et~al.}(1994){Rieutord}, {Brandenburg}, {Mangeney}, \&
  {Drossart}}]{RBMD94}
{Rieutord}, M., {Brandenburg}, A., {Mangeney}, A., \& {Drossart}, P. 1994,
  \aap, 286, 471

\bibitem[{{R\"udiger}(1980)}]{R80}
{R\"udiger}, G. 1980, Geophys. Astrophys. Fluid Dynam., 16, 239

\bibitem[{{R\"udiger}(1989)}]{R89}
{R\"udiger}, G. 1989, {Differential Rotation and Stellar Convection. Sun and
  Solar-type Stars} (Berlin: Akademie Verlag)

\bibitem[{{R{\"u}diger} {et~al.}(2005){R{\"u}diger}, {Egorov}, \&
  {Ziegler}}]{REZ05}
{R{\"u}diger}, G., {Egorov}, P., \& {Ziegler}, U. 2005, Astron. Nachr., 326,
  315

\bibitem[{{Saar} \& {Brandenburg}(1999)}]{SB99}
{Saar}, S.~H. \& {Brandenburg}, A. 1999, \apj, 524, 295

\bibitem[{{Schad} {et~al.}(2013){Schad}, {Timmer}, \& {Roth}}]{STR13}
{Schad}, A., {Timmer}, J., \& {Roth}, M. 2013, \apjl, 778, L38

\bibitem[{{Schou} {et~al.}(1998){Schou}, {Antia}, {Basu}, {Bogart}, {Bush},
  {Chitre}, {Christensen-Dalsgaard}, {di Mauro}, {Dziembowski}, {Eff-Darwich},
  {Gough}, {Haber}, {Hoeksema}, {Howe}, {Korzennik}, {Kosovichev}, {Larsen},
  {Pijpers}, {Scherrer}, {Sekii}, {Tarbell}, {Title}, {Thompson}, \&
  {Toomre}}]{Schouea98}
{Schou}, J., {Antia}, H.~M., {Basu}, S., {et~al.} 1998, \apj, 505, 390

\bibitem[{{Schrinner} {et~al.}(2005){Schrinner}, {R{\"a}dler}, {Schmitt},
  {Rheinhardt}, \& {Christensen}}]{SRSRC05}
{Schrinner}, M., {R{\"a}dler}, K.-H., {Schmitt}, D., {Rheinhardt}, M., \&
  {Christensen}, U. 2005, Astron. Nachr., 326, 245

\bibitem[{{Schrinner} {et~al.}(2007){Schrinner}, {R{\"a}dler}, {Schmitt},
  {Rheinhardt}, \& {Christensen}}]{SRSRC07}
{Schrinner}, M., {R{\"a}dler}, K.-H., {Schmitt}, D., {Rheinhardt}, M., \&
  {Christensen}, U.~R. 2007, Geophys. Astrophys. Fluid Dynam., 101, 81

\bibitem[{{Sch\"ussler}(1979)}]{Sch79}
{Sch\"ussler}, M. 1979, \aap, 72, 348

\bibitem[{{Stellingwerf}(1978)}]{stellingwerf}
{Stellingwerf}, R.~F. 1978, \apj, 224, 953

\bibitem[{{Strassmeier} {et~al.}(2003){Strassmeier}, {Kratzwald}, \&
  {Weber}}]{SKW03}
{Strassmeier}, K.~G., {Kratzwald}, L., \& {Weber}, M. 2003, \aap, 408, 1103

\bibitem[{{Thompson} {et~al.}(2003){Thompson}, {Christensen-Dalsgaard},
  {Miesch}, \& {Toomre}}]{TCDMT03}
{Thompson}, M.~J., {Christensen-Dalsgaard}, J., {Miesch}, M.~S., \& {Toomre},
  J. 2003, \araa, 41, 599

\bibitem[{{Tobias} {et~al.}(2001){Tobias}, {Brummell}, {Clune}, \&
  {Toomre}}]{TBCT01}
{Tobias}, S.~M., {Brummell}, N.~H., {Clune}, T.~L., \& {Toomre}, J. 2001, \apj,
  549, 1183

\bibitem[{{Warnecke} {et~al.}(2013){Warnecke}, {K{\"a}pyl{\"a}}, {Mantere}, \&
  {Brandenburg}}]{WKMB13}
{Warnecke}, J., {K{\"a}pyl{\"a}}, P.~J., {Mantere}, M.~J., \& {Brandenburg}, A.
  2013, \apj, 778, 141

\bibitem[{{Weber} {et~al.}(2005){Weber}, {Strassmeier}, \& {Washuettl}}]{WSW05}
{Weber}, M., {Strassmeier}, K.~G., \& {Washuettl}, A. 2005, Astron. Nachr.,
  326, 287

\bibitem[{{Zhao} {et~al.}(2013){Zhao}, {Bogart}, {Kosovichev}, {Duvall}, \&
  {Hartlep}}]{ZBKDH13}
{Zhao}, J., {Bogart}, R.~S., {Kosovichev}, A.~G., {Duvall}, Jr., T.~L., \&
  {Hartlep}, T. 2013, \apjl, 774, L29

\end{thebibliography}

\end{document}